\documentclass[aps,prd,twocolumn,a4paper,superscriptaddress,10pt,floatfix]{revtex4-1}
\usepackage{graphicx}
\usepackage{multirow}
\usepackage{latexsym}
\usepackage{hyperref}
\usepackage[british]{babel}
\usepackage{amsmath}
\usepackage{amsmath}
\usepackage{xcolor}

\begin{document}

\title{Domain wall evolution beyond quartic potentials: The Sine-Gordon and Christ-Lee potentials}
\author{R. Heilemann}
\email{Ricarda.Heilemann@astro.up.pt}
\affiliation{Centro de Astrof\'{\i}sica da Universidade do Porto, Rua das Estrelas, 4150-762 Porto, Portugal}
\affiliation{Faculdade de Ci\^encias, Universidade do Porto, Rua do Campo Alegre, 4150-007 Porto, Portugal}
\affiliation{Instituto de Astrof\'{\i}sica e Ci\^encias do Espa\c co, Universidade do Porto, Rua das Estrelas, 4150-762 Porto, Portugal}
\author{M. C. Rosa}
\email{manuel.crosa@gmail.com}
\affiliation{Centro de Astrof\'{\i}sica da Universidade do Porto, Rua das Estrelas, 4150-762 Porto, Portugal}
\author{J. R. C. C. C. Correia}
\email{jose.correia@helsinki.fi}
\affiliation{Department of Physics and Helsinki Institute of Physics, University of Helsinki, PL 64, FI-00014 Finland}
\author{C. J. A. P. Martins}
\email{Carlos.Martins@astro.up.pt}
\affiliation{Centro de Astrof\'{\i}sica da Universidade do Porto, Rua das Estrelas, 4150-762 Porto, Portugal}
\affiliation{Instituto de Astrof\'{\i}sica e Ci\^encias do Espa\c co, Universidade do Porto, Rua das Estrelas, 4150-762 Porto, Portugal}

\date{\today}

\begin{abstract}
Domain walls are the simplest type of topological defects formed at cosmological phase transitions, and one of the most constrained. Their studies typically assume a quartic double well potential, but this model is not fully representative of the range of known or plausible particle physics models. Here we study the cosmological evolution of domain walls in two other classes of potentials. The Sine-Gordon potential allows several types of walls, interpolating between different pairs of minima (which demands specific numerical algorithms to separately measure the relevant properties of each type). The Christ-Lee potential parametrically interpolates between sextic and quartic behavior. We use multiple sets of simulations in two and three spatial dimensions, for various cosmological epochs and under various choices of initial conditions, to discuss the scaling properties of these networks. In the Sine-Gordon case, we identify and quantify deviations from the usual scaling behavior. In the Christ-Lee case, we discuss conditions under which walls form (or not), and quantify how these outcomes depend on parameters such as the energy difference between the false and true vacua and the expansion rate of the Universe. Various biased initial conditions are also addressed in appendices. Finally, we briefly comment on the possible cosmological implications of our results.
\end{abstract}
\maketitle
\section{\label{chap:intro}Introduction} 

The formation of domain structures is a known phenomenon in gauge theories with spontaneous symmetry breaking. In cosmology, this was first studied by Kibble \cite{kibble-topology}, who pointed out that defects necessarily form at cosmological phase transitions, due to the presence of a horizon. In many cases these defects can persist throughout the evolution of the Universe and leave fossil relic imprints---which may or may not be observationally desirable. A detailed understanding of their evolution and observational consequences is essential for any credible effort to understand the early universe. Detecting these objects will open a new window into the very early universe, yielding unique clues on fundamental physics and possibly string theory itself. Domain walls, resulting from the breaking of discrete symmetries, are the simplest and arguably the most tightly constrained defects.

Defect networks are highly non-linear objects, so their study requires a combination of careful analytic modeling and complex numerical simulations \cite{cosmic-strings-and-other}. The former relies on developing a model that captures the thermodynamics of the network. This idea was first implemented by Kibble \cite{kibble-onescale} for a model of string networks with a single macroscopic correlation length which characterizes the dynamics of the system. The current state of the art, and the model that we adopt in the present work, is the velocity-dependent one-scale (VOS) model \cite{VOS-book}, initially developed for cosmic strings \cite{vos-strings1} and later extended to other defects \cite{VOS-walls,vos-monopoles}. The model provides fully quantitative dynamical equations for both the characteristic length scale (or equivalently, under the model assumptions, the correlation length) and the root mean square (RMS) velocity of the network. Depending on the required level of detail, this model contains two free parameters in its simplest version, or six in the extended and far more accurate version. In either case, these parameters must be calibrated using reliable numerical simulations. For domain walls with a standard quartic potential ($V\propto \phi^4$) this calibration procedure is carried out using high-resolution field theory simulations in \cite{extending-vos-walls}.

This brings us to the second method of exploring the evolution of topological defects, using high-resolution field theory numerical simulations. Those relying on standard CPUs are limited by memory and time, implying that only the simplest models are simulated---for domain walls, this means the quartic (double well) potential. This choice has the advantage of numerical simplicity, but does not fully represent the broad range of fundamental or phenomenological particle physics models, for the simple reason that many physical systems have more than two equilibrium points. Moreover, the number of simulations is fairly small, and often they are done at low resolution. Recent developments led to a new generation of GPU-based codes, which for domain walls are typically 30 to 100 times faster than analogous CPU ones, enabling higher resolution simulations in manageable amounts of time, as well as statistical gains accruing from large numbers of simulations \cite{gpu-implementation}. Similar gains have also been demonstrated for cosmic strings \cite{GPUstrings,Correia2}.

Here we take advantage of these developments to carry out detailed simulations of cosmological networks of domain walls in two model classes beyond the simplest quartic potential. The first has the Sine-Gordon potential \cite{Widrow}, whose shape emerges from specific symmetry breaking mechanisms, such as the schizon models \cite{schizon-model}. Further mathematical aspects of such potentials are discussed in \cite{SG1,SG2,SG3,SG4,SG5,SG6,SG7,SG8}. The second class is a broader one, concerning models with sextic potentials. These have been considered in the context of both condensed matter and high energy particle physics, including as a natural extension of the Ginzburg-Landau model \cite{phi6-6,phi6-8,phi6-5, phi6-2,phi6-7,phi6-3,phi6-4,phi6-1,phi6-0,Dorey0,Weigel}. Our focus will be on the Christ-Lee potential \cite{chris-lee,Demirkaya}, which can parametrically interpolate between the sextic and quartic behaviors. Kink collisions in this model have recently been studied in \cite{Dorey}, while our focus is on the cosmological dynamics of such networks. For both model classes, most simulations are done in two spatial dimensions, which has the advantage of a larger dynamic range for the same memory usage (as compared to simulations in three spatial dimensions), although we also report on some 3D results.

We start in Sect. \ref{chap:numerics} with a brief overview of standard $V\propto \phi^4$ domain walls, covering analytic modeling and its numerical implementation. These also provide a benchmark against which our subsequent results can be compared. The Sine-Gordon potential is then introduced in Sect. \ref{chap:chap3}. In this case one may have different types of domain walls, whose classification depends on the initial conditions; we identify these types and discuss the scaling properties of the corresponding wall networks evolving in two spatial dimensions, specifically comparing them to the quartic potential case. In Sect. \ref{newsims} we explore the robustness of these results, by carrying out several additional sets of simulations with physically or numerically relevant alternative choices of initial conditions, expansion rates and numbers of dimensions. In sect. \ref{chap:sextic} we briefly consider the simplest choice of a sextic potential, which turns out to be mostly uninteresting, since it does not generically lead to the formation of domain walls. Conversely, the Christ-Lee potential has a much richer phenomenology, which we explore in Sect. \ref{chap:christlee}, focusing on the impact of damping mechanisms of false vacuum decay. Finally, Sect. \ref{chap:conc} presents our conclusions. Some particular choices of biased initial conditions, for both model classes, are provided in two appendices.

\section{\label{chap:numerics}Standard walls: Analytic and numerical description} 

We start with a brief overview of the standard analytic and numerical methods for studying cosmological domain wall networks. This is not meant to be exhaustive, but only to introduce the key concepts and formalism, as well as to provide a benchmark against which we discuss the results of the Sine-Gordon and Christ-Lee cases, which are the focus of the rest of the article.

Let us consider a model with a real scalar field, with a Lagrangian
\begin{equation}
\label{real-scalar-lagrangian}
    \mathcal{L} = \frac{1}{2} \partial_\mu \phi \partial^\mu \phi - V\left( \phi \right)
\end{equation}
where the potential $V\left(\phi\right)$ has a discrete set of degenerate minima. Standard variational methods lead to the field equation of motion, expressed as a function of conformal time $\eta$
\begin{equation}
    \label{eom-conformal}
    \frac{\partial^2 \phi}{\partial \eta^2} + 2\left(\frac{d\mathrm{ln}a}{d\mathrm{ln}\eta}\right) \frac{1}{\eta} \frac{\partial\phi}{\partial \eta} - \nabla^2\phi = - a^2 \frac{\partial V}{\partial \phi}
\end{equation}
where the damping term $d\mathrm{ln}a/d\mathrm{ln}\eta$ depends solely on the rate of expansion of the universe. For a power-law evolution of the Universe given by $a(t)\propto t^\lambda$ we have
\begin{equation}
    \frac{d\mathrm{ln}a}{d\mathrm{ln}\eta} = \frac{\lambda}{1-\lambda}\,.
\end{equation}

The evolution of a domain wall network can be analytically described by the VOS model \cite{VOS-walls,extending-vos-walls}. This starts from the microscopic equations of motion for the walls and carries out a statistical average for the system assuming that the defects are randomly distributed at large scales. This leads to a nonlinear system for the evolution of the network density $\rho$, or equivalently the network characteristic length or correlation length $L$ (the two being related via $\rho=\sigma/L$, where $\sigma$ is the wall mass per unit area), and the RMS velocity of the walls $v$. In the simplest (two parameter) version of the model, this leads to the following evolution equations
\begin{eqnarray}
\frac{dL}{dt}&=&(1+3v^2)HL + c_w v \label{vos-eq1}\\
\frac{dv}{dt}&=&(1-v^2)\left(\frac{k_w}{L}-3Hv\right)\,, \label{vos-eq2}\\
\end{eqnarray}
where $c_w$ and $k_w$ are the model's energy loss and curvature parameters. In this work we will make the simplifying assumption that these two parameters are constant; while this is not fully accurate, it is sufficient for the (comparatively) small numerical simulations carried out in this preliminary study. A detailed description of the physical interpretation of the model parameters, together with their behavior in an extended version of the VOS model (suitable for higher resolution simulations, as well as any credible observational predictions) can be found in \cite{extending-vos-walls}.

Moreover, we neglect the effect of the energy density of the walls on the background (specifically, on the Friedmann equations), and assume homogeneous and isotropic universes. Even though a cosmological wall network energy will at some point end up dominating the universe \cite{cosmic-strings-and-other,VOS-book}, the above is a good approximation in the context of the numerical simulations we report on, which assume a constant expansion rate $a \propto t^\lambda$, as further discussed below. Under these assumptions the VOS model attractor is a linear scaling solution
\begin{eqnarray}
    \label{length-scaling-eq}
    L &=& \sqrt{\frac{k_w\left(k_w+c_w\right)}{3\lambda\left(1-\lambda\right)}} \, t\\
    \label{vel-scaling-eq}
    v^2 &=& \frac{1-\lambda}{3\lambda} \frac{k_w}{k_w+c_w}\,.
\end{eqnarray}
Since $c_w$ and $k_w$ are free parameters of the model, they need to be calibrated numerically using either bootstrapping techniques or more robust methods such as MCMC \cite{MCMC-calib,Correia2}. 

\subsection{\label{sec:phi4}The quartic potential}

\begin{figure*}
  \begin{center}
    \leavevmode
    \includegraphics[width=1.0\columnwidth]{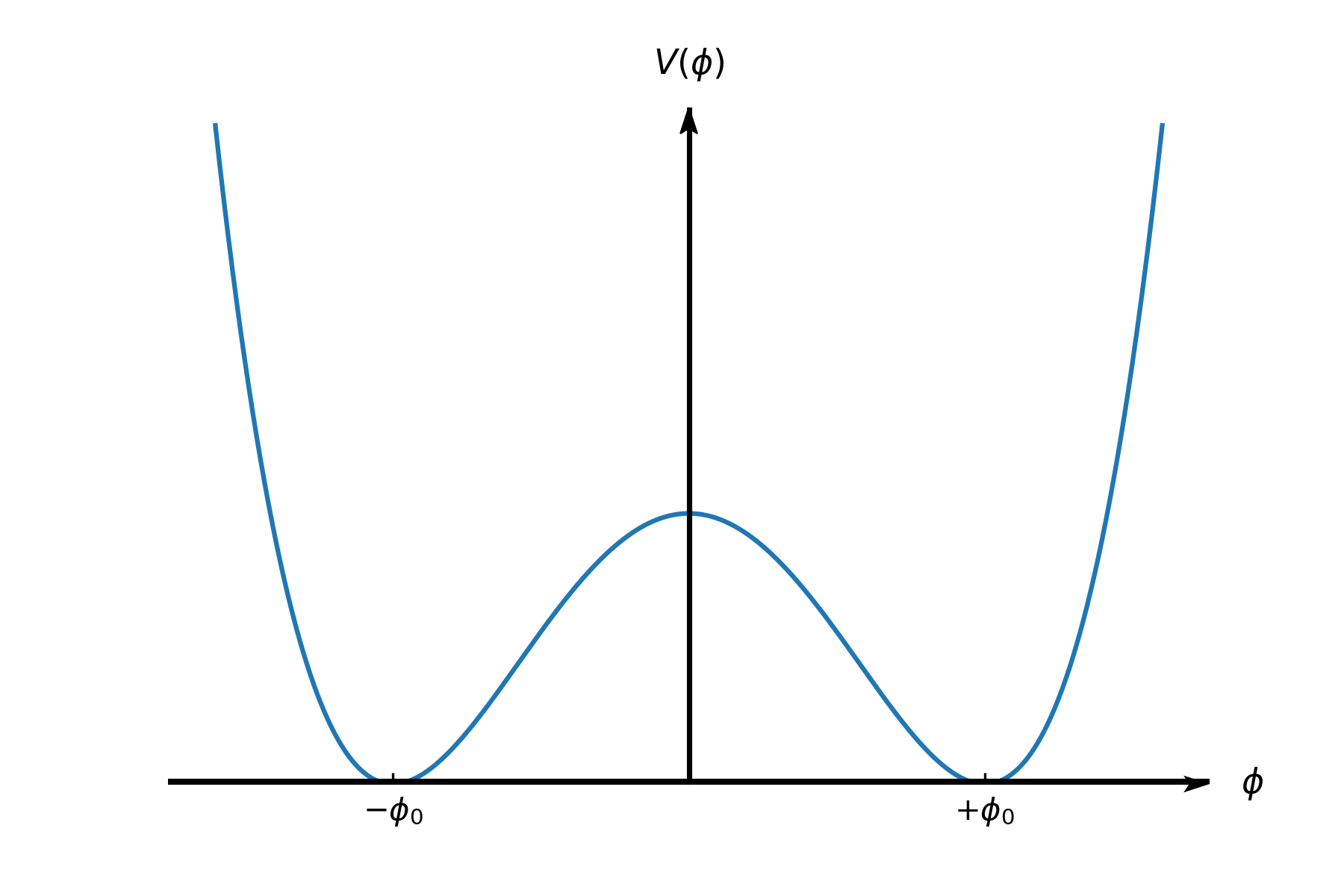}
    \includegraphics[width=1.0\columnwidth]{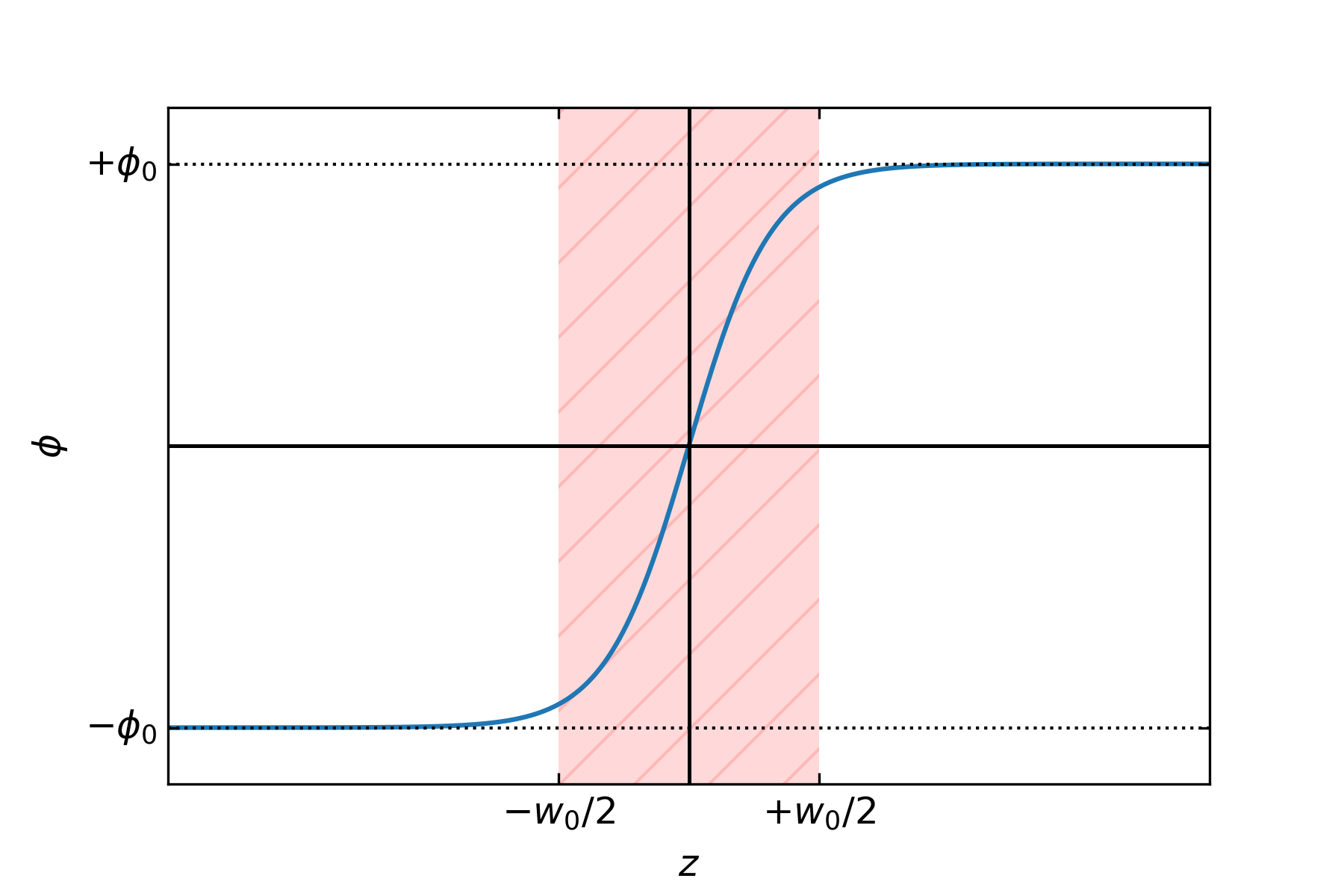}
    \caption{Left panel: General shape of the canonical (quartic) domain wall potential. Right panel: Shape of a planar wall at rest with a localized kink at $z=0$ and a thickness of $w_0$ that interpolates between the two minima localized at $\phi=\pm\phi_0$.}
    \label{fig01}
  \end{center}
\end{figure*}

The prototypical model for numerical studies of cosmological domain walls has a quartic potential
\begin{equation}
    \label{phi4-potential}
     V\left( \phi \right) = V_0 \left(1-\frac{\phi^2}{\phi_0^2}\right)^2\,.
\end{equation}
Here $\phi_0$ is the field value at the potential energy minimum and $V_0$ is the height of the potential barrier; for subsequent comparison, the general shape of the potential is depicted in the left panel of Fig.~\ref{fig01}. 

\begin{figure*}
\centering
    \includegraphics[width=1.0\columnwidth]{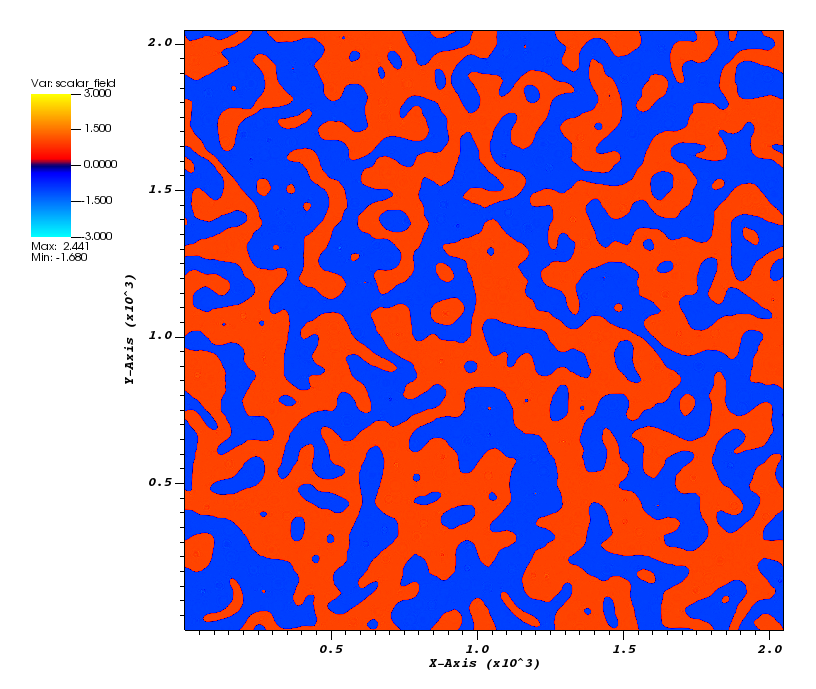}
    \includegraphics[width=1.0\columnwidth]{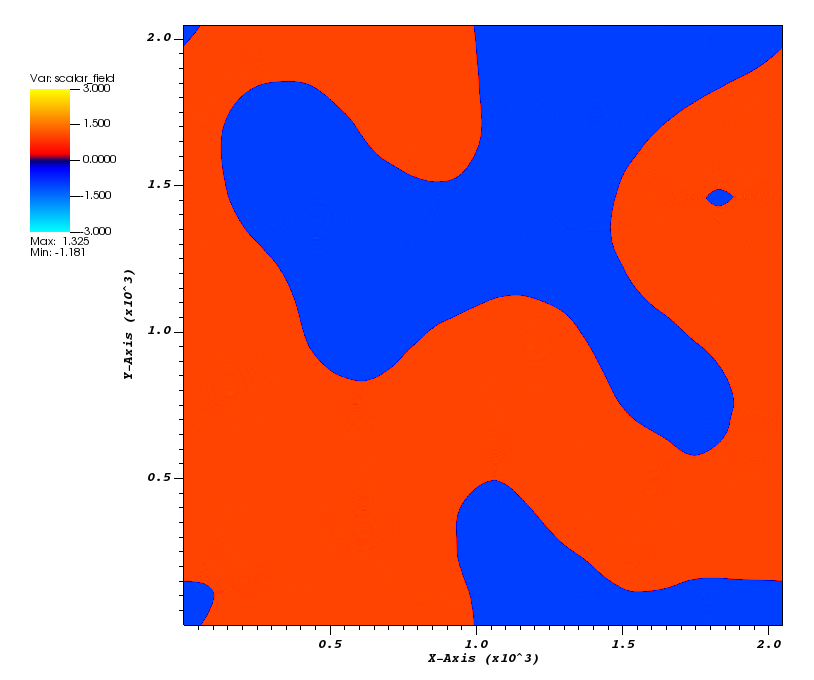}
    \includegraphics[width=1.0\columnwidth]{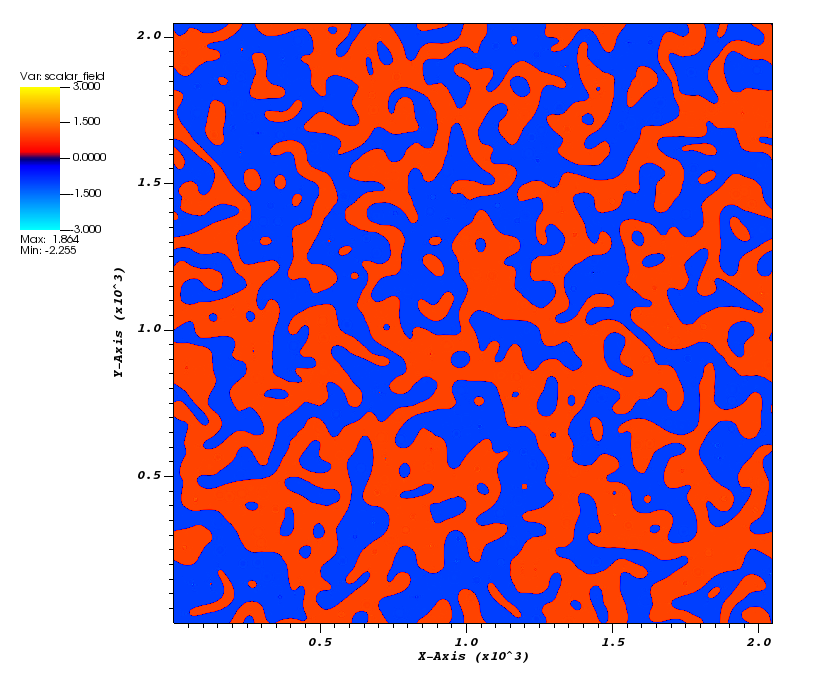}
    \includegraphics[width=1.0\columnwidth]{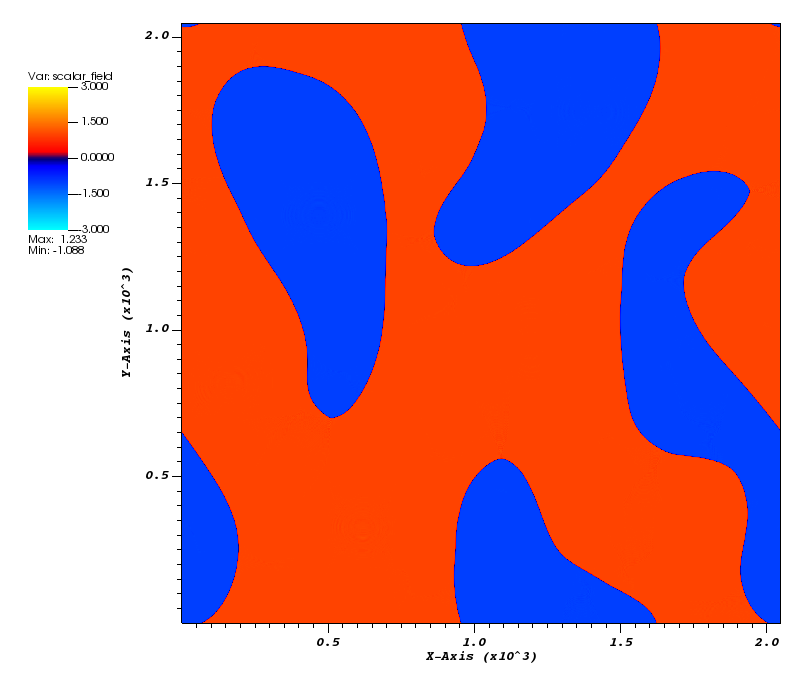}
  \caption{Snapshots of a domain wall network with a quartic potential in a $2048^2$ grid, for $\lambda=1/2$ (top) and $\lambda=2/3$ (bottom). The color map represents the value of the field $\phi$. The snapshots were taken for conformal times $\eta=101$ (left) and $\eta=751$ (right).}\label{fig02}
\end{figure*}

\begin{figure*}
  \begin{center}
    \includegraphics[width=1.0\columnwidth]{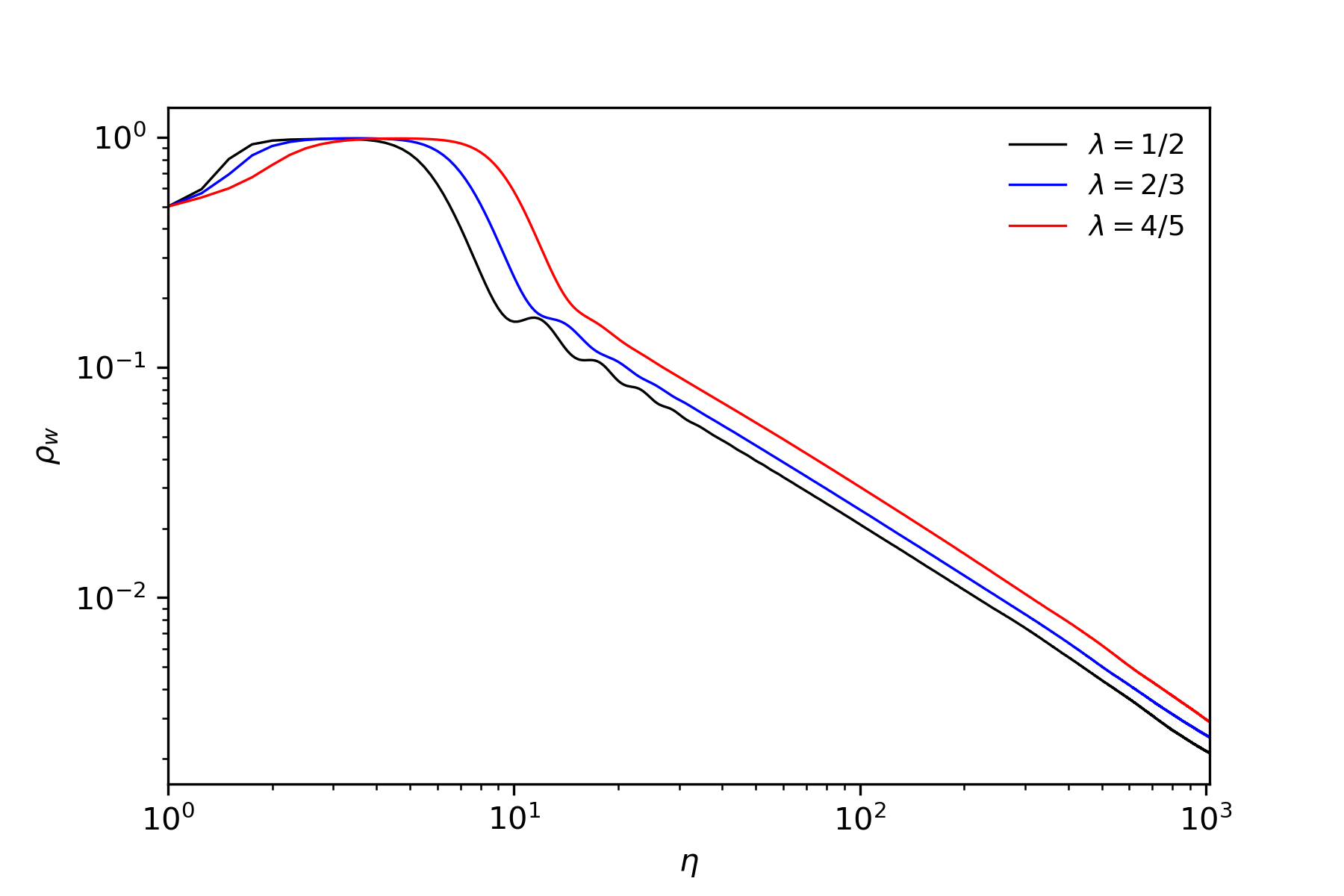}
    \includegraphics[width=1.0\columnwidth]{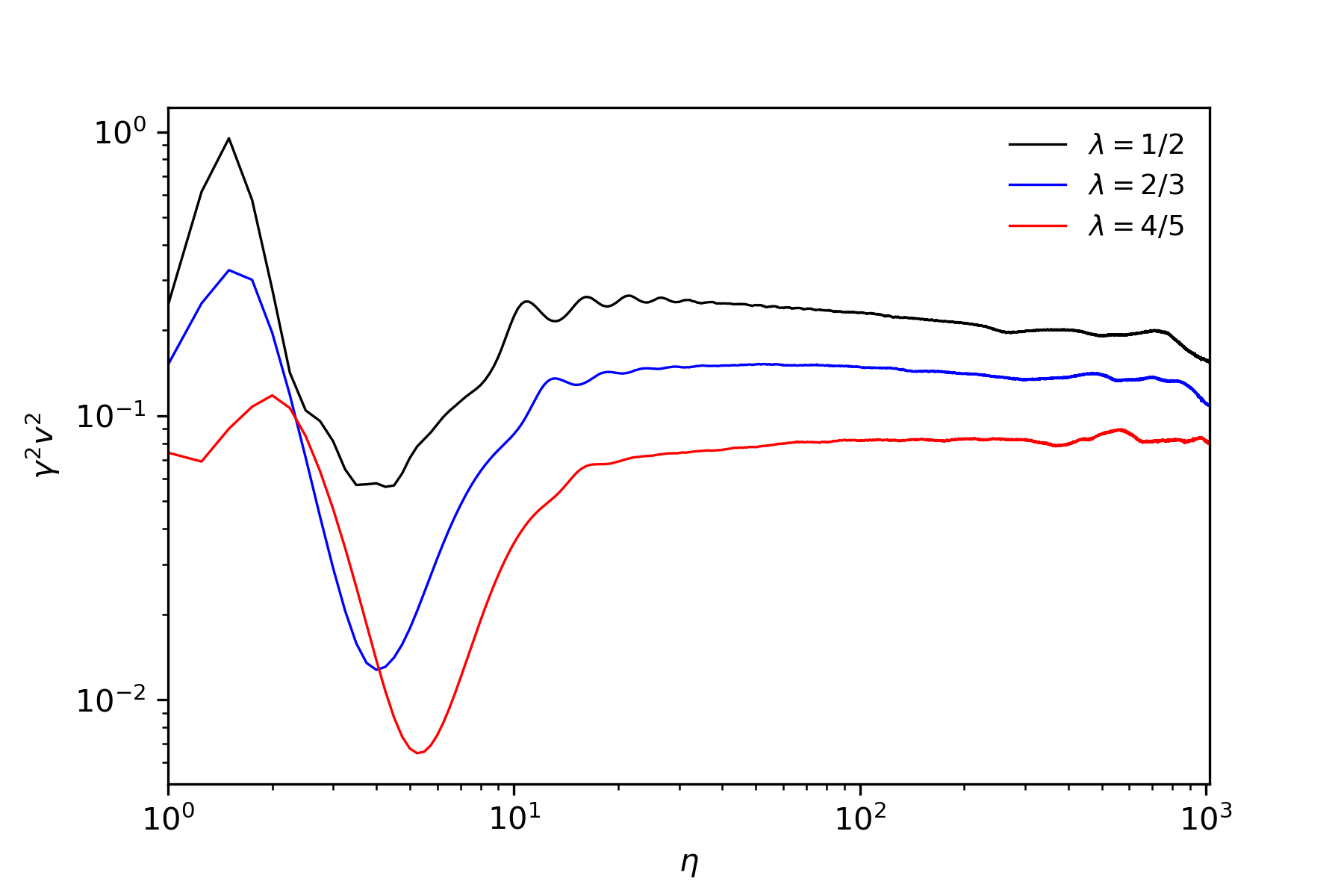}
    \caption{Evolution of the domain wall network with a $\phi^4$ potential as a function of conformal time for a box size of $2048^2$ and three different expansion rates. The plotted curves are averaged over 10 different simulations with different seeds, identical for each value of $\lambda$. The left panel shows the wall density, while the right panel shows the wall velocity (specifically $\gamma^2 v^2$).}
    \label{fig03}
  \end{center}
\end{figure*}

Defining the characteristic wall thickness
\begin{equation}
    w_0 \equiv \frac{\pi \phi_0}{\sqrt{2V_0}}\,,
    \label{wall-thick}
\end{equation}
an exact solution for an adiabatically static universe is
\begin{equation}
    \phi\left(z\right) = \phi_0\, \mathrm{tanh}\left(\frac{\pi}{w_0}z\right)
\end{equation}
which corresponds to a planar wall at rest with a localized kink centered at $z=0$ and is depicted in the right panel of Fig.~\ref{fig01}. Furthermore, it can be shown that its surface tension is 
\begin{equation}
    \sigma = \frac{4\sqrt{2}}{3}\, \phi_0\, \sqrt{V_0}\,,
\end{equation}
and that this surface tension in the two tangential directions is also equal to $\sigma$. Numerically, and for consistency throughout this work, we will choose potential parameters to ensure that the curvature of the potential at the minima is the same in all the models simulated.

\subsection{\label{prs}Numerical implementation}

Our numerical implementation builds upon the method developed by Press, Ryden \& Spergel (PRS) \cite{dynamical-evolution-of-domain-walls}. When evolving a domain wall network in an expanding universe, the wall thickness decreases as $a^{-1}$. This is a numerical bottleneck since, for later times, we lose resolution on the wall and can easily encounter apparent energy losses purely due to numerical limitations. The PRS method solves this issue by introducing a new equation of motion which preserves the wall dynamics but maintains a constant wall comoving thickness. To do this, consider the general equation
\begin{equation}
    \label{PRS-eom1}
    \frac{\partial^2 \phi}{\partial \eta^2} + \alpha \left(\frac{d \mathrm{ln} a}{d\mathrm{ln}\eta}\right) \frac{\partial\phi}{\partial \eta}- \nabla^2\phi = - a^\beta\frac{\partial V}{\partial \phi}\,,
\end{equation}
where $\alpha$ and $\beta$ are constants. Eq.~(\ref{eom-conformal}) is the particular case where $\alpha=\beta=2$. A fortunate result is that setting $\alpha=3$ and $\beta=0$ in these equations gives us the same dynamics as in the original system but with the benefit of maintaining a constant comoving wall thickness \cite{dynamical-evolution-of-domain-walls}, and this is therefore the choice we adopt, in common with most of the literature.

The method used to discretize the equation of motion is also a standard one. Here we describe the general 3D case, although most of our simulations will be done in 2D, which has the advantage of a larger dynamic range for the same memory usage. A simple finite differences scheme with a 7-point stencil is used to compute the 3-dimensional Laplace operator:
\begin{eqnarray}
    \label{numerical2}
    \left(\nabla^2\phi\right)_{ijk} &\equiv& \phi_{i+1,j,k} + \phi_{i-1,j,k} + \phi_{i,j+1,k}+\\ \nonumber
    &+& \phi_{i,j-1,k} + \phi_{i,j,k+1} + \phi_{i,j,k-1} - 6\phi_{i,j,k}\,,
\end{eqnarray}
while time derivatives rely on a staggered leapfrog finite differences scheme for the second-order term and Crank-Nicolson for the first-order term
\begin{equation}
    \label{numerical3}
    \dot{\phi}^{n+1/2}_{ijk} = \frac{\left(1-\delta\right)\dot{\phi}^{n-1/2}_{ijk} + \Delta\eta\left(\nabla^2\phi^{n}_{ijk} - \partial V/\partial\phi^{n}_{ijk}\right)} {1+\delta}\,.
\end{equation}
Here, $\dot{\phi} \equiv \partial\phi/\partial\eta$, $\Delta\eta$ is the discrete time step and the damping term $\delta$ is given by the expression 
\begin{equation}
    \label{numerical1}
    \delta \equiv \frac{1}{2} \alpha \frac{\Delta\eta}{\eta}\frac{d\mathrm{ln}a}{d\mathrm{ln}\eta}\,.
\end{equation}
Using a central difference scheme to compute $\dot{\phi}^{n+1/2}$, the evolution of the field is given by 
\begin{equation}
    \label{numerical4}
    \phi^{n+1}_{ijk} = \phi^{n}_{ijk} + \Delta\eta \dot{\phi}^{n+1/2}_{ijk}\,.
\end{equation}

The highly parallel nature of steps of Eqs. (\ref{numerical2},\ref{numerical3},\ref{numerical4}) makes this algorithm a prime candidate for GPGPU implementation. This was firstly done by Correia and Martins \cite{gpu-implementation}, using Open Computing Language (OpenCL) 1.2 as specified by the Khronos Consortium \cite{openCL12}, which allowed us to more efficiently run several simulations on a single machine. The machine used for the simulations in the present work was equipped with a NVIDIA Quadro P5000 with 2560 CUDA cores clocked at 1607$MHz$. It also packed 16$GB$ of total video memory clocked at 1126$MHz$.

The first relevant numerical diagnostic to be measured is the wall density, which is simply given by
\begin{equation}
    \rho_w = \frac{A}{V}\,,
\end{equation}
where $A$ is the comoving area of the wall and $V$ is the volume of the box; $\rho_w$ is related to the correlation length as discussed above. For this, we use the value of $\phi$ on each point of the grid to determine if it is located at a maximum of $V\left(\phi\right)$ within a certain margin $\delta\phi$, and we divide it by the total number of points in the grid. The other important quantity to measure is the wall RMS velocity, more specifically $\gamma v$ (where $\gamma$ is the the Lorentz factor) measured as described in \cite{VOS-walls}, where this quantity is shown to be given by
\begin{equation}
    \left< \gamma^2 v^2 \right> = \frac{1}{2N} \sum \frac{\dot{\phi}^2}{V\left(\phi\right)}\,.
\end{equation}

As a baseline, our initial conditions assume a uniformly distributed field between the two potential minima $-\phi_0$ and $+\phi_0$ and vanishing initial field speed ($\partial\phi/\partial\eta=0$), but alternative choices will also be studied. This initially leads to large energy gradients, which need some time to dissipate; to mitigate this, one may introduce a cooling mechanism to soften these gradients. All simulations assume $\phi_0=1$ and $w_0=5$, start at a conformal time $\eta_0=1$ and evolve in time steps of $\Delta\eta=0.25$ until a conformal time equal to half the box size. The exception is that while the cooling mechanism is active (for $\eta<1$), we use smaller time steps, $\Delta\eta_{cooling}=\Delta\eta/30$. All the results we present are averages from 10 separate runs with identical initial conditions except for the random seed (for consistency, the same set of 10 random seeds was used in the various cases).

To test the validity of the code we first implemented it for the canonical quartic potential. This also provides some benchmark results to be compared to our simulations of the other potentials, reported in the following section. By using the aforementioned parameter choices, Eq.~(\ref{phi4-potential}) becomes
\begin{equation}
    V\left( \phi \right) = \frac{\pi^2}{50} \left(1-\phi^2\right)^2\,.
\end{equation}
We will use three different expansion rates: a radiation-dominated Universe with $\lambda=1/2$, a matter-dominated Universe with $\lambda=2/3$ and a faster expanding one with $\lambda=4/5$. For the benchmark case in this section, we limit ourselves to $2048^2$ simulation boxes.

For illustration we depict in Fig.~\ref{fig02} snapshots of simulations in radiation- and matter-dominated Universes for two distinct time steps. Fig.~\ref{fig03} shows the evolution of the wall density and velocity computed from our simulations.  As expected, after an initial period until $\eta\sim30$, the network converges towards a scaling solution predicted by the VOS model. To quantify this relationship one can look for the best fit to the power laws
\begin{eqnarray}
    \label{power-law1}    \rho_w &\propto& \eta^\mu\\
    \label{power-law2}    (\gamma v) &\propto& \eta^\nu\,.
\end{eqnarray}
For a scale invariant behavior, we asymptotically expect $\mu=-1$ and $\nu=0$. There needs to be special care to only fit the data in the relevant dynamic range which, in this case, we conservatively chose to be $\eta\ge100$. We also report the asymptotic values of $(\rho_w \eta)^{-1}$ and $\gamma v$ on the later stages of evolution where we assumed to have achieved scaling. These quantities can then be used to calculate the parameters $c_w$ and $k_w$ in the analytic model.

\begin{table}
  \centering
  \caption{Values of the scaling exponents $\mu$ and $\nu$ for the $\phi^4$ potential and three different expansion rates in a box size of $2048^2$. Each value was taken by averaging 10 simulations and fitting the data in the range $\eta=\left[100,1024\right]$. The fourth and fifth column show the asymptotic values for $(\rho_w \eta)^{-1}$ and $\gamma v$ which are described by the VOS model. One-sigma statistical uncertainties are given throughout.}
\begin{tabular}{| c | c | c | c | c |}
\hline
$\lambda$ & $\mu$ & $\nu$ & $(\rho_w \eta)^{-1}$ & $\gamma v$ \\ \hline \hline
$1/2$ & $-0.938\pm 0.030$ & $-0.058\pm 0.020$ & $0.49\pm 0.10$ & $0.39\pm 0.03$ \\ \hline
$2/3$ & $-0.942\pm 0.025$ & $-0.034\pm 0.029$ & $0.41\pm 0.08$ & $0.33\pm 0.02$ \\ \hline
$4/5$ & $-0.949\pm 0.020$ & $-0.004\pm 0.040$ & $0.34\pm 0.05$ & $0.28\pm 0.04$ \\ \hline
\end{tabular}
  \label{tab1}
\end{table}

\begin{figure}
  \begin{center}
    \leavevmode
    \includegraphics[width=1.0\columnwidth]{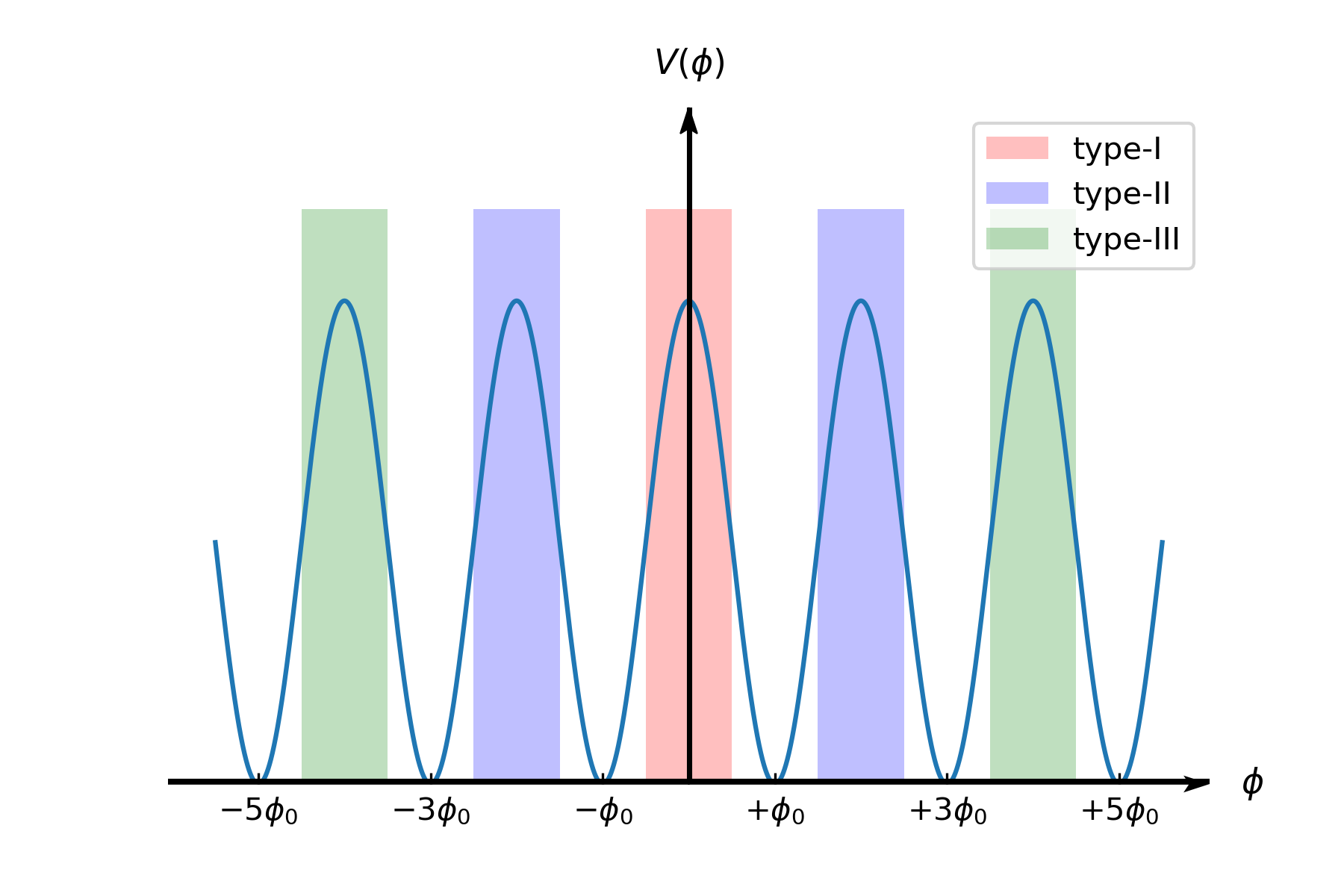}
    \caption{General shape of the Sine-Gordon potential, illustrating the terminology used in the text for the different types of walls which can form across each of its maxima. Note that the standard choice of initial conditions for the field evolution is a uniform distribution in the range  $\pm \, \phi_0$, but some alternative choices will also be used.}
    \label{fig04}
  \end{center}
\end{figure}

Table~\ref{tab1} shows the results of our analysis. These are consistent with the analytic model and with the numerical simulations previously done by Leite \& Martins \cite{scaling-properties-of-domain-walls} who have used an independent (CPU based) version of the code and ran it for box sizes and exponents which are identical to those of the present work. The small deviations of $\mu$ with respect to the linear scaling exponent are known to be due to the relatively small dynamical range of the boxes, which does not enable a complete relaxation to the scaling attractor. This can be confirmed with larger simulations, as has been explored in \cite{extending-vos-walls}. In any case, we note that the one-sigma statistical uncertainties included in the table show that the deviations from $\mu=-1$ and $\nu=0$ are not statistically significant.

\section{\label{chap:chap3}Sine-Gordon Potential}

A potential shape which emerges from specific symmetry breaking mechanisms, such as the so-called schizon models \cite{schizon-model}, is the Sine-Gordon potential
\begin{equation}
    \label{SG-potential}
     V\left( \phi \right) = V_{0,\mathrm{SG}} \left[1+\mathrm{cos}\left( \pi  \frac{\phi}{\phi_0} \right)\right]
\end{equation}
which has periodic minima at $\phi_{\rm min}=\left(2n+1\right)\phi_0$ for any integer $n$, as depicted in Fig. \ref{fig04}. 

\begin{figure*}
\centering
    \includegraphics[width=1.0\columnwidth]{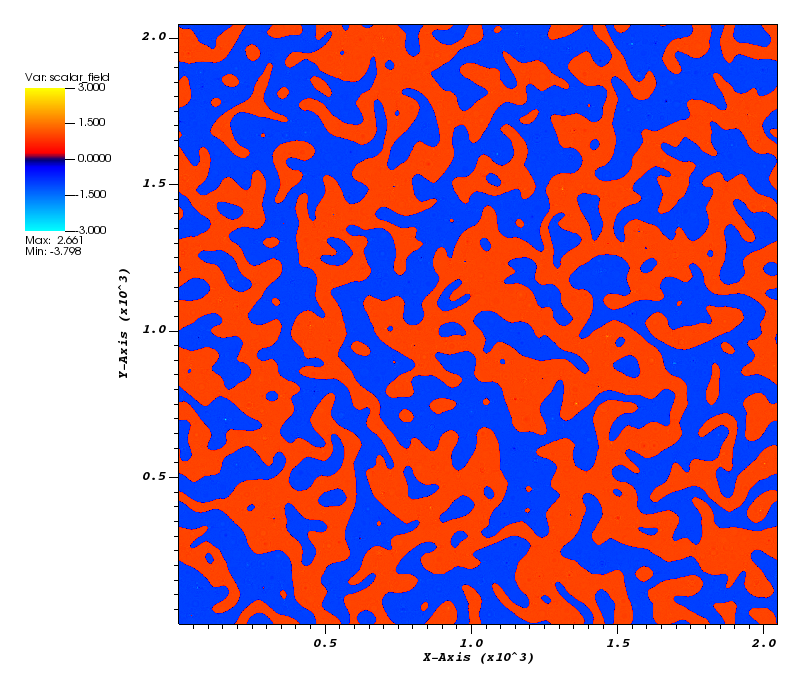}
    \includegraphics[width=1.0\columnwidth]{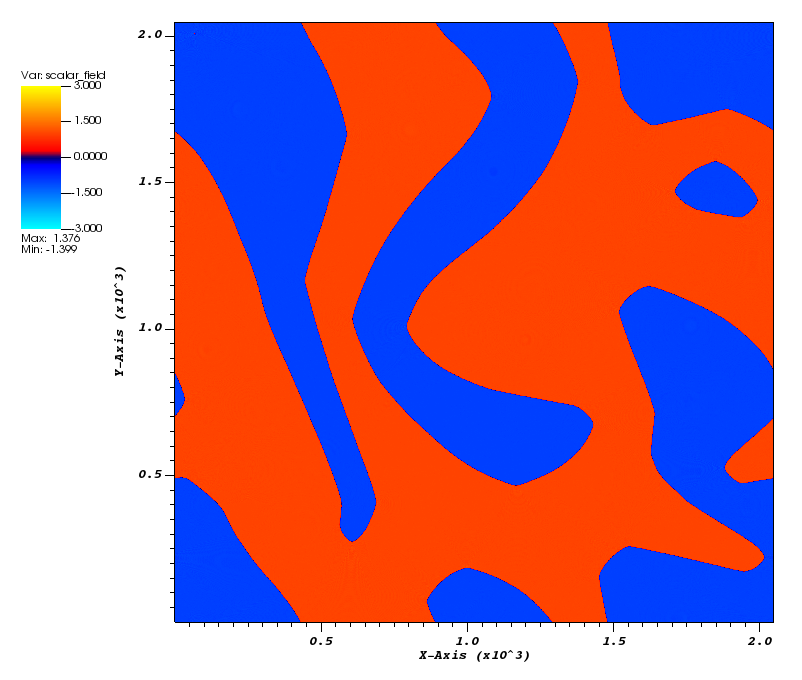}
    \includegraphics[width=1.0\columnwidth]{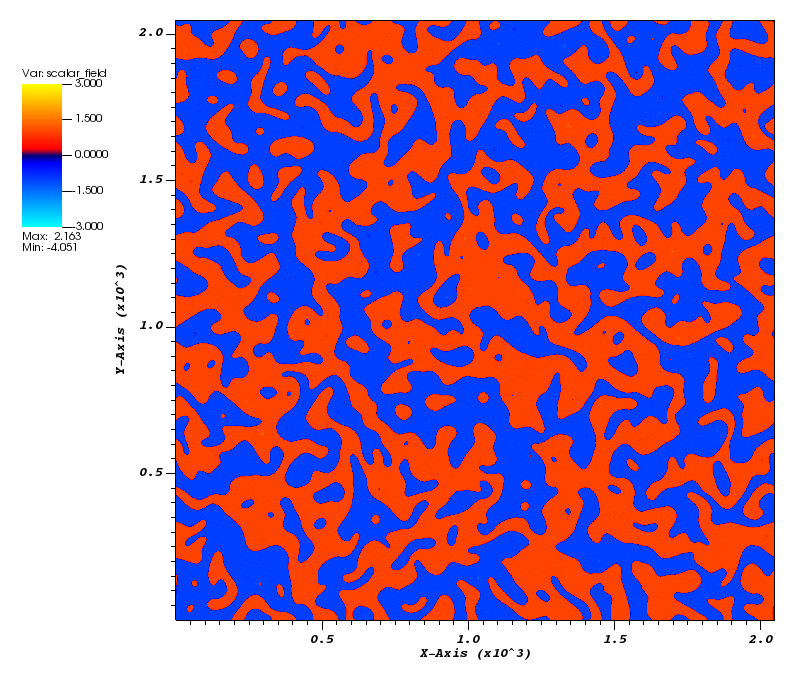}
    \includegraphics[width=1.0\columnwidth]{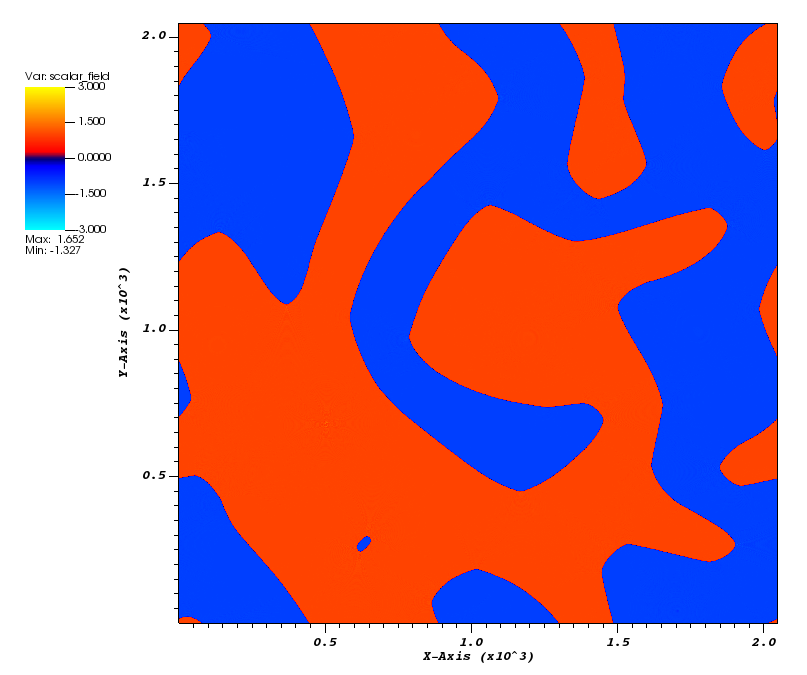}
  \caption{Snapshots of a domain wall network with a Sine-Gordon potential in a $2048^2$ grid, for $\lambda=1/2$ (top) and $\lambda=2/3$ (bottom). The color map represents the value of the field $\phi$. Snapshots were taken for conformal times $\eta=101$ (left) and $\eta=751$ (right).}\label{fig05}
\end{figure*}

For consistency through this work, it is important to relate $V_0$ with the wall thickness $w_0$, which has a fixed value in our simulations. To ensure that the field has the same dynamics in the vacuum as in the $\phi^4$ case, we must choose a value of $V_0$ such that the local curvature at the minima is the same in both models. To ensure this, we calculate the second derivative of both the $\phi^4$ and Sine-Gordon potentials, cf. Eqs.~(\ref{phi4-potential}) and (\ref{SG-potential}), finding
\begin{equation}
    \left. \frac{d^2 V_{\phi^4}}{d\phi^2} \right|_{\phi = \pm\phi_0} =  \frac{8V_0}{\phi_0^2} 
\end{equation}    
\begin{equation}    
    \qquad \quad , \qquad \qquad 
    \left. \frac{d^2 V_{\mathrm{SG}}}{d\phi^2} \right|_{\phi = \pm\phi_0} =  \frac{\pi^2 V_{0,\mathrm{SG}}}{\phi_0^2}\,.
\end{equation}
Imposing that they must have the same local curvature in the minima, one obtains
\begin{equation}
    V_{0,\mathrm{SG}} = \frac{8}{\pi^2} V_0\,.
\end{equation}
Using this relation and substituting the same values for $\phi_0$ and $w_0$ as we used in the previous section, we can rewrite the potential of Eq.~(\ref{SG-potential}) as
\begin{equation}
    \label{SG-potential2}
     V\left( \phi \right) = \frac{4}{25}\left[1+\mathrm{cos}\left( \pi \phi \right)\right]
\end{equation}
These choices ensure similar local field dynamics in the minima while at the same time having the same fixed wall thickness as in the $\phi^4$ case.

We have followed the same numerical procedure (including the same initial conditions) as in the previous section to evolve the network of domain walls emerging from this potential. Initially, the only walls present are the ones separating the minima at $\pm \, \phi_0$, which we henceforth denote type-I walls; with this standard choice of initial conditions, these dominate the subsequent dynamics. For comparison, we show in Fig.~\ref{fig05} snapshots of the field for $\lambda=1/2$ and $\lambda=2/3$. The subsequent dynamics of the model does lead to the formation of some walls of other types, which we shall denote type-II and type-III walls; these are schematically identified in Fig.~\ref{fig04}. Note that this terminology identifies how far each type of wall is from the range of field values which is included in the standard initial conditions. In other words, although this classification of wall types depends on a particular convention for the initial conditions, once this convention is defined one can separately and unambiguously identify them in the simulations and explore the dynamics of the various types. In the following section, we will explore the impact of alternative choices of initial conditions.

\begin{figure}
  \begin{center}
    \leavevmode
    \includegraphics[width=1.0\columnwidth]{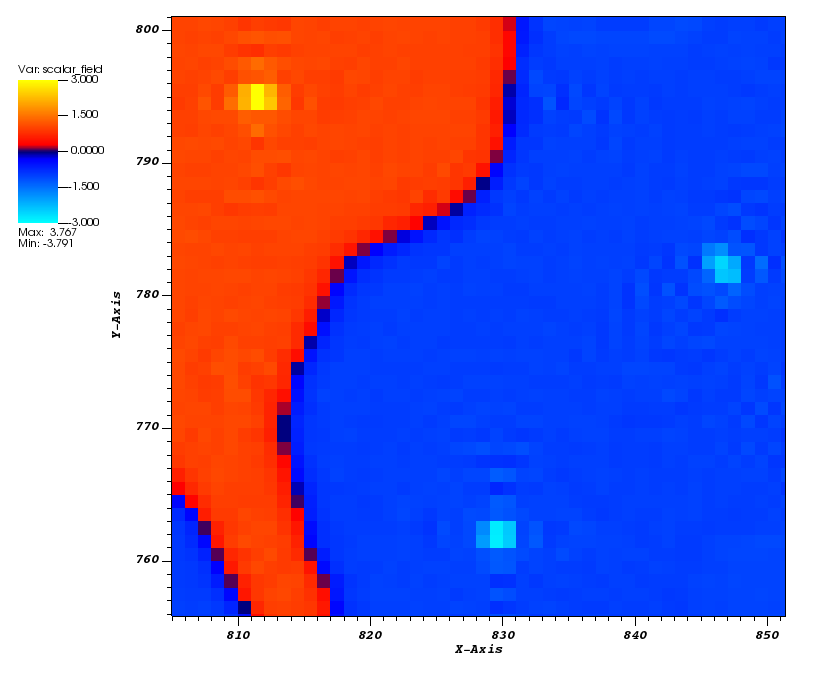}
    \caption{Zoom-in snapshot of the field values, detailing the emergence of domains outside the range of initial conditions in a Sine-Gordon potential giving rise to type-II walls. The color map represents the value of the field $\phi$. This snapshot was taken from a simulation for a box size of $2048^2$ and $\lambda=1/2$ at conformal time $\eta=26$.}
    \label{fig06}
  \end{center}
\end{figure}

Since the field can explore several adjacent minima, it will naturally do so predominantly in the earlier time steps where it still has a comparatively large energy. (Recall that the expansion provides a minimum unavoidable damping mechanism.) Figure \ref{fig06} illustrates how these domains emerge in the simulation and cause the formation of different types of walls. It also shows that it is important to separately quantify how the network's energy density is distributed among the various types of wall, the RMS velocity of each type of wall, and how these distributions change as the network evolves. This can be done through straightforward extensions of the numerical diagnostic algorithms introduced in the previous section.

\begin{figure*}
\centering
    \includegraphics[width=1.0\columnwidth]{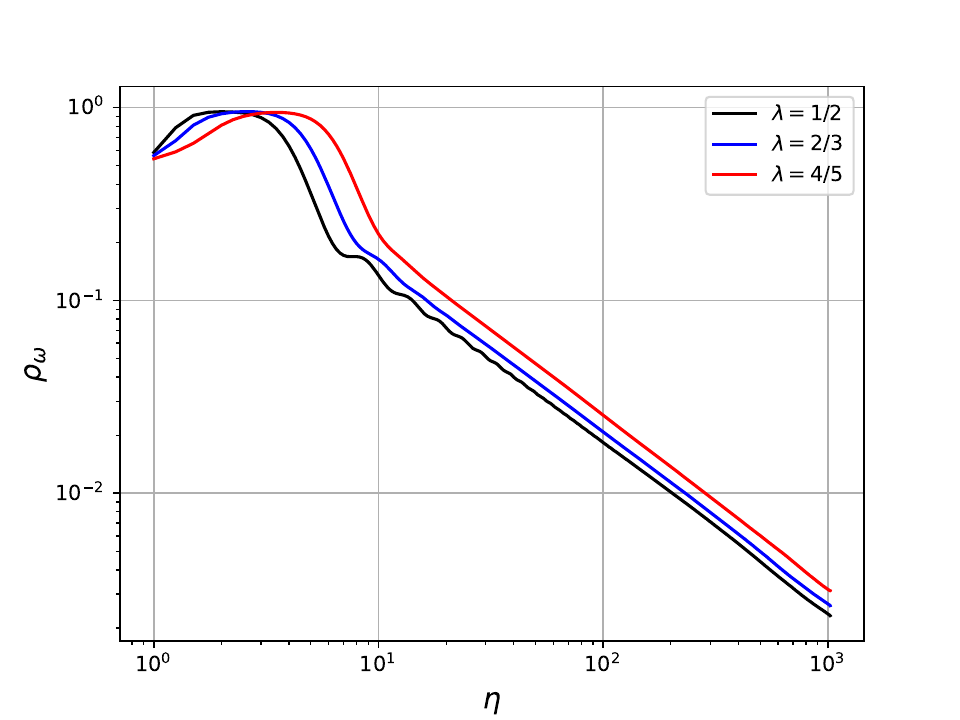}
    \includegraphics[width=1.0\columnwidth]{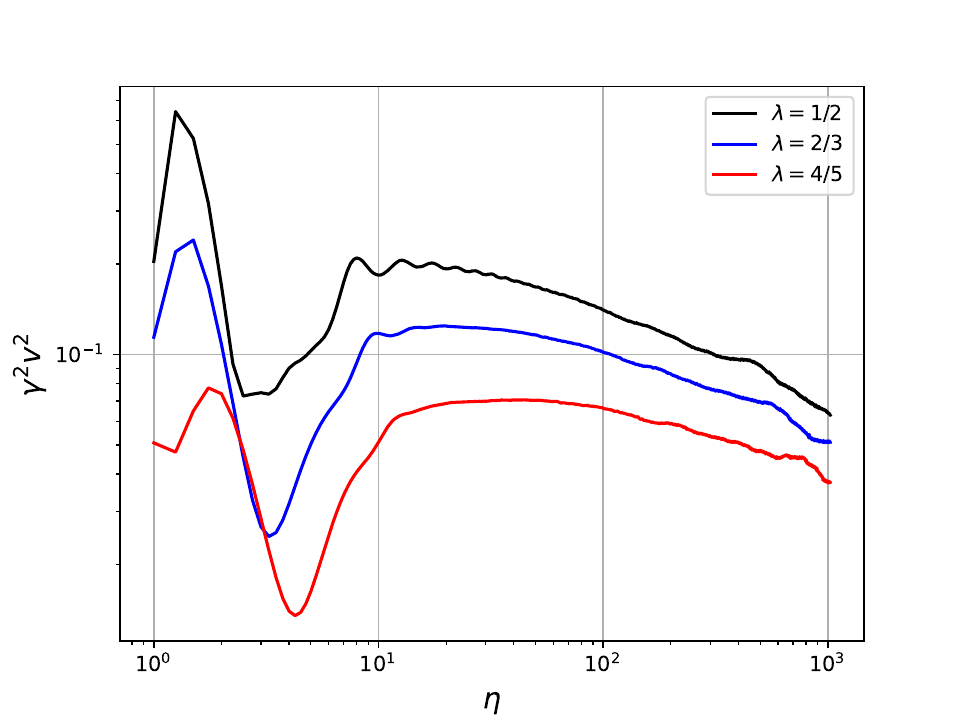}
    \includegraphics[width=1.0\columnwidth]{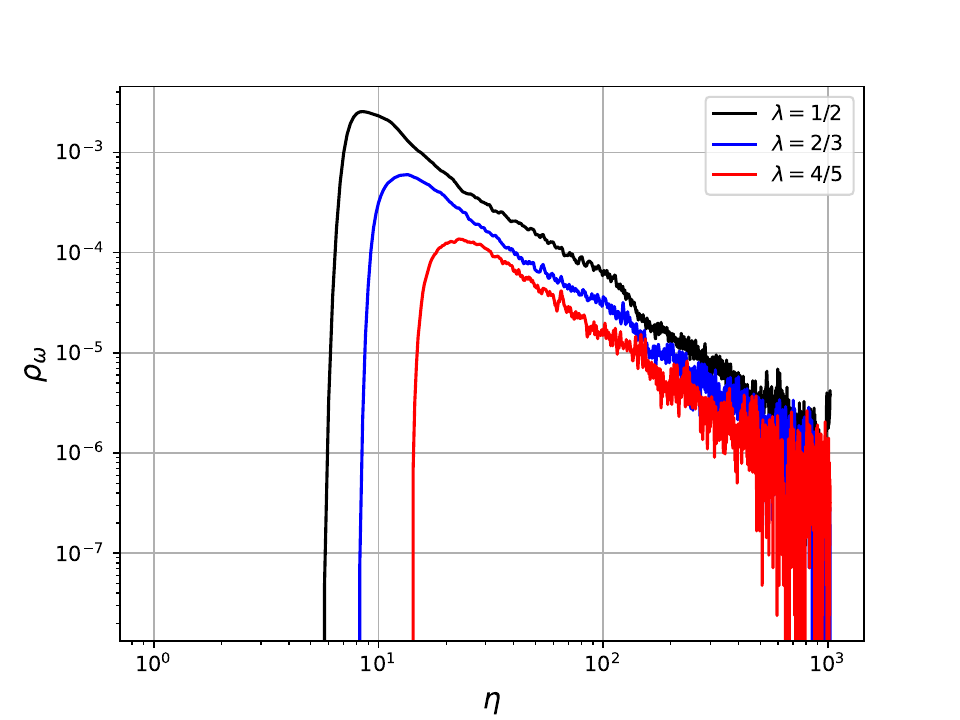}
    \includegraphics[width=1.0\columnwidth]{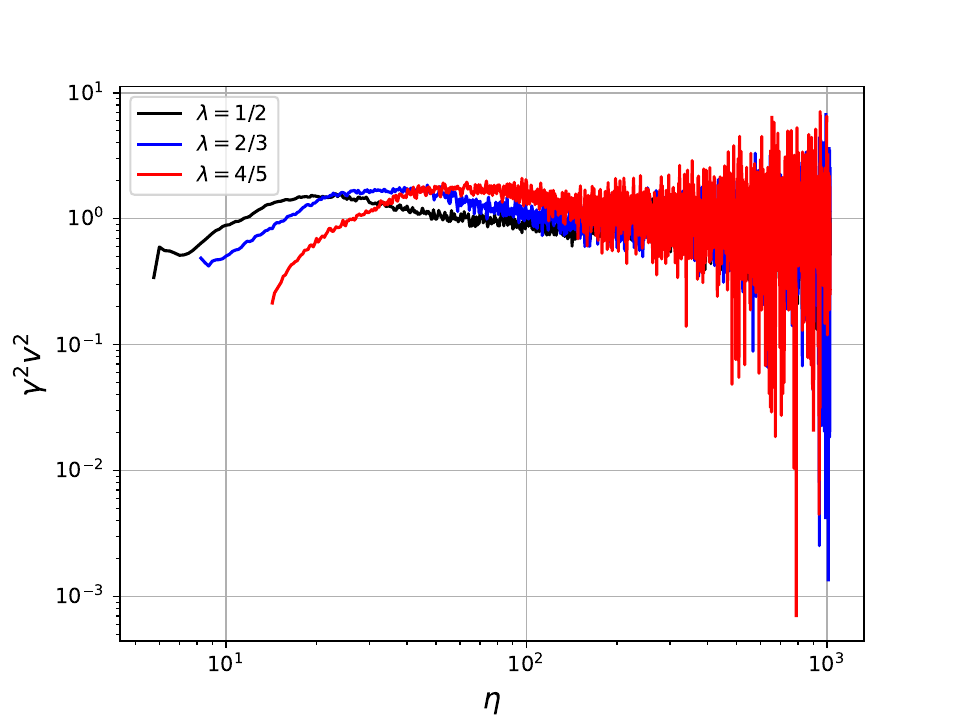}
    \includegraphics[width=1.0\columnwidth]{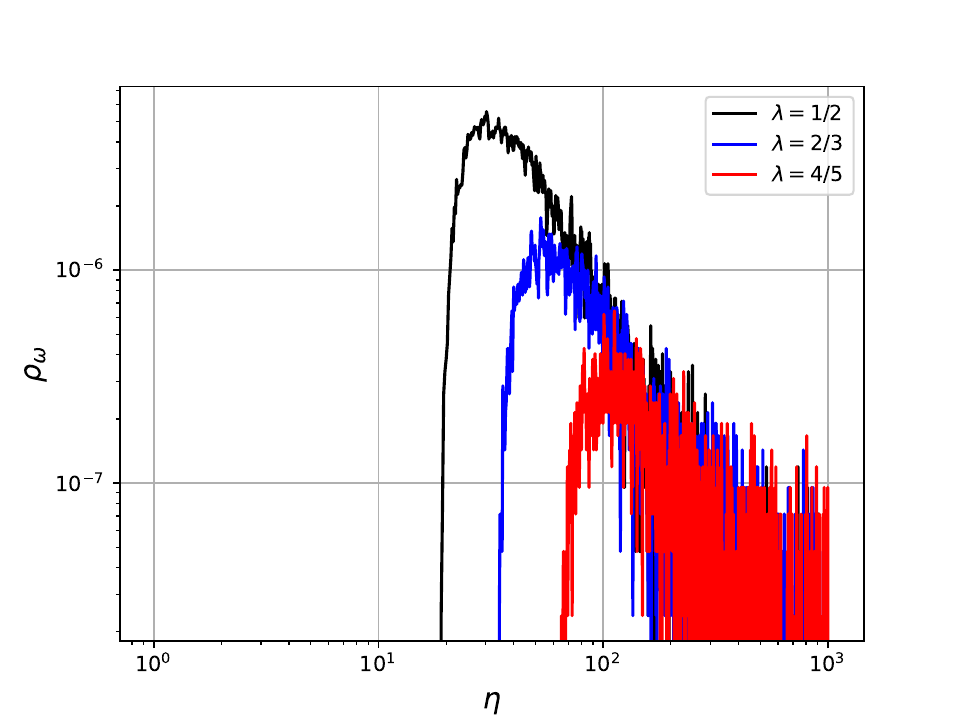}
    \includegraphics[width=1.0\columnwidth]{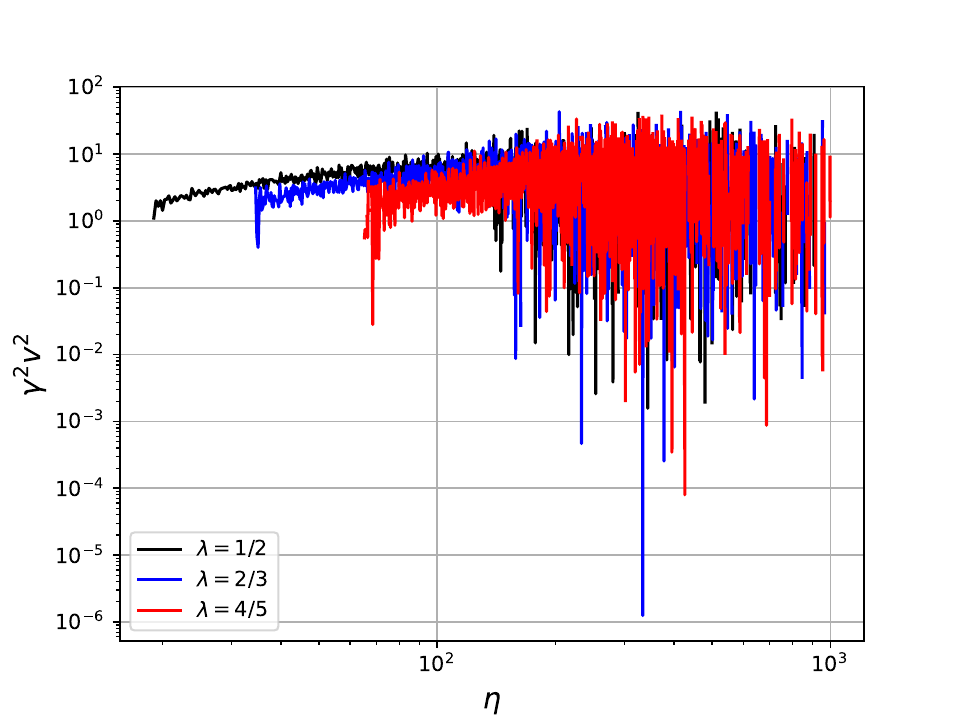}
   \caption{Evolution of the density ($\rho_w$, left side panels) and velocity ($\gamma^2v^2$, right side panels) of different types of domain walls in a Sine-Gordon potential as a function of conformal time for a box size of $2048^2$. The top, middle and bottom plots correspond respectively to the type-I, type-II and type-III walls, as defined in Fig. \ref{fig04}. The plotted values are averages over sets of 10 simulations, with initial conditions generated from a set of 10 different random seeds.}\label{fig07}
\end{figure*}

The measured values of the density and velocity of each type of wall, averaged for sets of ten different simulations, are plotted in Fig.~\ref{fig07}. These confirm that there is enough energy in the box to form a small fraction of type-II walls and, subsequently, a much smaller fraction of type-III walls. The plots also show that these fractions decrease for faster expansion rates. This is an expected result since a smaller Hubble damping provides more opportunities for the field to jump over the maximum of the potential and into a neighboring minimum. For our chosen box sizes, initial conditions and expansion rates, we do not identify any domain walls beyond type-III. Indeed, the bottom panels of Fig.~\ref{fig07} show that even type-III walls are very sparse and short-lived given these conditions, especially for the faster expansion rate, $\lambda=4/5$. 

Figure \ref{fig07} also shows that the velocities of type-I walls decrease as the network evolves. This is again not surprising, since the regions of the box where the field has larger kinetic energies are the ones that are more likely to jump over the maximum of the potential. It is noteworthy that this decrease in velocity is not apparent in the case of type-II walls. Unfortunately, for type-III walls, no statistically significant conclusion can be drawn, given their sparsity.

\begin{figure*}
\centering
    \includegraphics[width=1.0\columnwidth]{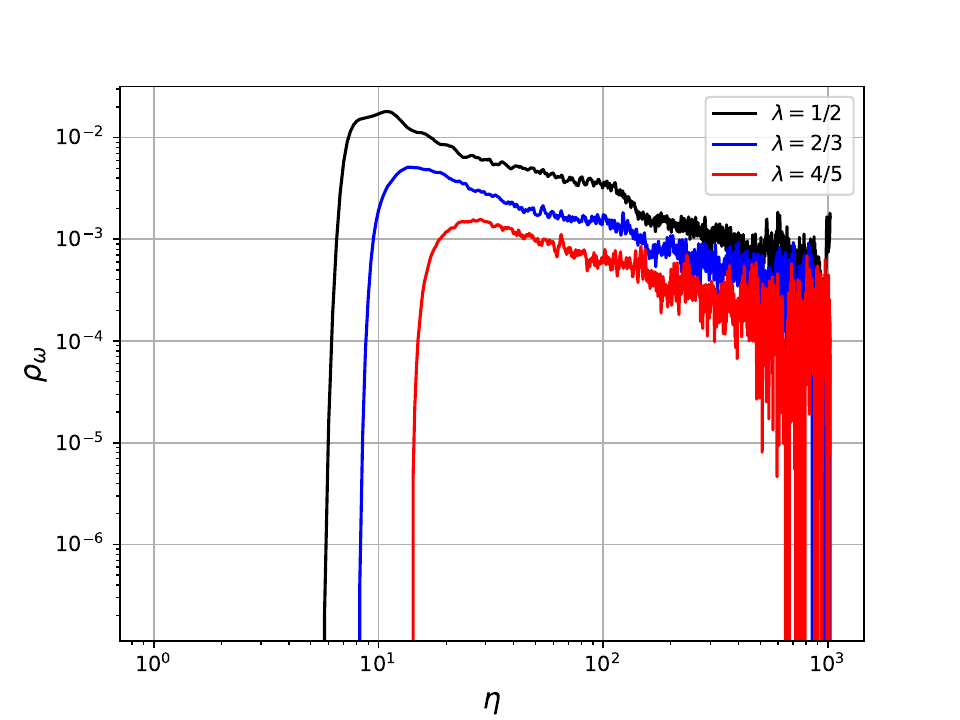}
    \includegraphics[width=1.0\columnwidth]{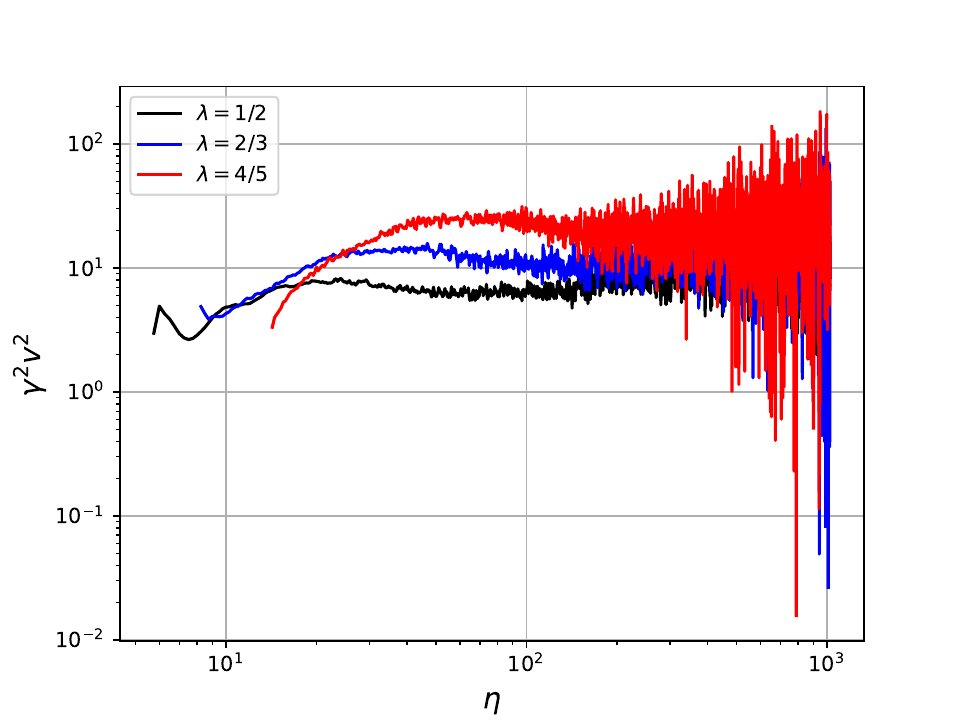}
    \includegraphics[width=1.0\columnwidth]{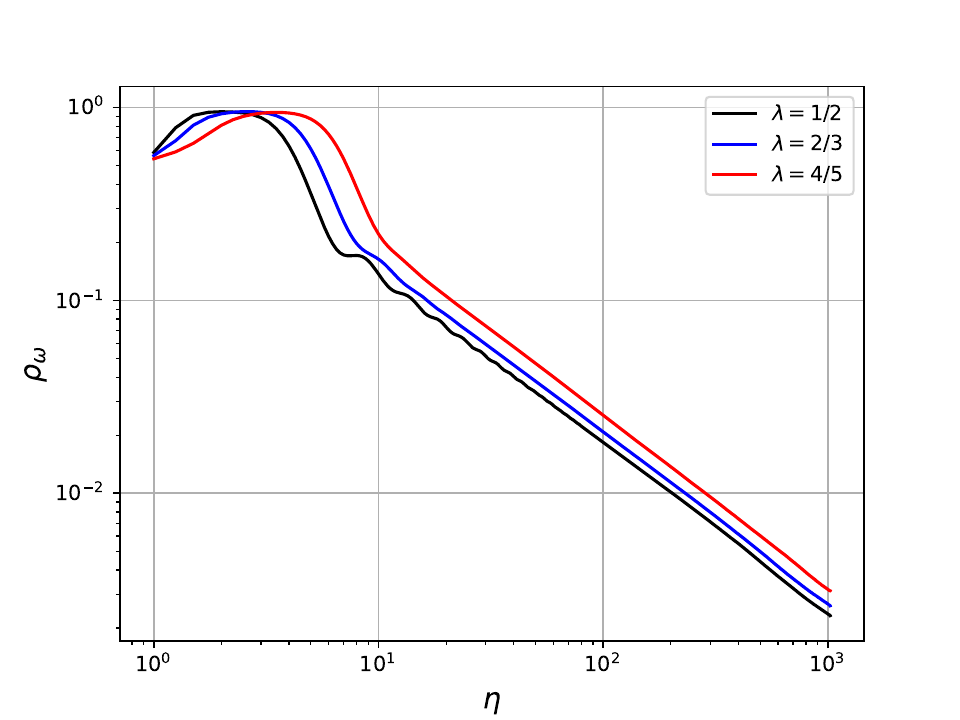}
    \includegraphics[width=1.0\columnwidth]{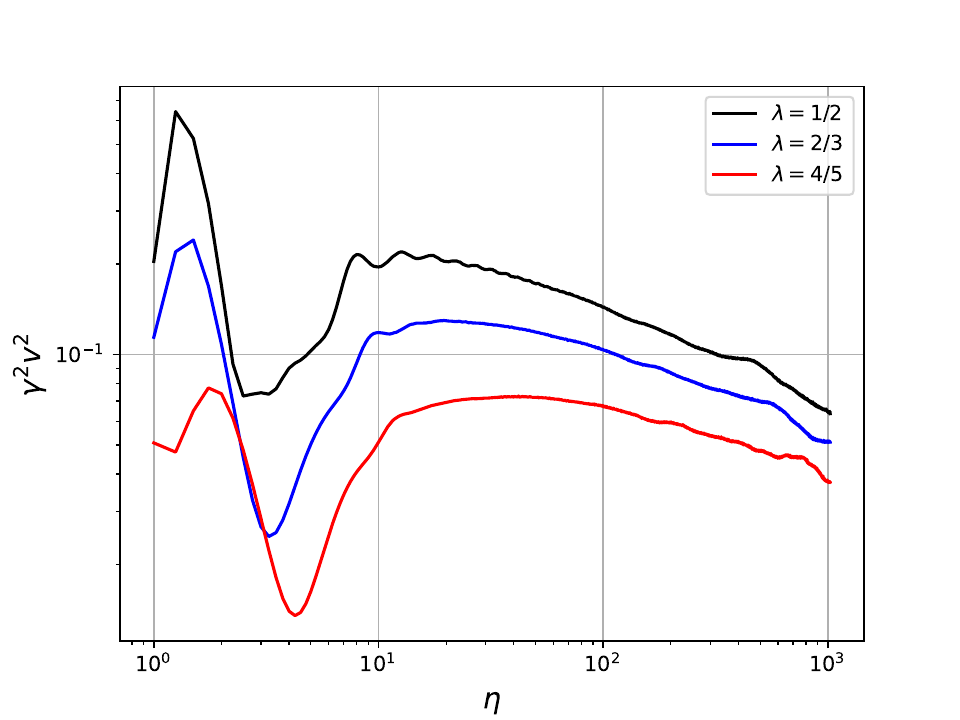}
  \caption{Top panels: Ratio of the densities (left panel) and velocities (right panel) of type-I and type-II walls in a Sine-Gordon potential. The data is the same as in Fig. \ref{fig07}.Bottom panels: Total density of walls considering type-I, type-II and type-III walls (left panel) and average velocities among all three types of walls (right panel).}\label{fig08}
\end{figure*}

It is also instructive to study the evolution of the ratio of the densities and velocities of the type-II and type-I walls, which can be found in the top panels of Fig.~\ref{fig08}. (The very noisy type-III walls data again precludes an analogous plot for that case.) It is clear that, once formed, type-II walls decay faster than those of type-I. Moreover, the ratio of the velocities increases with the expansion rate. We approximately find
\begin{equation}
\left[\frac{(\gamma v)^2_{II}}{(\gamma v)^2_{I}}\right]_{\lambda=1/2}\sim7.1
\end{equation}
\begin{equation}
\left[\frac{(\gamma v)^2_{II}}{(\gamma v)^2_{I}}\right]_{\lambda=2/3}\sim10.7
\end{equation}
\begin{equation}
\left[\frac{(\gamma v)^2_{II}}{(\gamma v)^2_{I}}\right]_{\lambda=4/5}\sim22.5\,,
\end{equation}
respectively in the radiation, matter and fast expansion eras. Again, this is to be expected: since the field needs more energy to overcome the potential barrier if the expansion rate is faster, if it does so it naturally must be further along the tail end of the velocity distribution, leading to larger ratios of the two velocities. This simultaneously explains why these walls become less common for faster expansion rates: only field configurations at the far end of the velocity distribution may lead to these walls.

Overall, the main result of this analysis is that the evolution of these networks is somewhat different from the linear scaling solution of the canonical $\phi^4$ walls. This can be confirmed by calculating the scaling exponents $\mu$ and $\nu$, previously defined in Eqs.~(\ref{power-law1}--\ref{power-law2}). The results of this analysis, for type-I walls, are shown in Table~\ref{tab2}, to be compared with the quartic potential wall results in Table \ref{tab1}; the table also lists the asymptotic values for $(\rho_w \eta)^{-1}$ and $\gamma v$. The two scaling exponents are clearly different in both cases, with $\mu$ being larger and $\nu$ smaller for the Sine-Gordon case, and there are also noticeable changes in the asymptotic values of the density and velocity, with the latter being considerably smaller for Sine-Gordon walls, for the reasons already discussed.

\begin{table*}
  \centering
  \caption{The scaling exponents $\mu$ and $\nu$ calculated for three different expansion rates for type-I walls in a Sine-Gordon potential. Each value was obtained by averaging over $10$ simulations, with one-sigma statistical uncertainties given throughout. The fifth and sixth column show the asymptotic values for $(\rho_w \eta)^{-1}$ and $\gamma v$. For the $2048^2$ simulation boxes, these values should be compared to the values obtained for $\phi^4$ walls, cf. Table \ref{tab1}.}
\begin{tabular}{| c | c | c | c | c | c |}
\hline 
box size & $\lambda$ & $\mu$ & $\nu$ & $(\rho_w \eta)^{-1}$ & $\gamma v$ \\ \hline \hline
$2048^2$ & $1/2$ & $-0.842\pm 0.018$ & $-0.217\pm 0.025$ & $0.42\pm 0.04$ & $0.29\pm 0.03$ \\ \hline
$2048^2$ & $2/3$ & $-0.853\pm 0.016$ & $-0.184\pm 0.016$ & $0.38\pm 0.03$ & $0.27\pm 0.01$ \\ \hline
$2048^2$ & $4/5$ & $-0.873\pm 0.017$ & $-0.143\pm 0.021$ & $0.33\pm 0.03$ & $0.24\pm 0.02$ \\ \hline \hline
$8192^2$ & $1/2$ & $-0.856\pm 0.017$ & $-0.190\pm 0.021$ & $0.43\pm 0.02$ & $0.25\pm 0.02$ \\ \hline
$8192^2$ & $2/3$ & $-0.872\pm 0.016$ & $-0.177\pm 0.022$ & $0.40\pm 0.04$ & $0.23\pm 0.01$ \\ \hline
$8192^2$ & $4/5$ & $-0.891\pm 0.014$ & $-0.139\pm 0.017$ & $0.36\pm 0.03$ & $0.22\pm 0.01$ \\ \hline \hline
$32768^2$ & $1/2$ & $-0.860\pm 0.030$ & $-0.196\pm 0.018$ & $0.37\pm 0.04$ & $0.22\pm 0.01$ \\ \hline
$32768^2$ & $2/3$ & $-0.865\pm 0.015$ & $-0.343\pm 0.067$ & $0.36\pm 0.03$ & $0.20\pm 0.01$ \\ \hline
$32768^2$ & $4/5$ & $-0.901\pm 0.017$ & $-0.269\pm 0.115$ & $0.32\pm 0.02$ & $0.18\pm 0.01$ \\ \hline
\end{tabular}
  \label{tab2}
\end{table*}

Note that for $2048^2$ simulations of the $\phi^4$ model, the small deviations of $\mu$ with respect to the linear scaling exponent ($\mu=-1$) are known to be due to the relatively small dynamical range of such simulations, as demonstrated in \cite{extending-vos-walls}. For Sine-Gordon walls one not only has much larger deviations from the $\mu=-1$ and $\nu=0$ behavior (which are clearly inconsistent, in a statistical sense, with the scale invariant behavior), but our results show that they persist for box sizes up to $32768^2$. Increasing the box size one sees a statistically significant decrease of the scaling exponent $\nu$ (which becomes slightly closer to zero) for the larger boxes, but for the scaling exponent $\mu$ there is no statistically significant difference between $2048^2$ and $32768^2$ boxes for any of the expansion rates which we have explored.

In fact the behaviors of the two scaling exponents, $\mu$ and $\nu$, are related to one another. Unlike in the standard quartic case, the high-velocity tail of type-I walls has the opportunity to migrate into type-II configurations, leading to a lower average velocity of type-I walls. This reduction in velocity decreases the likelihood of interactions and annihilations among type-I walls and therefore contributes to a slower decay of their energy density compared to the quartic case. Additionally, the periodic structure of the Sine Gordon potential facilitates energy redistribution across adjacent minima, allowing walls to transition between types rather than annihilate outright. This mechanism preserves the total wall population while altering its composition, suppressing the decay rate of type-I wall energy density.

\begin{table}
  \centering
  \caption{The scaling exponents $\mu$ and $\nu$ calculated for three different expansion rates for type-II walls in a Sine-Gordon potential. Each value was obtained by averaging over $10$ simulations, with one-sigma statistical uncertainties given throughout. The fifth column shows the asymptotic values for $\gamma v$. }
\begin{tabular}{| c | c | c | c | c |}
\hline 
box size & $\lambda$ & $\mu$ & $\nu$ & $\gamma v$ \\ \hline \hline
$2048^2$ & $1/2$ & $-1.585\pm 0.018$ & $-0.22\pm 0.029$ & $0.94\pm 0.07$ \\ \hline
$2048^2$ & $2/3$ & $-1.498\pm 0.030$ & $-0.51\pm 0.040$ & $1.18\pm 0.14$ \\ \hline
$2048^2$ & $4/5$ & $-1.830\pm 0.042$ & $-0.43\pm 0.095$ & $1.65\pm 0.17$ \\ \hline \hline
$8192^2$ & $1/2$ & $-1.571\pm 0.012$ & $-0.19\pm 0.070$ & $0.84\pm 0.08$ \\ \hline
$8192^2$ & $2/3$ & $-1.482\pm 0.025$ & $-0.59\pm 0.050$ & $1.13\pm 0.18$ \\ \hline
$8192^2$ & $4/5$ & $-1.861\pm 0.021$ & $-0.45\pm 0.040$ & $1.61\pm 0.16$ \\ \hline \hline
$32768^2$ & $1/2$ & $-1.678\pm 0.001$ & $-0.054\pm 0.070$ & $0.92\pm 0.21$ \\ \hline
$32768^2$ & $2/3$ & $-1.504\pm 0.003$ & $-0.107\pm 0.030$ & $0.97\pm 0.28$ \\ \hline
$32768^2$ & $4/5$ & $-1.547\pm 0.002$ & $-0.236\pm 0.050$ & $1.02\pm 0.23$ \\ \hline
\end{tabular}
  \label{tab3}
\end{table}

Table \ref{tab3} shows the analogous scaling diagnostics for type-II walls. In this case the asymptotic value for $(\rho_w \eta)^{-1}$ is not provided since for some late time steps no type-II walls are identified in the box, which prevents a meaningful statistical analysis and a fair comparison between the various cases. On the other hand, for the asymptotic velocities, we benefit from their near-constant behavior to provide the value for the latest time step for which it can be meaningfully calculated. The most noteworthy point is that the density scaling exponent is $\mu<-1$ (in fact, it is consistent with $\mu=-1.5$), confirming that the walls are decaying quite fast. We also confirm that these walls move significantly faster than their type-I counterparts.

Last but not least, it is interesting to consider the global evolution of the wall network, i.e. without separating the walls into the various types. To compute this global behavior, we sum the energy densities of all wall types and calculate the average velocity, weighted by the respective densities of each type. 
The results are shown in the bottom panels of Fig.~\ref{fig08}. At first glance, it is evident that the evolution of the global density and velocity closely mirrors that of the type-I walls alone. This observation is confirmed by calculating the scaling exponents $\mu$ and $\nu$ and comparing them to the values listed in Table~\ref{tab2}. In all three expansion scenarios, the scaling exponents for the global quantities agree with those of the type-I walls up to $\mathcal{O}\left(10^{-3}\right)$. This can be understood by noting that most energy in the box is walls of type-I, so when taking an average of all types these will dominate.

\section{\label{newsims}Scaling solution robustness tests}

In this section, we present results of Sine-Gordon domain wall simulations in several other numerical settings. The main purpose is to provide some tests of the robustness of the results of the previous section. The choices of these settings are partially due to physical reasons, but mainly due to the convenience of their numerical implementation. For each such case, we briefly present the changed numerical setting and report the corresponding results.

One common aspect among these initial conditions, as well as those of the previous section, is that they are symmetric with respect to $\phi_0=0$. We will relax this assumption in Appendix \ref{sineics}. On the other hand, we do not simulate explicitly biased initial conditions, since it is clear from previous work \cite{biases1,biases2} that in this case the networks will quickly decay.

\subsection{Wider and split initial conditions}

A simple numerical change is to have a wider range of initial conditions for the scalar field. Specifically, we have carried out simulations with the field uniformly distributed in the range $[-3\phi_0,+3\phi_0]$, as opposed to the standard $[-\phi_0,+\phi_0]$, so initially we have both type-I and type-II walls (and indeed twice as many type-II walls as type-I). We simulate the same three expansion rates ($\lambda=1/2$, $\lambda=2/3$, $\lambda=4/5$), and restrict ourselves to $2048^2$ boxes.

\begin{figure*}
\centering
    \includegraphics[width=1.0\columnwidth]{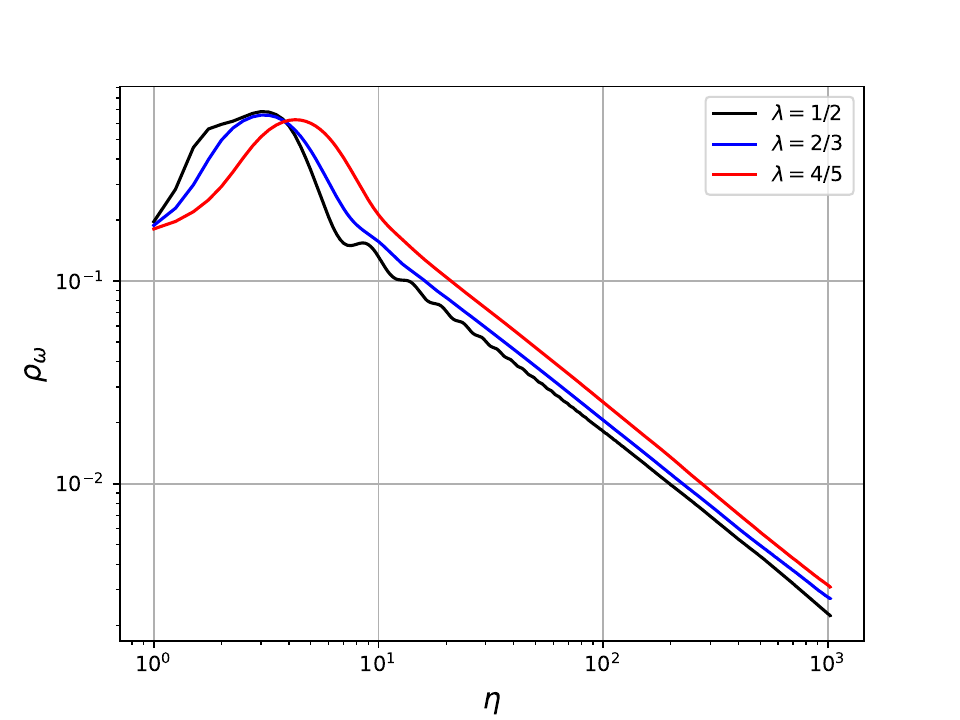}
    \includegraphics[width=1.0\columnwidth]{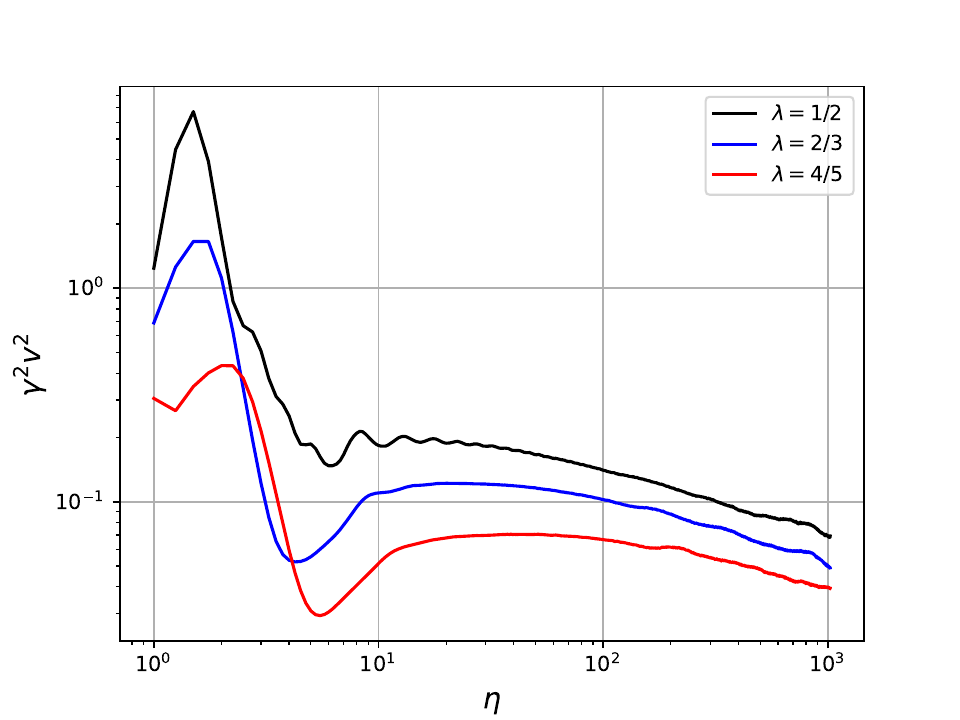}
    \includegraphics[width=1.0\columnwidth]{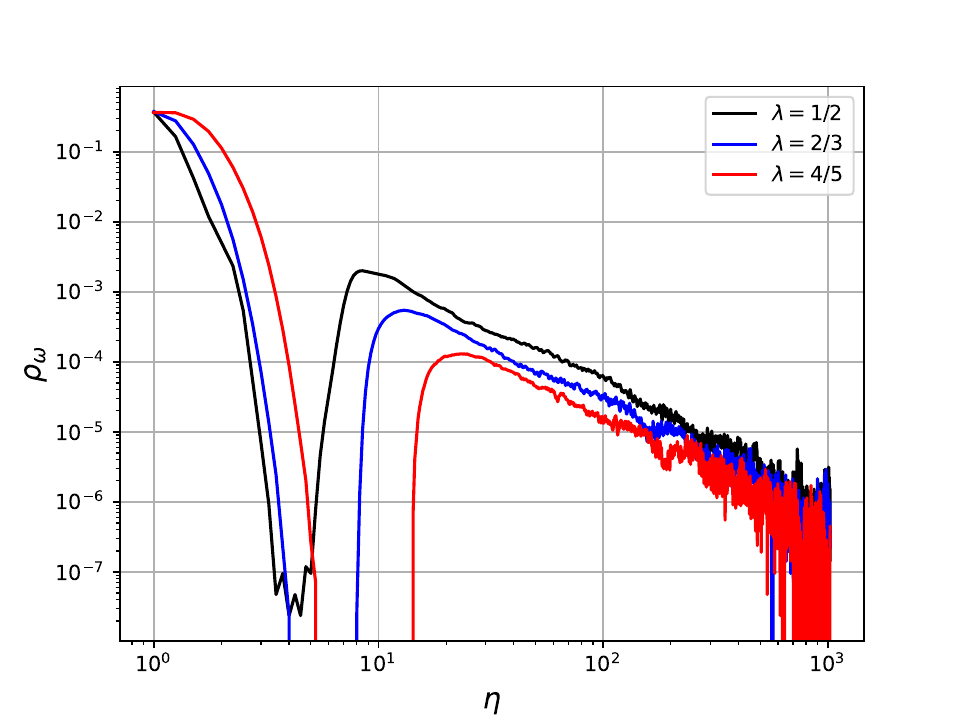}
    \includegraphics[width=1.0\columnwidth]{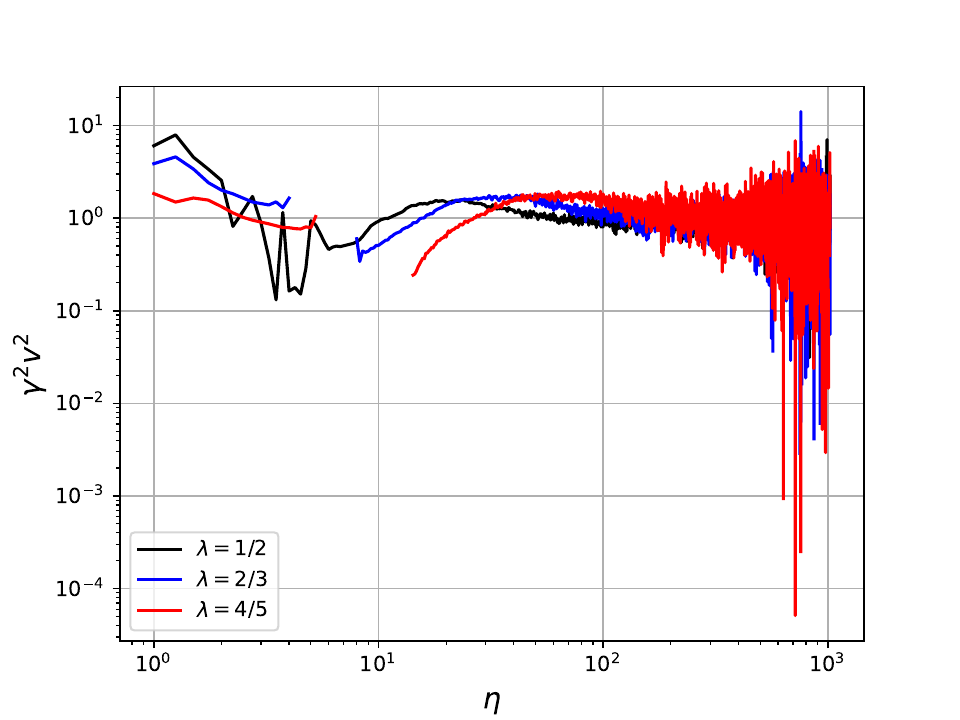}
    \includegraphics[width=1.0\columnwidth]{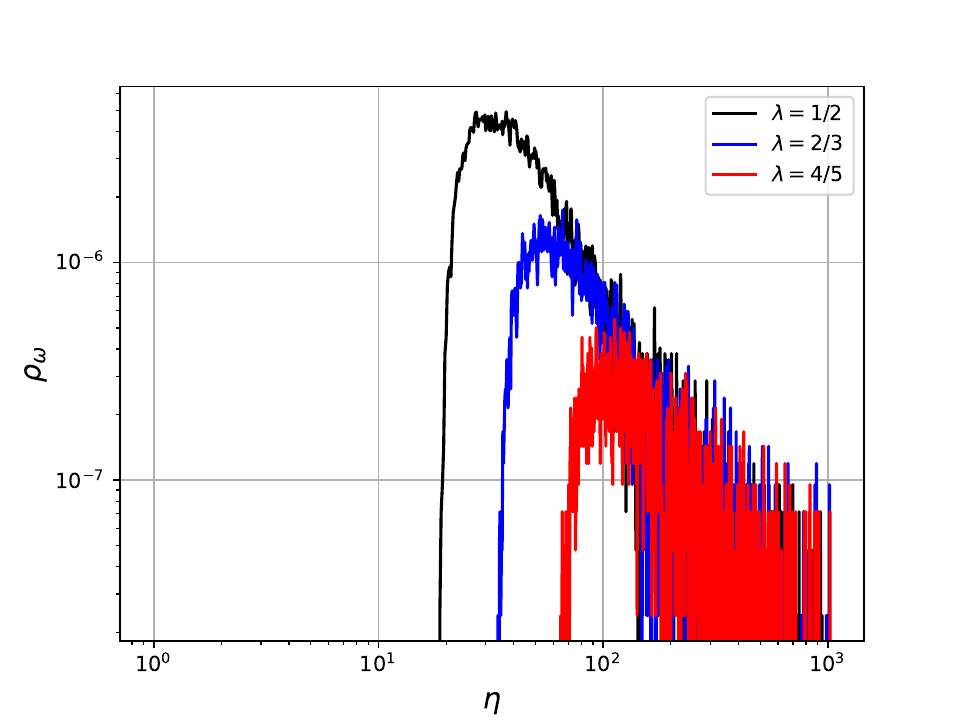}
    \includegraphics[width=1.0\columnwidth]{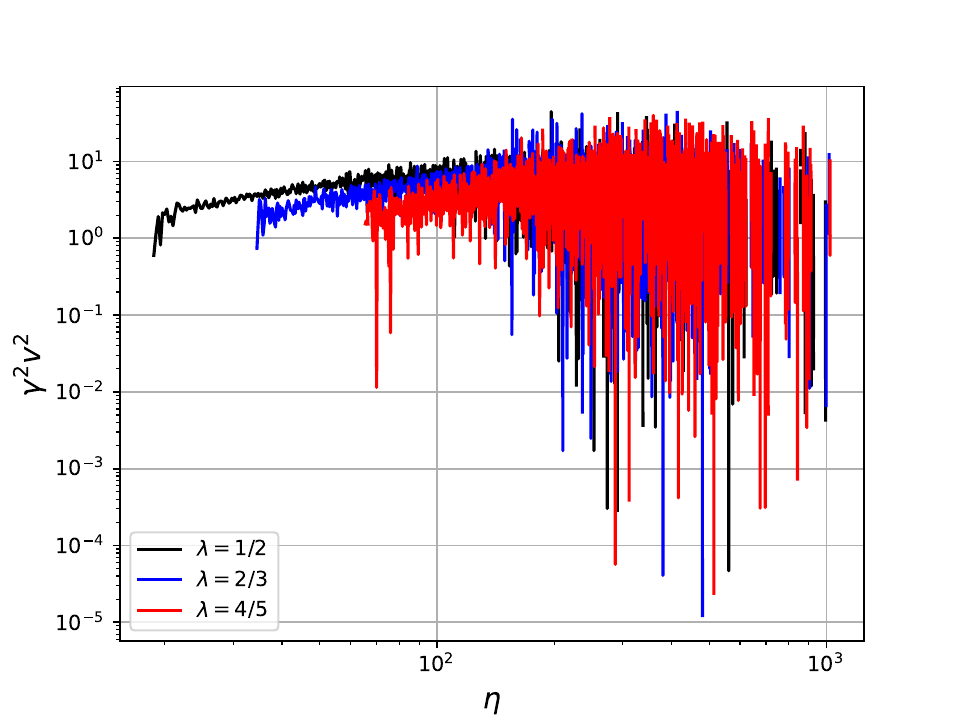}
   \caption{Evolution of the density ($\rho_w$, left side panels) and velocity ($\gamma^2v^2$, right side panels) of different types of domain walls in a Sine-Gordon potential as a function of conformal time for a box size of $2048^2$ and wide (specifically $\phi \in [-3\phi_0, 3\phi_0]]$) initial conditions. The top, middle and bottom plots correspond respectively to the type-I, type-II and type-III walls, as defined in Fig. \ref{fig04}. The plotted values are averages over sets of 10 simulations, with initial conditions generated from a set of 10 different random seeds.}\label{fig09}
\end{figure*}

\begin{table}
  \centering
  \caption{The scaling exponents $\mu$ and $\nu$ calculated for three different expansion rates for type-I and type-II walls in a Sine-Gordon potential. The results come from $2048^2$ simulation boxes with wide initial conditions. Each value was obtained by averaging over $10$ simulations, with one-sigma statistical uncertainties given throughout.}
\begin{tabular}{| c | c | c | c | c |}
\hline 
Case & Box size & $\lambda$ & $\mu$ & $\nu$  \\ \hline \hline
Type-I & $2048^2$ & $1/2$ & $-0.857\pm0.005$ & $-0.267\pm 0.064$ \\ \hline
Type-I & $2048^2$ & $2/3$ & $-0.897\pm0.005$ & $-0.203\pm 0.113$ \\ \hline
Type-I & $2048^2$ & $4/5$ & $-0.916\pm0.005$ & $-0.103\pm 0.141$\\ \hline \hline
Type-II & $2048^2$ & $1/2$ & $-1.598\pm0.016 $ & $ -0.211\pm0.022 $ \\ \hline
Type-II & $2048^2$ & $2/3$ & $-1.190\pm0.008 $ & $ -0.582\pm0.038 $ \\ \hline
Type-II & $2048^2$ & $4/5$ & $-1.415\pm0.016 $ & $ -0.840\pm0.057 $ \\ \hline
\end{tabular}
  \label{tab4}
\end{table}

The results are summarized in Fig.~\ref{fig09} and Table~\ref{tab4}. Although the early evolution of the simulations is quite different, while the relaxation of the initial conditions takes place, the late-time evolution is broadly in agreement with our earlier results, for the standard initial conditions. This includes the scaling exponents $\mu$ and $\nu$, with one exception: in this case the velocities of type-II walls now decrease (instead of remaining constant), the reason being that in this case some energy is shifting from type-II to type-I walls. At late times, most of the energy is in type-I walls.

An analogous modification consists of using initial conditions with the field uniformly distributed in two separate ranges, $[-3\phi_0,-\phi_0]$ and $[+\phi_0,+3\phi_0]$. This means that initially we nominally only have type-II walls. We have simulated the same three expansion rates ($\lambda=1/2$, $\lambda=2/3$, $\lambda=4/5$), with $2048^2$ boxes.

\begin{figure*}
\centering
    \includegraphics[width=1.0\columnwidth]{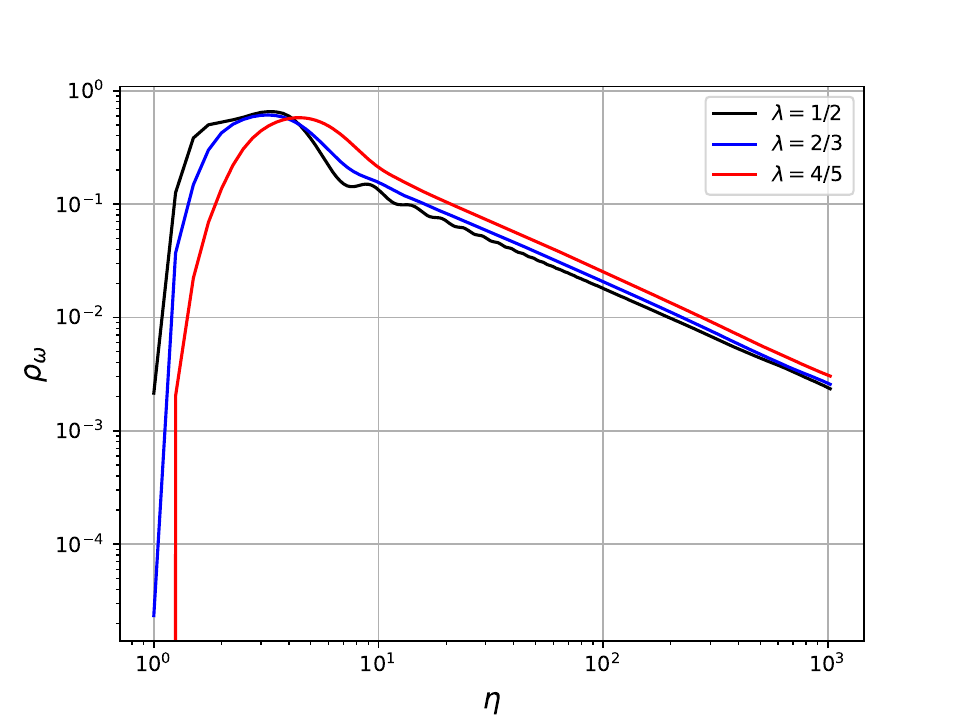}
    \includegraphics[width=1.0\columnwidth]{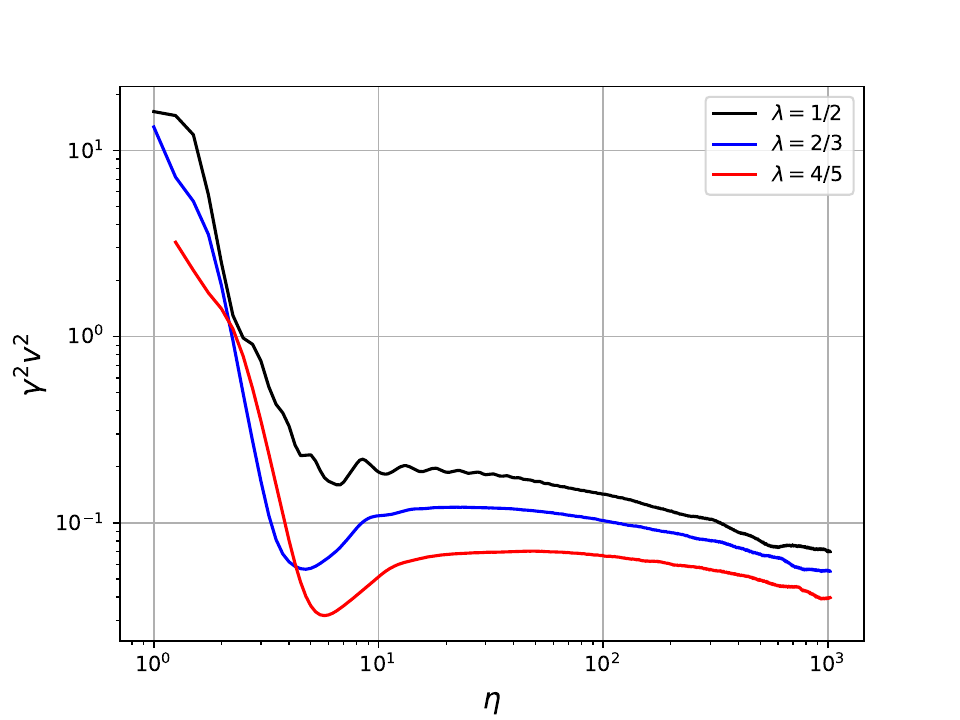}
    \includegraphics[width=1.0\columnwidth]{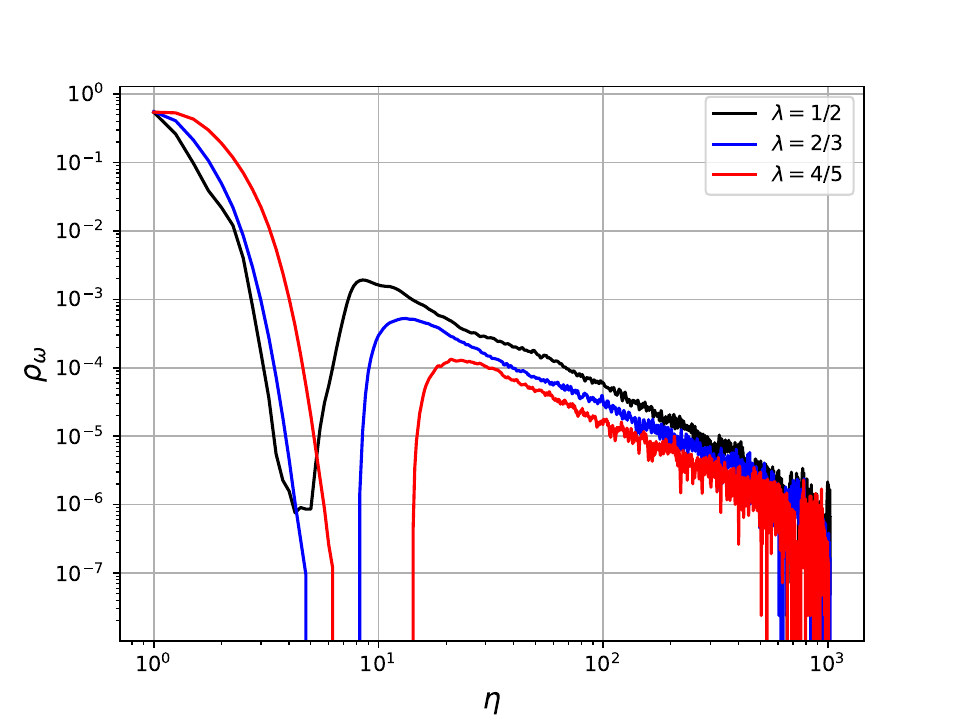}
    \includegraphics[width=1.0\columnwidth]{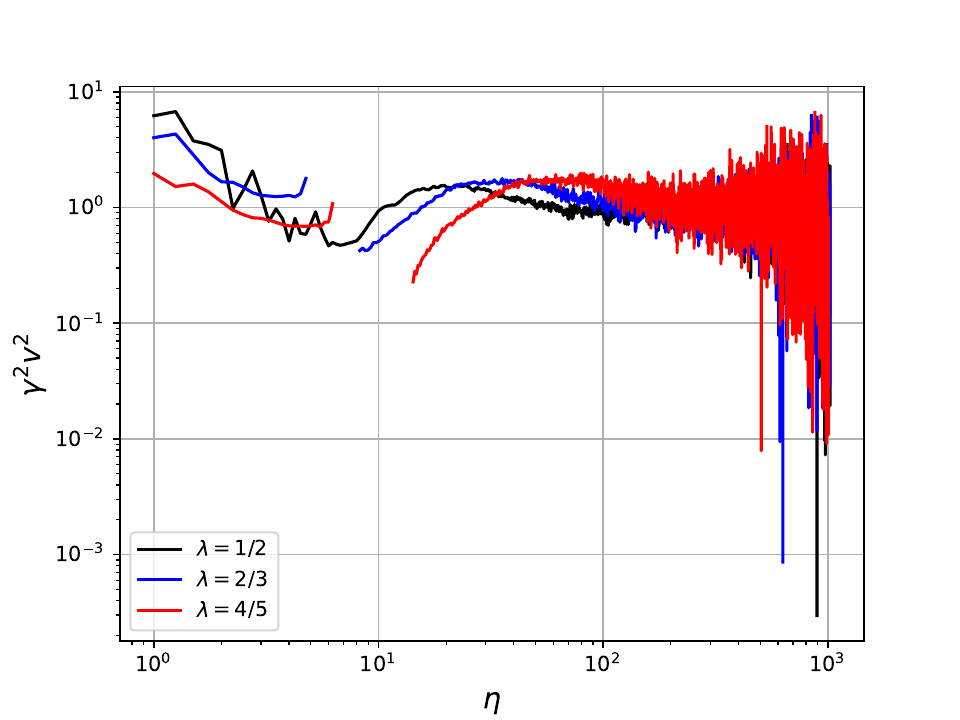}
    \includegraphics[width=1.0\columnwidth]{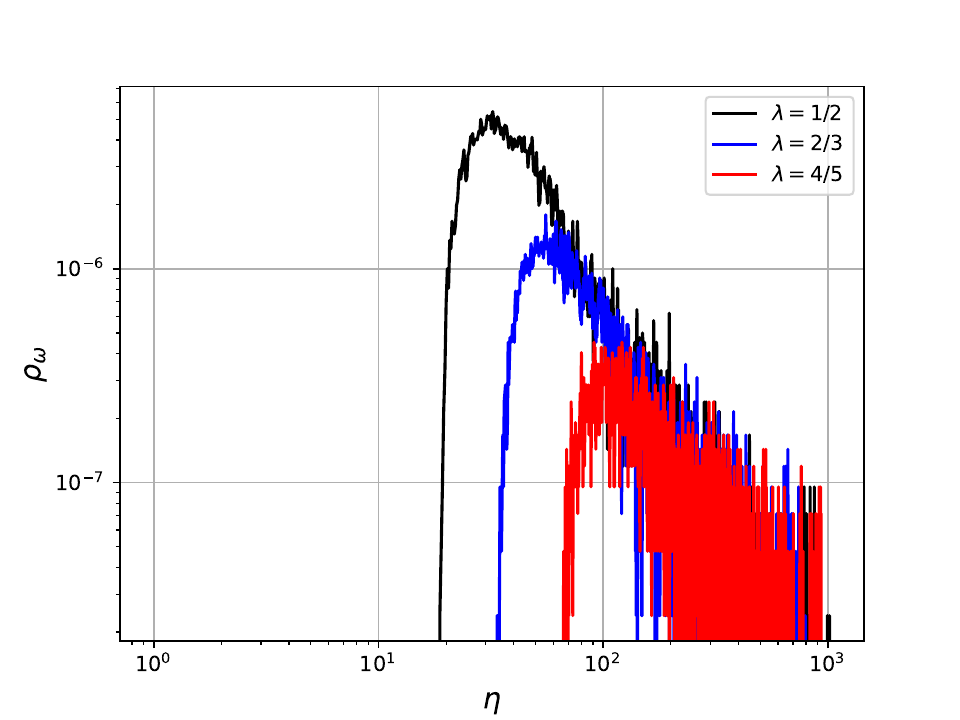}
    \includegraphics[width=1.0\columnwidth]{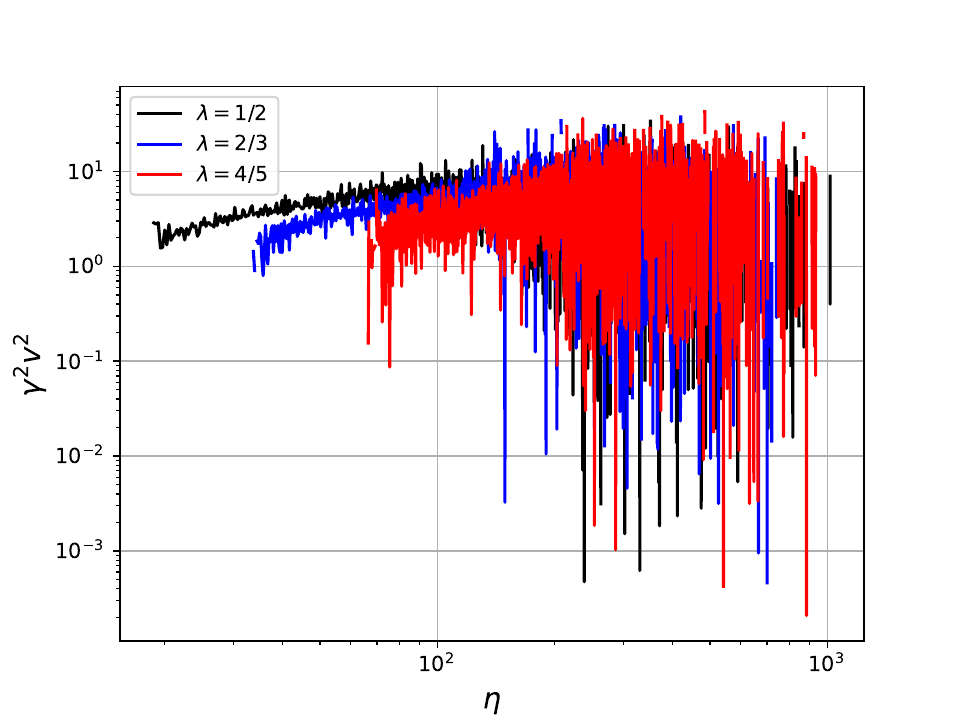}
   \caption{Evolution of the density ($\rho_w$, left side panels) and velocity ($\gamma^2v^2$, right side panels) of different types of domain walls in a Sine-Gordon potential as a function of conformal time for a box size of $2048^2$ and split initial conditions ($\phi \in [-3\phi_0, -\phi_0] \cup [+\phi_0, +3 \phi_0]$). The top, middle and bottom plots correspond respectively to the type-I, type-II and type-III walls, as defined in Fig. \ref{fig04}. The plotted values are averages over sets of 10 simulations, with initial conditions generated from a set of 10 different random seeds.}\label{fig10}
\end{figure*}

\begin{table}
  \centering
  \caption{The scaling exponents $\mu$ and $\nu$ calculated for three different expansion rates for type-I and type-II walls in a Sine-Gordon potential. The results come from $2048^2$ simulation boxes with split initial conditions. Each value was obtained by averaging over $10$ simulations, with one-sigma statistical uncertainties given throughout.}
\begin{tabular}{| c | c | c | c | c |}
\hline 
Case & box size & $\lambda$ & $\mu$ & $\nu$ \\ \hline \hline
Type-I & $2048^2$ & $1/2$ & $-0.854\pm0.003$ & $-0.294\pm 0.072$ \\ \hline
Type-I & $2048^2$ & $2/3$ & $-0.889\pm0.001$ & $-0.209\pm 0.148$ \\ \hline
Type-I & $2048^2$ & $4/5$ & $-0.909\pm0.003$ & $-0.092\pm 0.164$ \\ \hline \hline
Type-II & $2048^2$ & $1/2$ & $-1.554\pm0.072$ & $ -0.148\pm0.014 $ \\ \hline
Type-II & $2048^2$ & $2/3$ & $-1.314\pm0.013$ & $ -0.473\pm0.026 $ \\ \hline
Type-II & $2048^2$ & $4/5$ & $-1.576\pm0.016$ & $ -0.978\pm0.068 $ \\ \hline \hline
\end{tabular}
  \label{tab5}
\end{table}

The results are summarized in Fig.~\ref{fig10} and Table~\ref{tab5}. It is clear that they are, in a statistical sense, identical with those of the wide initial conditions case. Comparing the three choices of initial conditions so far, one may surmise that the preference for Type-I walls stems from the symmetry of the initial conditions with respect to $\phi_0=0$. This issue is further addressed in Appendix \ref{sineics}.

\subsection{Slower expansion rates}

Since high speeds are important to enable the field to jump across the maxima of the potential, it is also of interest to simulate slower expansion rates, and we consider six such expansion rates, ($\lambda=0$, $\lambda=1/100$, $\lambda=1/20$, $\lambda=1/10$, $\lambda=1/5$, $\lambda=1/3$). In the first of these there is no expansion (i.e., the network evolves in Minkowski space), which provides a numerical test of the code, especially for what concerns energy conservation. In these simulations we use standard initial conditions, with the field uniformly distributed in the range $[-\phi_0,+\phi_0]$, and we simulate $2048^2$ boxes.

\begin{figure*}
\centering
    \includegraphics[width=1.0\columnwidth]{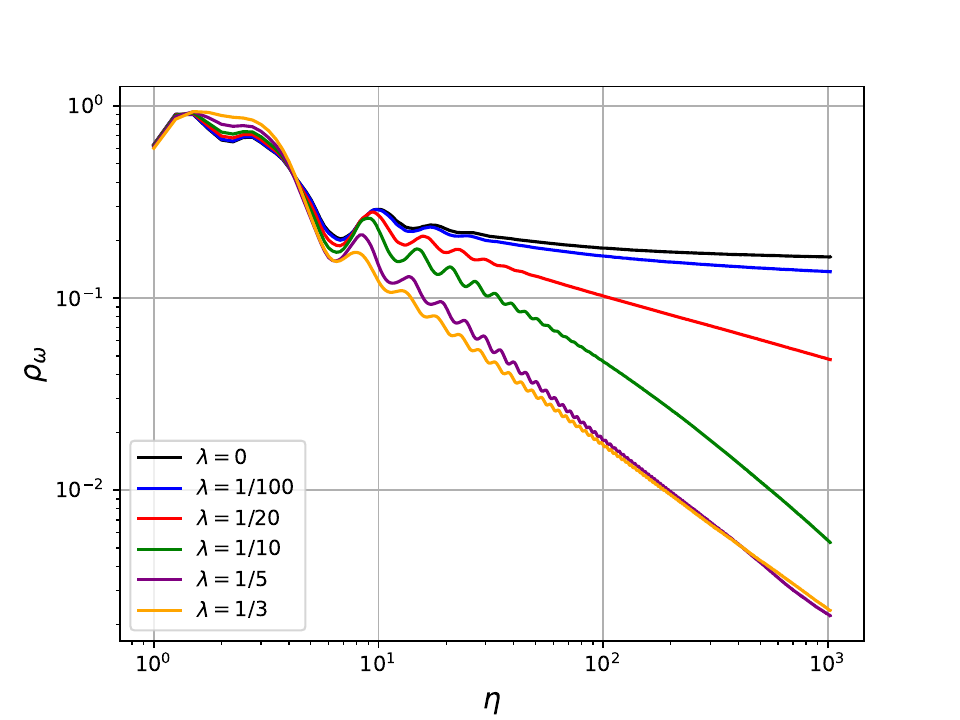}
    \includegraphics[width=1.0\columnwidth]{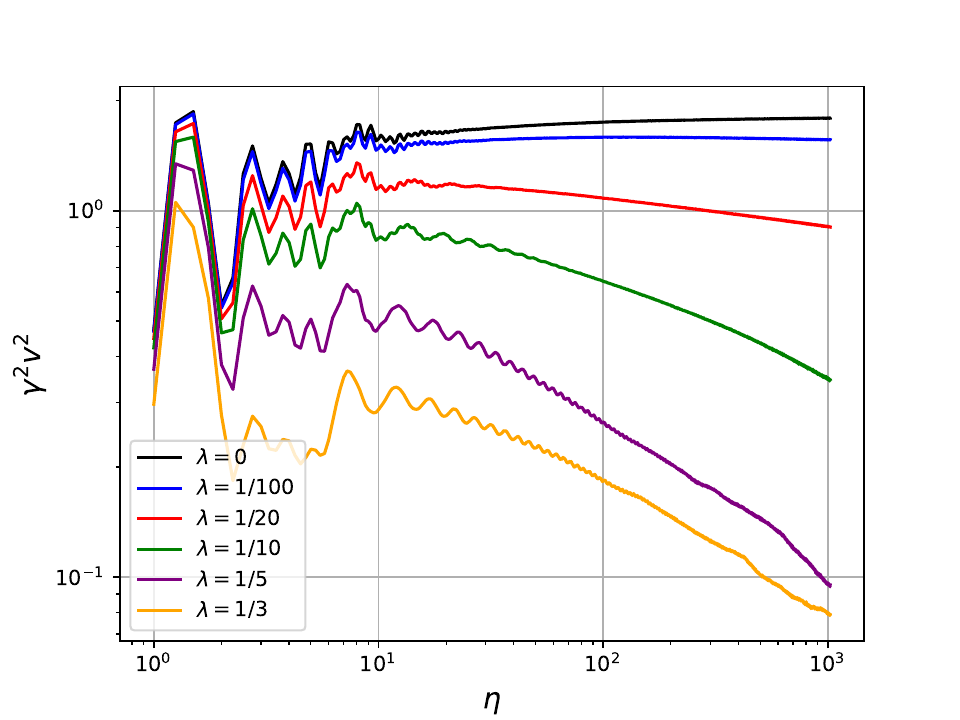}
    \includegraphics[width=1.0\columnwidth]{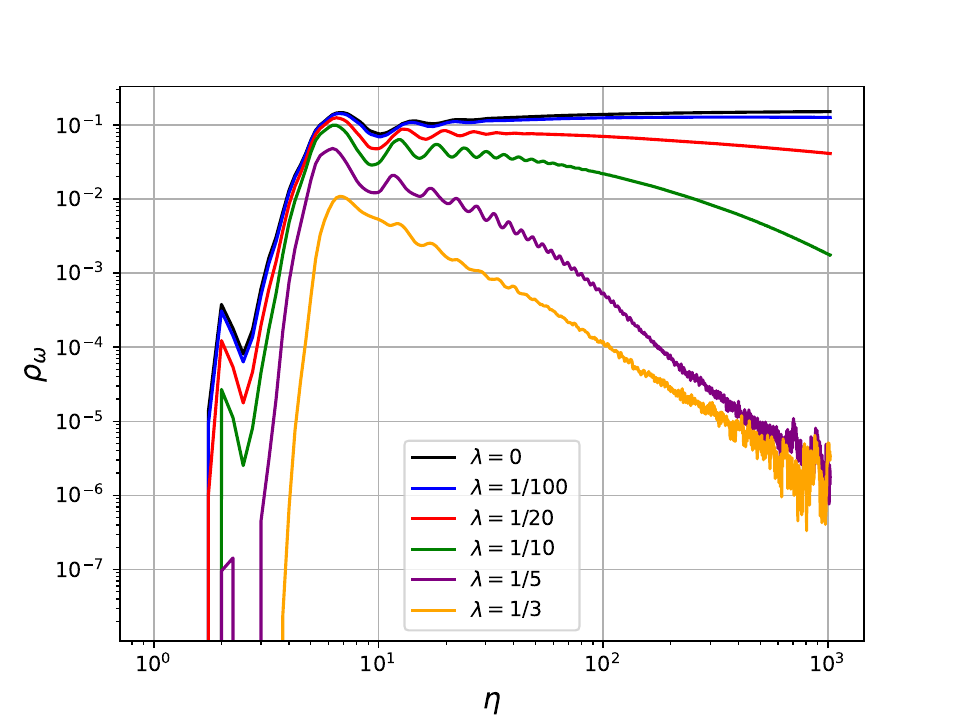}
    \includegraphics[width=1.0\columnwidth]{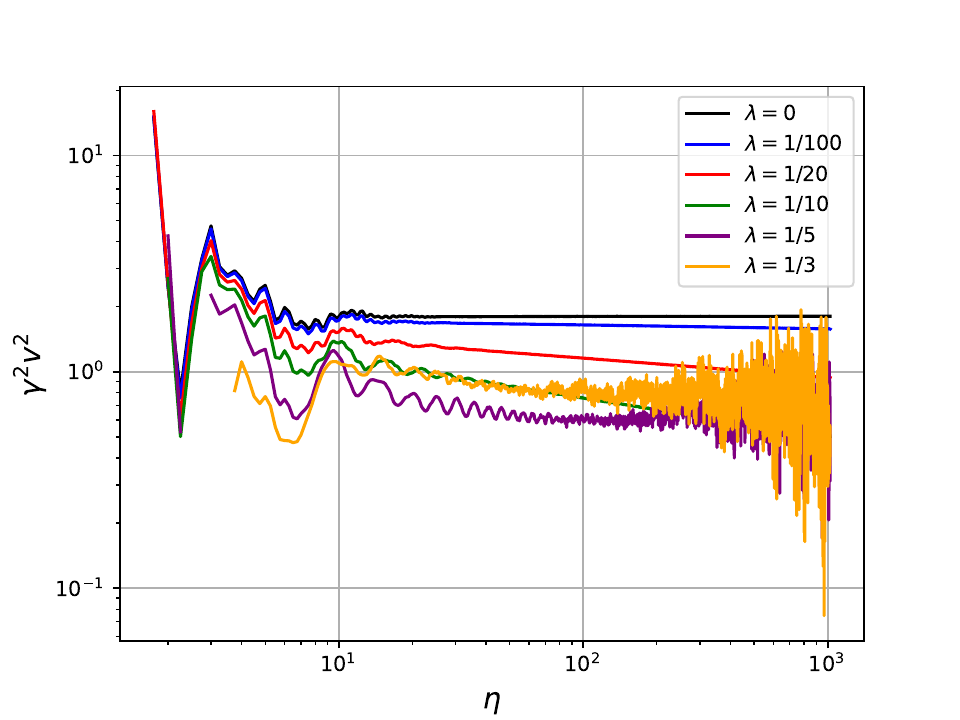}
    \includegraphics[width=1.0\columnwidth]{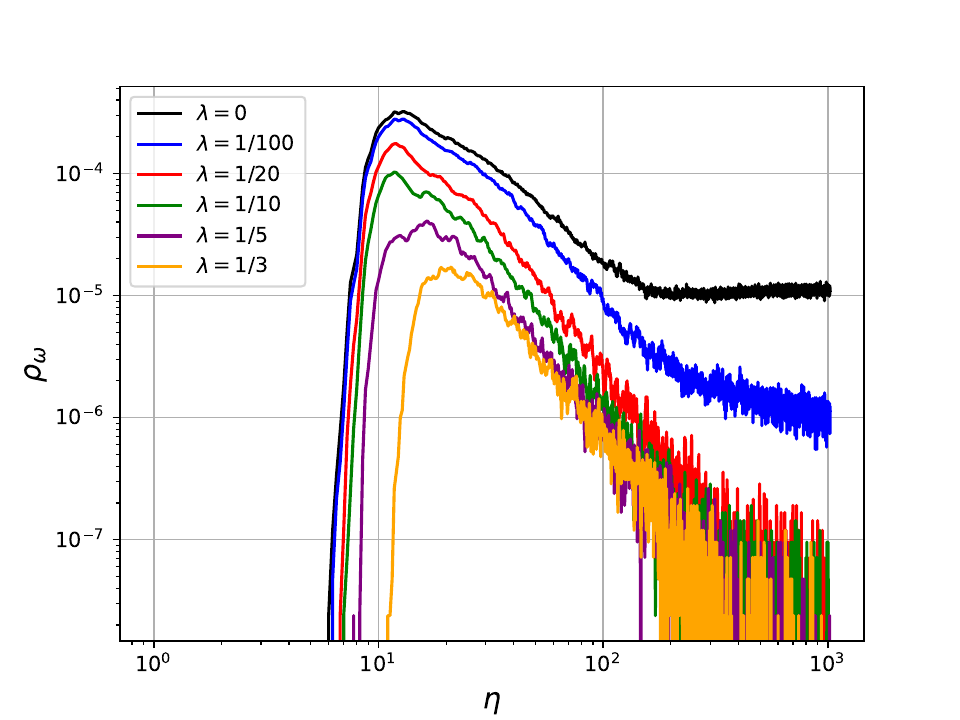}
    \includegraphics[width=1.0\columnwidth]{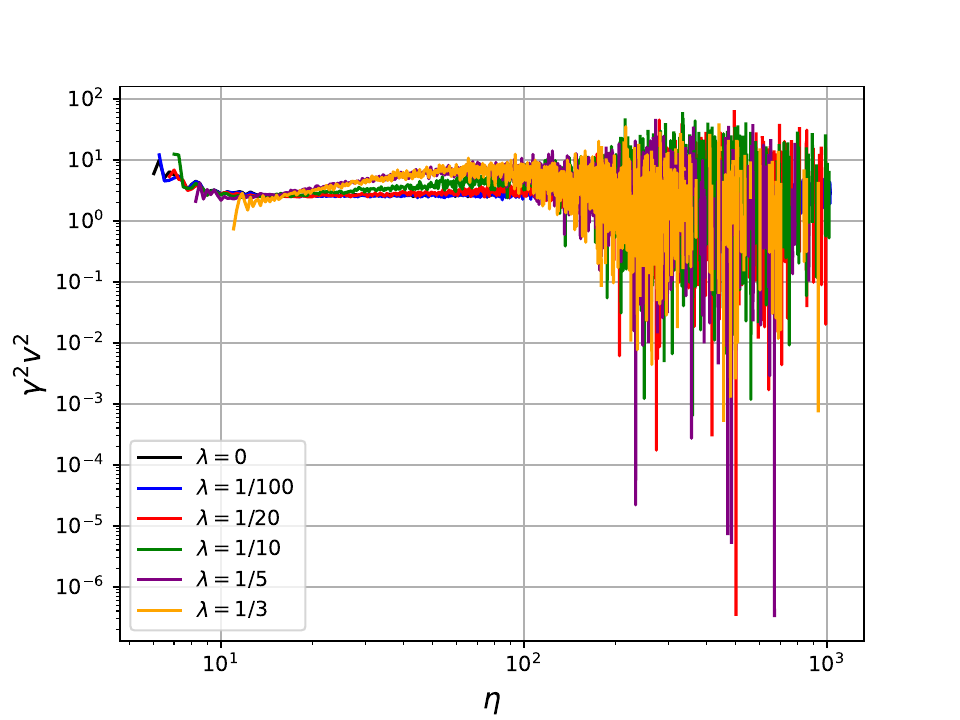}
   \caption{Evolution of the density ($\rho_w$, left side panels) and velocity ($\gamma^2v^2$, right side panels) of different types of domain walls in a Sine-Gordon potential as a function of conformal time for a box size of $2048^2$ and standard initial conditions in Minkowski and with slow expansion rates. The top, middle and bottom plots correspond respectively to the type-I, type-II and type-III walls, as defined in Fig. \ref{fig04}. The plotted values are averages over sets of 10 simulations, with initial conditions generated from a set of 10 different random seeds.}\label{fig11}
\end{figure*}

\begin{table}
  \centering
  \caption{The scaling exponents $\mu$ and $\nu$ calculated for two different low expansion rates and for a Minkowski background, for type-I and type-II walls in a Sine-Gordon potential. The results come from $2048^2$ simulation boxes with standard initial conditions. Each value was obtained by averaging over $10$ simulations, with one-sigma statistical uncertainties given throughout.}
\begin{tabular}{| c | c | c | c | c |}
\hline 
Case & box size & $\lambda$ & $\mu$ & $\nu$ \\ \hline \hline
Type-I & $2048^2$ & $0$   & $-0.041\pm0.001$ & $0.010\pm0.001$ \\ \hline
Type-I & $2048^2$ & $1/100$ & $-0.077\pm0.001$ & $-0.008\pm0.001$ \\ \hline
Type-I &  $2048^2$ & $1/20$ & $-0.327\pm0.001$ & $-0.079\pm0.001$ \\ \hline
Type-I & $2048^2$ & $1/10$ & $-0.913\pm0.006$ & $-0.278\pm0.004$ \\ \hline
Type-I & $2048^2$ & $1/5$ & $-0.918\pm0.003$ & $-0.439\pm0.007$ \\ \hline
Type-I & $2048^2$ & $1/3$ & $-0.858\pm0.002$ & $-0.385\pm0.005$\\ \hline \hline
Type-II & $2048^2$ & $0$   & $0.032\pm0.001$ & $0.001\pm0.001$ \\ \hline
Type-II & $2048^2$ & $1/100$ & $0.002\pm0.001$ & $-0.018\pm0.001$ \\ \hline
Type-II & $2048^2$ & $1/20$ & $-0.239\pm0.003$ & $-0.091\pm0.001$ \\ \hline
Type-II & $2048^2$ & $1/10$ & $-1.092\pm0.017$ & $-0.169\pm0.002$ \\ \hline
Type-II & $2048^2$ & $1/5$ & $-2.591\pm0.046$  & $-0.317\pm0.164$ \\ \hline
Type-II & $2048^2$ & $1/3$ & $-2.051\pm0.049$ & $-0.689\pm0.263$ \\ \hline 
\end{tabular}
  \label{tab6}
\end{table}

Fig.~\ref{fig11} and Table~\ref{tab6} summarize the results. We note that the slower expansion rates imply that there will be additional scalar radiation in the simulation boxes, which makes the identification of the domain walls, and therefore the determination of the densities and velocities and the corresponding scaling exponents, more challenging. In other words, in this case there could be systematic biases impacting the estimators. Such biases are difficult to quantify numerically (doing so is beyond the scope of this work), but these would lead to systematic uncertainties in the scaling exponents, which would add to the statistical uncertainties which we do report.

With that caveat in mind, we nevertheless find that for the cases with moderate expansion ($\lambda=1/5$ and $\lambda=1/3$) the $\mu$ scaling exponent for type-I walls is compatible with the one found earlier for the fast expansion rates, while $\nu$ is slightly more negative (i.e., velocities are decreasing slightly faster here, plausibly due to the enhanced presence of scalar radiation). For type-II walls we have the opposite behavior: the value of $\nu$ is approximately the same as in the faster expansion cases, while $\mu$ is more negative here (i.e., the ratio of energies in type-II and type-I walls decreases faster here).

On the other hand, in Minkowski space ($\lambda=0$) we find $\mu\sim\nu\sim0$ for both type-I and type-II walls, as expected. Moreover, this is also (approximately) the case for type-III walls, although again their scarcity precludes a meaningful quantitative determination of the two exponents in this case. Indeed in this case we do observe a small but persistent population of type-III walls: the fraction of the network's energy density in these walls is approximately constant (though with large fluctuations), while in the previously simulated cases the damping due to the universe's expansion makes such a population transient.

The simulations with very small expansion rates ($\lambda=1/100$, $\lambda=1/20$, $\lambda=1/10$) show that the domain wall evolution properties do not undergo a sudden change when going from the Minkowski to the expanding case. The physically distinctive feature of the $\lambda=0$ is that there is no damping, which does exist for any positive value thereof. Obviously, a very small $\lambda$ means very little damping, so a behavior relatively similar to $\lambda=0$, effectively leading to a smooth transition between the two regimes.
As was done for the faster expansion rates, it is useful to compare the asymptotic ratios of the velocities of type-II and type-I walls
\begin{equation}
\left[\frac{(\gamma v)^2_{II}}{(\gamma v)^2_{I}}\right]_{\lambda=0}\sim 1.0
\end{equation}
\begin{equation}
    \left[\frac{(\gamma v)^2_{II}}{(\gamma v)^2_{I}}\right]_{\lambda=1/10}\sim 1.5\,
\end{equation}
\begin{equation}
\left[\frac{(\gamma v)^2_{II}}{(\gamma v)^2_{I}}\right]_{\lambda=1/5}\sim 2.5
\end{equation}
\begin{equation}
\left[\frac{(\gamma v)^2_{II}}{(\gamma v)^2_{I}}\right]_{\lambda=1/3}\sim 4.5\,,
\end{equation}
and for Minkowski space simulations we can also extract this ratio for type-III walls
\begin{equation}
\left[\frac{(\gamma v)^2_{III}}{(\gamma v)^2_{I}}\right]_{\lambda=0}\sim 1.7
\end{equation}
These ratios are significantly smaller than those for faster expansion rates, confirming the role of the different amounts of damping present in each case. We do not explicitly report the values of these ratios for the very small expansion rates ($\lambda=1/100$, $\lambda=1/20$) because they are, within numerical uncertainties, similar to the ones obtained for $\lambda=0$.

\begin{figure*}
\centering
    \includegraphics[width=1.0\columnwidth]{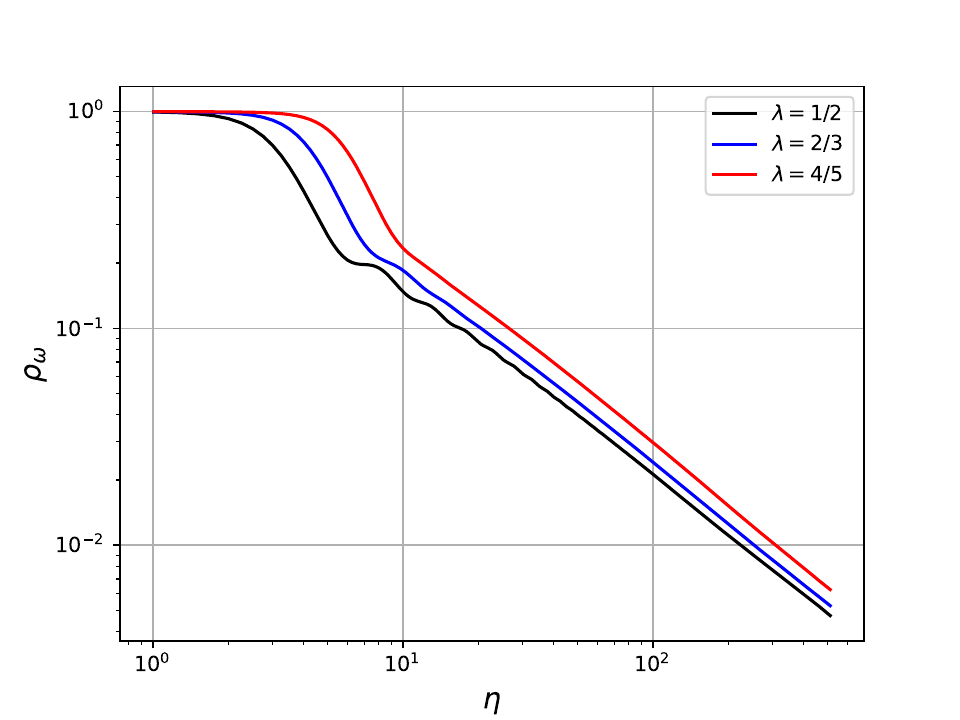}
    \includegraphics[width=1.0\columnwidth]{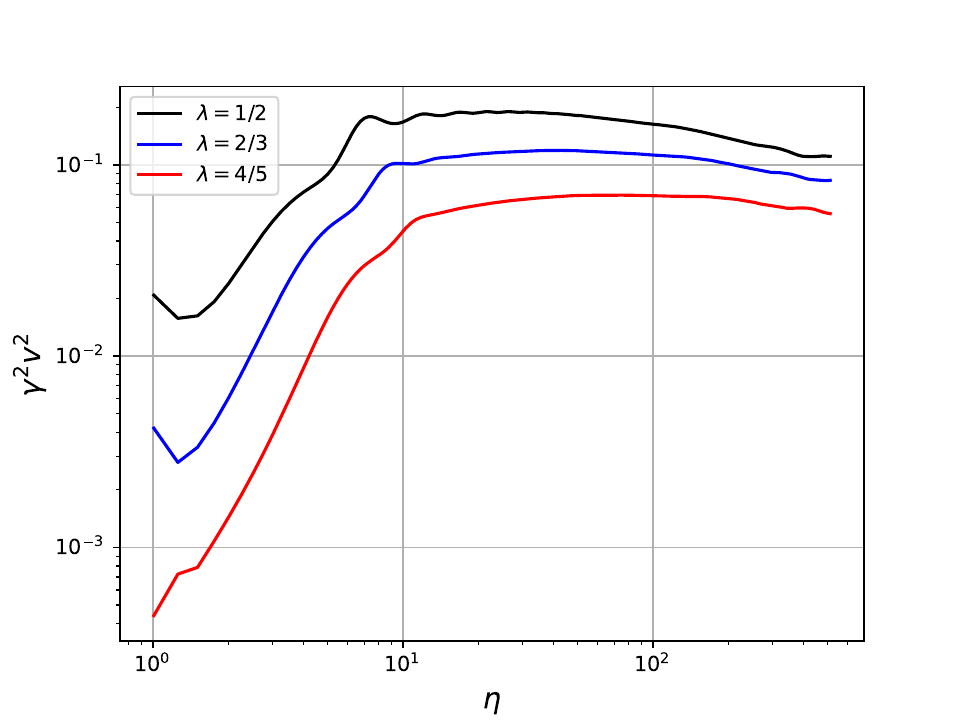}
    \includegraphics[width=1.0\columnwidth]{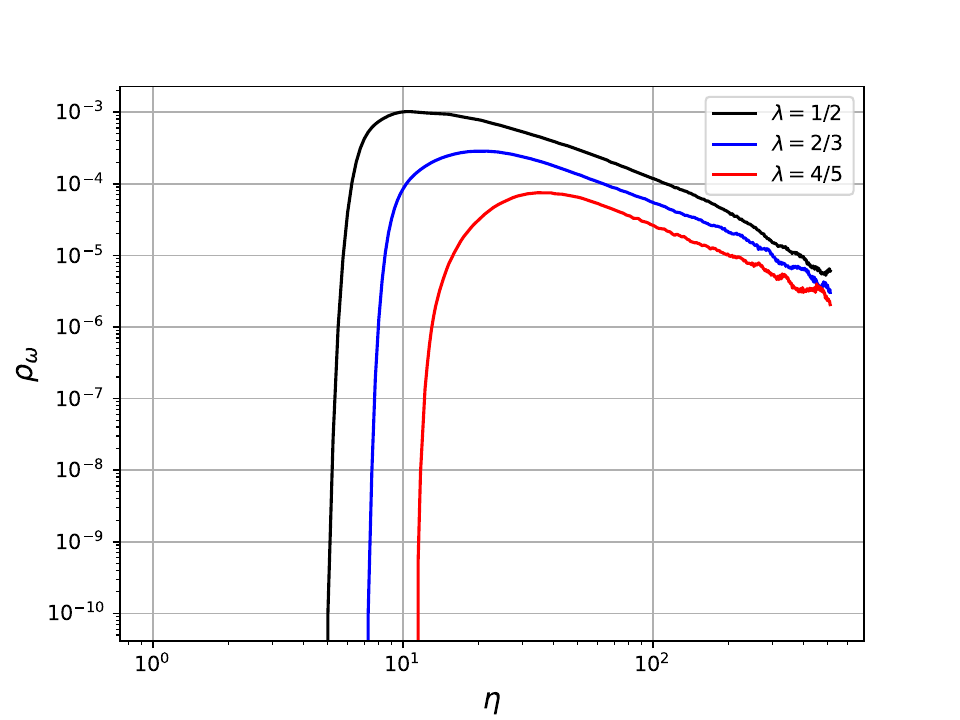}
    \includegraphics[width=1.0\columnwidth]{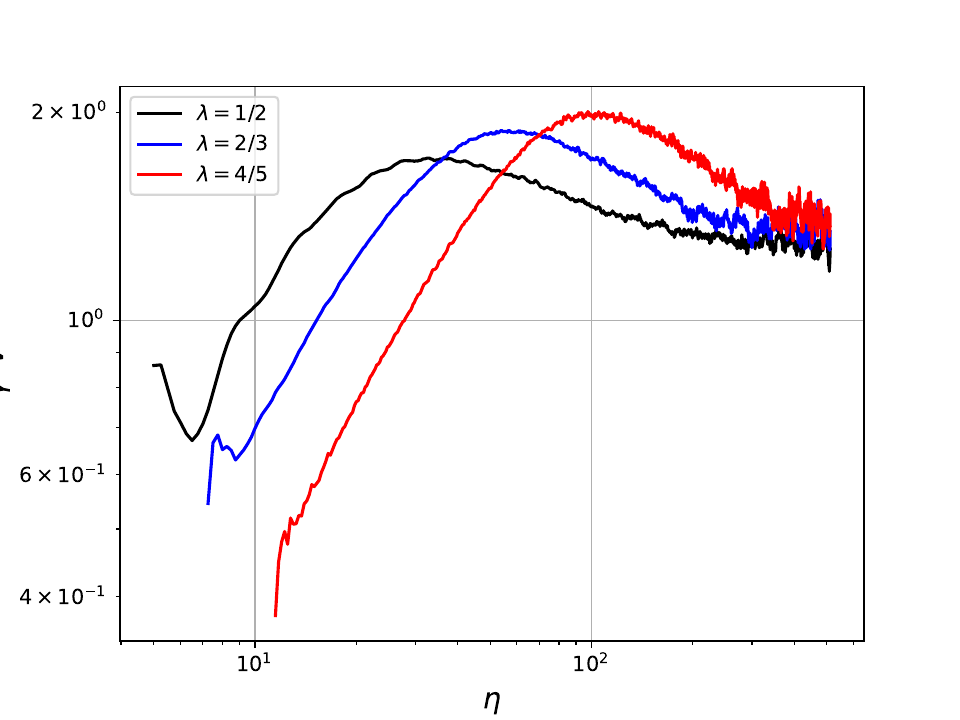}
    \includegraphics[width=1.0\columnwidth]{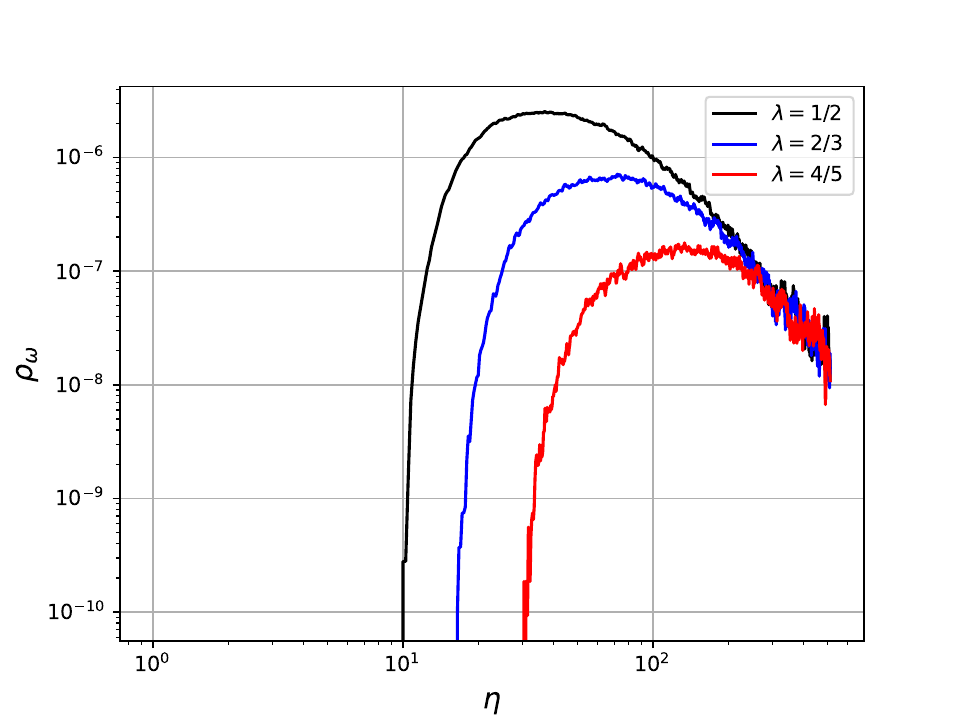}
    \includegraphics[width=1.0\columnwidth]{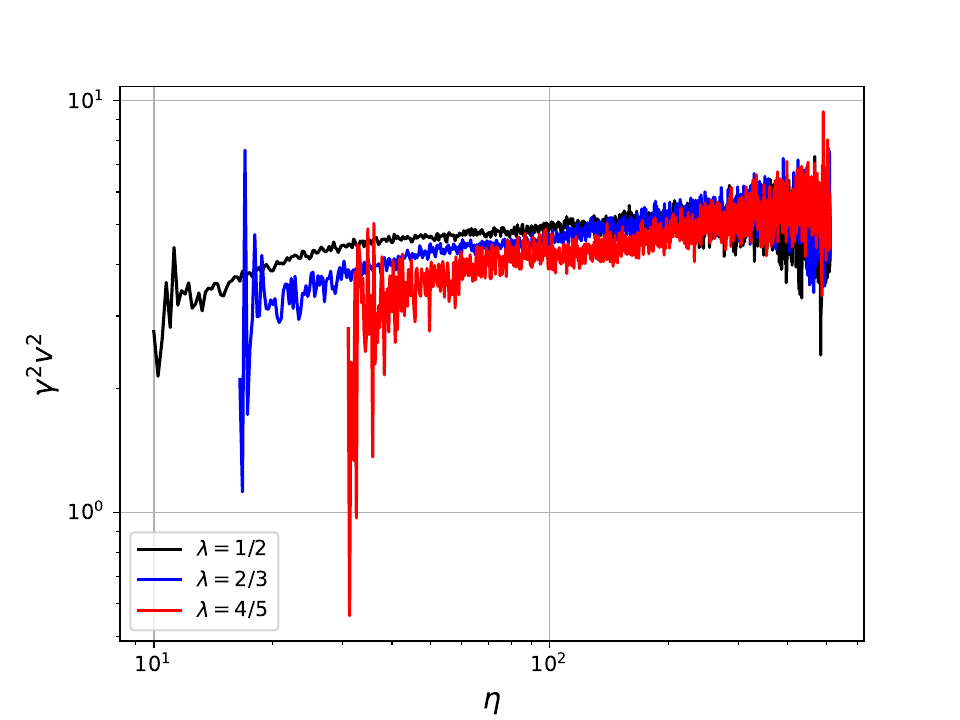}
   \caption{Evolution of the density (left side panels) and velocity ($\gamma^2v^2$, right side panels) of different types of domain walls in a Sine-Gordon potential as a function of conformal time for a box size of $1024^3$ and standard initial conditions in radiation, matter and with fast expansion epochs and with an initial cooling time of $\eta_\mathrm{cooling}=1$. The top, middle and bottom plots correspond respectively to the type-I, type-II and type-III walls, as defined in Fig.~\ref{fig04}. The plotted values are averages over sets of 10 simulations, with initial conditions generated from a set of 10 different random seeds.}\label{fig12}
\end{figure*}

\begin{table}
  \centering
  \caption{The scaling exponents $\mu$ and $\nu$ calculated for three different expansion rates for type-I and type-II walls in a Sine-Gordon potential. The results come from $1024^3$ 3D simulation boxes with standard initial conditions. Each value was obtained by averaging over $10$ simulations, with one-sigma statistical uncertainties given throughout.}
\begin{tabular}{| c | c | c | c | c |}
\hline 
Case & box size & $\lambda$ & $\mu$ & $\nu$ \\ \hline \hline
Type-I & $1024^3$ & $1/2$ & $-0.917\pm0.003$ & $-0.290\pm0.008$ \\ \hline
Type-I & $1024^3$ & $2/3$ & $-0.929\pm0.004$ & $-0.215\pm0.012$ \\ \hline
Type-I & $1024^3$ & $4/5$ & $-0.951\pm0.003$ & $-0.126\pm0.011$ \\ \hline \hline
Type-II & $1024^3$ & $1/2$ & $-1.990\pm0.047$ & $-0.033\pm0.026$ \\ \hline
Type-II & $1024^3$ & $2/3$ & $-1.701\pm0.072$ & $-0.077\pm0.028$ \\ \hline
Type-II & $1024^3$ & $4/5$ & $-1.497\pm0.071$ & $-0.252\pm0.028$ \\ \hline 
\end{tabular}
  \label{tab7}
\end{table}

\subsection{Three-dimensional simulations}

Finally, we check whether our results depend on the fact that we have carried out simulations with two spatial dimensions (which are convenient since they provide a large dynamic range for the available amount of memory). To this end, we have also carried out $1024^3$ simulations, for our baseline expansion rates ($\lambda=1/2$, $\lambda=2/3$, $\lambda=4/5$) and using our standard initial conditions, with the field uniformly distributed in the range $[-\phi_0,+\phi_0]$.

Fig.~\ref{fig12} and Table~\ref{tab7} summarize the results. Here we note that these simulations have a smaller dynamic range than the smallest of our two-dimensional simulations. Nevertheless, the fitted scaling exponents are in broad agreement with the ones obtained in the earlier simulations, with the exception of $\mu$ for type-II walls in the radiation and matter eras (for the faster expansion rate, $\lambda=4/5$, the results are fully consistent). Again it is useful to compare the asymptotic ratios of the velocities of type-II and type-I walls
\begin{equation}
\left[\frac{(\gamma v)^2_{II}}{(\gamma v)^2_{I}}\right]_{\lambda=1/2}\sim 10.5
\end{equation}
\begin{equation}
\left[\frac{(\gamma v)^2_{II}}{(\gamma v)^2_{I}}\right]_{\lambda=2/3}\sim 15.2
\end{equation}
\begin{equation}
\left[\frac{(\gamma v)^2_{II}}{(\gamma v)^2_{I}}\right]_{\lambda=4/5}\sim 25.3\,,
\end{equation}
These ratios are slightly larger than those for two-dimensional simulations with the same expansion rates $\lambda$, but it is not clear that these differences are statistically significant, since obtaining accurate error bars for each of the ratios is not straightforward due to the sparsity of the data.

\section{\label{chap:sextic}A Simple Sextic Potential}

\begin{figure}
  \begin{center}
    \leavevmode
    \includegraphics[width=1.0\columnwidth]{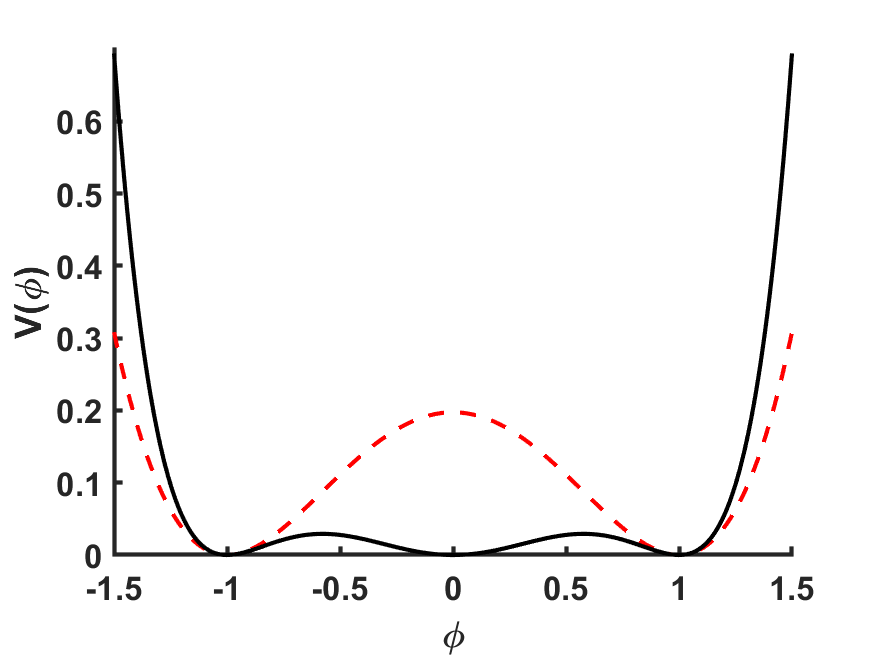}
    \caption{The sextic potential of Eq. (\ref{pot6}) (black solid line), compared to the quartic potential of Eq. (\ref{phi4-potential}) (red dashed line). Note that the local curvature of the potential is the same at their common minima.}
    \label{fig13}
  \end{center}
\end{figure}

Let us now consider a simple triple well potential
\begin{equation}
    \label{pot6}
     V\left( \phi \right) = \frac{\pi^2}{50} \; \phi^2 \left(1-\phi^2 \right)^2\,.
\end{equation}
This can be visually compared to the quartic potential in Fig.~\ref{fig13}. The common normalization of the two potentials ensures that they both have the same local curvature at the common minima, $\phi=\pm1$, as in the previously discussed Sine-Gordon case. This ensures that the field has the same dynamics in the vacuum as in the $\phi^4$ case. On the other hand, as is evident from the figure, this choice makes the height of the local maxima of the potential much smaller in the sextic case.

\begin{figure}
  \begin{center}
    \leavevmode
    \includegraphics[width=1.0\columnwidth]{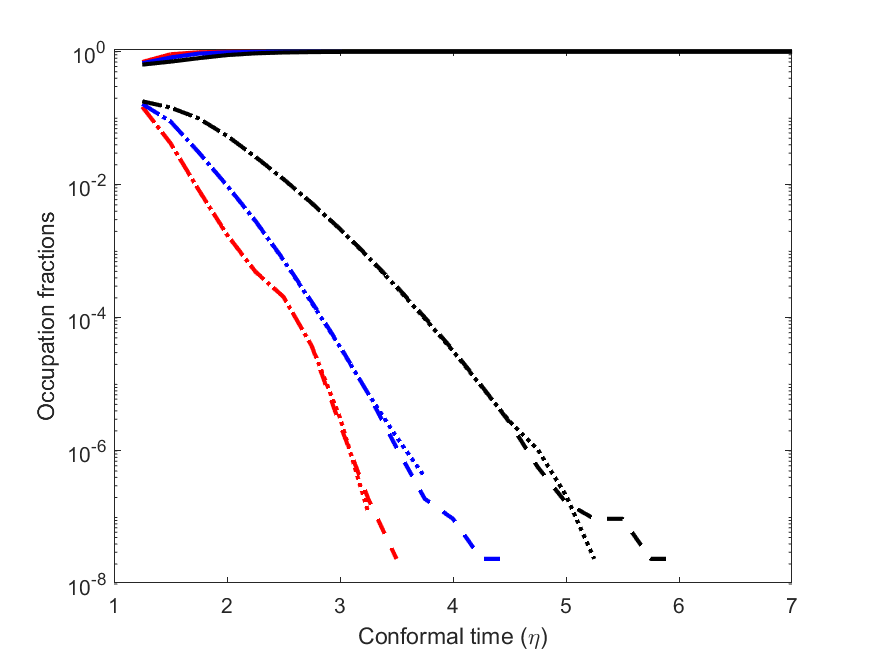}
    \caption{Evolution of the fraction of the box in which the field is located in the middle minimum ($|\phi|\le1/\sqrt{3}$, solid lines), the left minimum ($\phi<-1/\sqrt{3}$,  dashed lines), and the right minimum ($\phi>1/\sqrt{3}$, dotted lines) for a $\phi^6$ potential with symmetric initial conditions, $|\phi|\le1$. The data comes from  $2048^2$ simulations, but only the non-trivial early time steps are plotted.  Red, blue and back curves correspond to $\lambda=1/2$, $\lambda=2/3$ and $\lambda=4/5$ respectively. The plotted curves are averaged over 10 different simulations with different random initial conditions seeds.}
    \label{fig14}
  \end{center}
\end{figure}

If one chooses initial conditions with a uniformly distributed field between the three potential minima, that is $|\phi|\le1$, the outcome is predictably uninteresting: no stable defects form for any expansion rate, since the field rapidly evolves towards the middle potential minimum, $\phi=0$. Nevertheless, this decay is slightly slower for faster expansion rates, since typical field speeds are smaller in this case. Fig.~\ref{fig14} shows the outcome of simulations, expressed as the fraction of the grid points in the region of each of the three minima. Note that the boundary between them---in other words, the local maxima of the potential---occurs at $\phi=\pm1/\sqrt{3}$. It is remarkable that the decay only takes a dozen or so simulation time steps.

\begin{figure}
  \begin{center}
    \leavevmode
    \includegraphics[width=1.0\columnwidth]{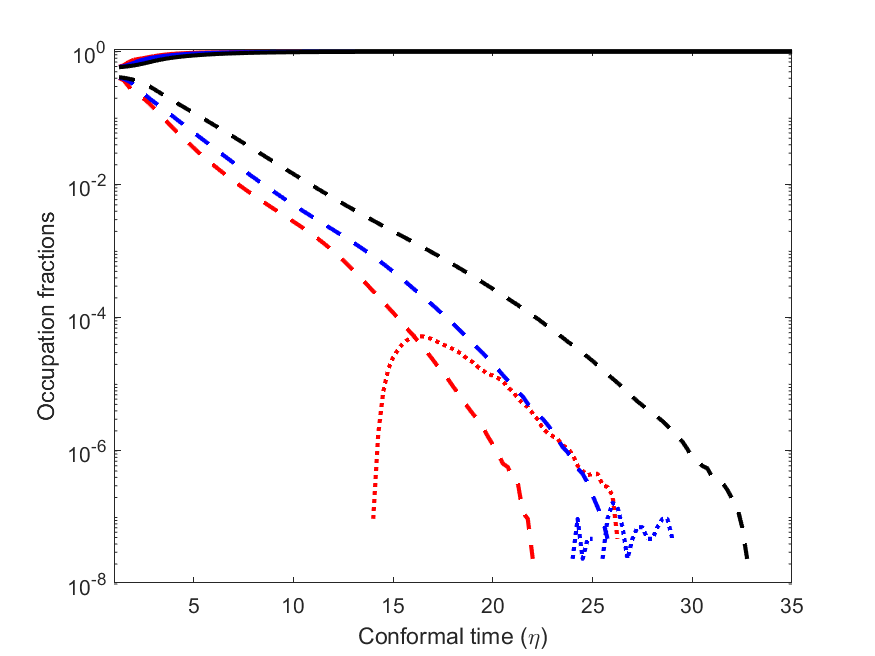}
    \caption{Same as Fig. \ref{fig14}, but with biased initial conditions, $-1\le\phi\le0$. Note the different range on the horizontal axis, and the temporary occupation of the rightmost minimum in the radiation era case (dotted red curve).}
    \label{fig15}
  \end{center}
\end{figure}

Still, it is interesting to check whether this behavior can be changed by other choices of initial conditions, as was checked for the Sine-Gordon case, and specifically by biased initial conditions, as has been done in detail for the standard $\phi^4$ case \cite{biases1,biases2}. Asymmetries do occur in nature, such as in the case of the open baryon asymmetry problem, and one may therefore investigate whether the introduction of biased initial conditions in a $\phi^6$ potential can lead to domain walls without the introduction of additional constraints in the potential itself.

Towards this end, we carried out an analogous set of simulations with the initial field distribution being uniform only in the leftmost half of the potential, that is $-1\le\phi\le0$, the results of which are shown in Fig.~\ref{fig15}. Clearly the end result is the same, although the decay towards the middle minimum is noticeably slower, by about a factor of six in conformal time. It is worthy of note that in the radiation era, which is the slowest depicted expansion rate, the rightmost minimum (which is initially unoccupied) does become temporarily occupied, as the result of the large oscillations that are a feature of the numerical choice of initial conditions. However, this only occurs for a small number of simulation time steps, and is fully eliminated by the increased damping associated with faster expansion rates. For our $2048^2$ box size simulations, only a handful of lattice sites are temporarily found in the rightmost minimum in the matter era simulations, and none are found therein for the faster expansion rate.

These results therefore suggest that, for this specific potential, the only conceivable scenario in which domain walls could form and survive for a cosmologically nontrivial amount of time would be if they are overdamped. Otherwise, the symmetry of the potential makes the central minimum effectively preferred, a full occupation of the central minimum being energetically preferred to equipartition between the three minima. Although the initial conditions of our simulations included vanishing field speeds, adding such speeds would clearly not have changed the final outcome. In the following section, we explore how these results may change for a more generic sextic triple well potential.

\section{\label{chap:christlee}The Christ-Lee Potential}

We now investigate the possibility of wall formation in a triple well potential by introducing a parameter in the potential which makes the central minimum metastable. A convenient choice is a Christ-Lee type potential \cite{chris-lee,Demirkaya},
\begin{equation}
    \label{potSG}
     V\left( \phi \right) = \frac{\pi^2}{50} \left(\frac{\phi^2+\varepsilon^2} {1+\varepsilon^2}\right) \left(1-\phi^2\right)^2\,.
\end{equation}
The parameter $\varepsilon$ interpolates the potential between $\phi^4$ (when $\varepsilon\to\infty$) and $\phi^6$ (when $\varepsilon=0$).

\begin{figure*}
  \begin{center}
    \leavevmode
    \includegraphics[width=1.0\columnwidth]{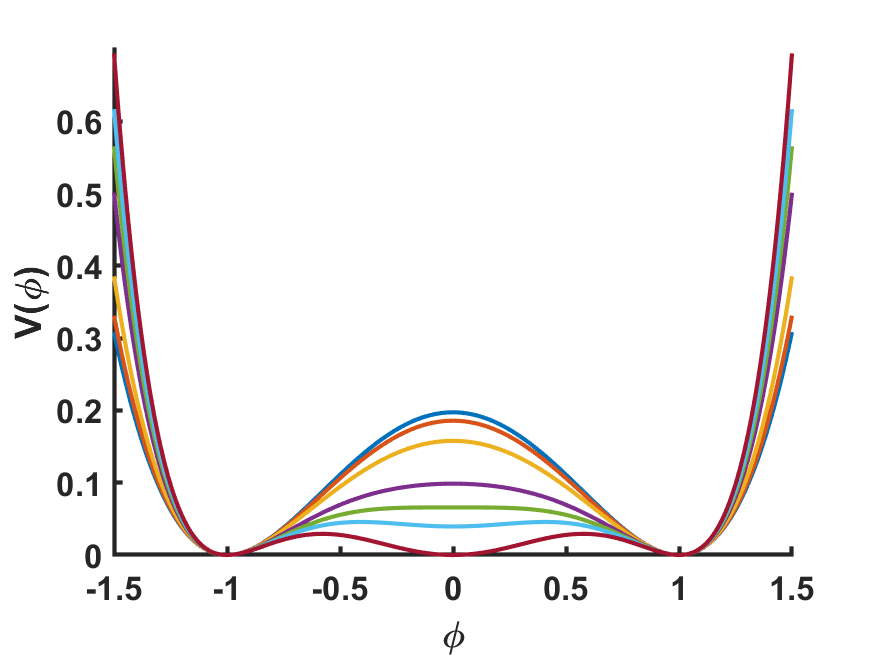}
    \includegraphics[width=1.0\columnwidth]{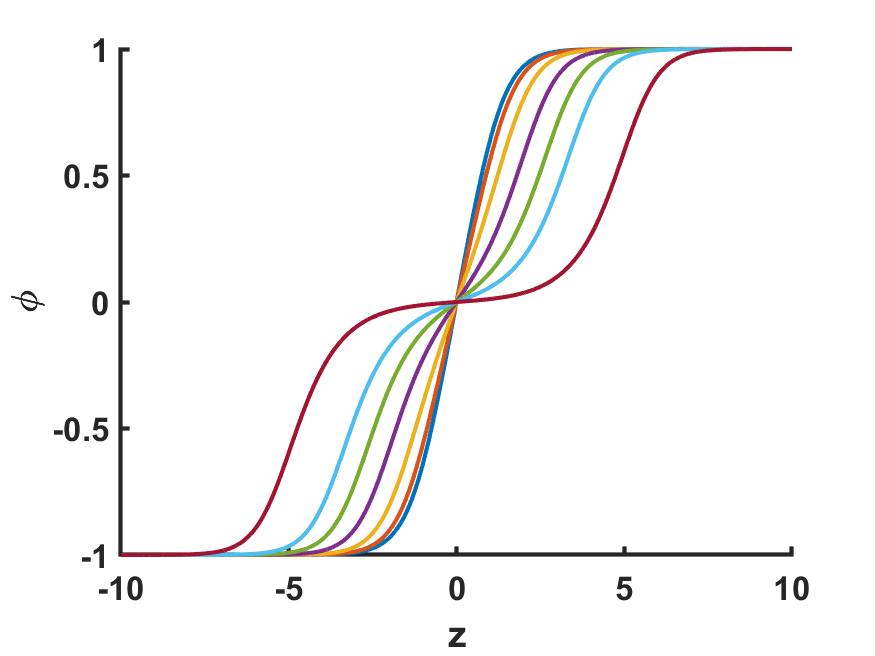}
    \caption{Left: The Christ-Lee potential of Eq. (\ref{potSG}), for the choices $\varepsilon=\infty$, $\varepsilon=4$, $\varepsilon=2$, $\varepsilon=1$, $\varepsilon=1/\sqrt{2}$, $\varepsilon=1/2$, and $\varepsilon=0$, from top to bottom at $\phi\sim0$ respectively. Note that the first and last of these values correspond to the quartic and sextic potentials of Fig. \ref{fig13}, respectively. Right: The kink solution of Eq.(\ref{kinkSG}), for the choices $\varepsilon=1$, $\varepsilon=1/\sqrt{2}$, $\varepsilon=0.4$, $\varepsilon=0.2$, $\varepsilon=0.1$, $\varepsilon=0.05$, and $\varepsilon=0.01$, from top to bottom at $z>0$, respectively.}
    \label{fig16}
  \end{center}
\end{figure*}

The left panel of Figure \ref{fig16} depicts the shape of this potential for various relevant choices of $\varepsilon$. It is worthy of note that the curvature of the side minima does not depend on the parameter $\varepsilon$ and therefore, since we keep our previous normalization factor, it is the same as in the pure $\phi^4$ and $\phi^6$ cases. 

We will separately consider the quartic and sextic limits, but before doing so we note that a stationary kink solution exists for any value of the parameter $\varepsilon$, namely \cite{chris-lee,Demirkaya,Dorey}
\begin{equation}
    \label{kinkSG}
         \phi(z) \propto \frac{\varepsilon\sinh{(z)}}{\sqrt{1+\varepsilon^{2}\cosh^2{(z)}}}\,,
\end{equation}
which in the sextic limit can be envisaged as two simple kinks glued together, as can be seen in the right panel of Fig.~\ref{fig16}.

\subsection{\label{sub4}The quartic limit}

For sufficiently large values of $\varepsilon$ (specifically, for $\varepsilon>1/\sqrt{2}$) the potential has only two degenerate minima, at $\phi=\pm1$, and one should therefore expect results similar to those of the canonical quartic potential. Our simulations, summarized in Table~\ref{tab8}, confirm that this is the case. 

\begin{table}
  \centering
  \caption{Values for the exponents $\mu$ and $\nu$ calculated for the Christ-Lee potential, with different values of $\varepsilon$, in three different expansion rates with a box size of $2048^2$. Each value was taken by averaging over 10 simulations and fitting the data in the range $\eta=\left[31,1024\right]$. The fifth and sixth column show the asymptotic values for $(\rho_w \eta)^{-1}$ and $\gamma v$.}
\begin{tabular}{| c | c | c | c | c | c |}
\hline \hline
$\varepsilon$ & $\lambda$ & $\mu$ & $\nu$ & $(\rho_w \eta)^{-1}$ & $\gamma v$ \\ \hline \hline
$4$ & $1/2$ & $-0.95 \pm 0.03$ & $-0.06\pm 0.02$ & $0.62$ & $0.41$ \\ \hline
$4$ & $2/3$ & $-0.96 \pm 0.03$ & $-0.03\pm 0.03$ & $0.54$ & $0.35$ \\ \hline
$4$ & $4/5$ & $-0.96 \pm 0.02$ & $+0.03\pm 0.04$ & $0.44$ & $0.29$ \\ \hline 
\hline
$2$ & $1/2$ & $-0.94 \pm 0.03$ & $-0.02\pm 0.04$ & $0.57$ & $0.43$ \\ \hline
$2$ & $2/3$ & $-0.94 \pm 0.02$ & $-0.02\pm 0.04$ & $0.46$ & $0.34$ \\ \hline
$2$ & $4/5$ & $-0.95 \pm 0.01$ & $-0.00\pm 0.05$ & $0.37$ & $0.26$ \\ \hline
\hline
$1$ & $1/2$ & $-0.93 \pm 0.02$ & $-0.05\pm 0.05$ & $0.41$ & $0.33$ \\ \hline
$1$ & $2/3$ & $-0.95 \pm 0.02$ & $-0.02\pm 0.04$ & $0.37$ & $0.27$ \\ \hline
$1$ & $4/5$ & $-0.94 \pm 0.01$ & $-0.01\pm 0.05$ & $0.30$ & $0.21$ \\ \hline
\end{tabular}
  \label{tab8}
\end{table}

\begin{figure*}
\centering
    \includegraphics[width=1.0\columnwidth]{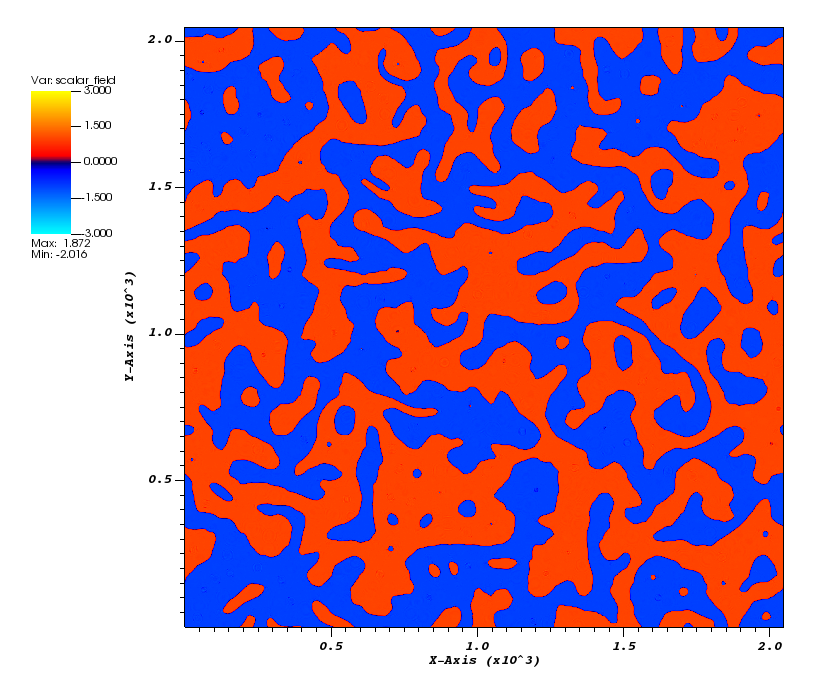}
    \includegraphics[width=1.0\columnwidth]{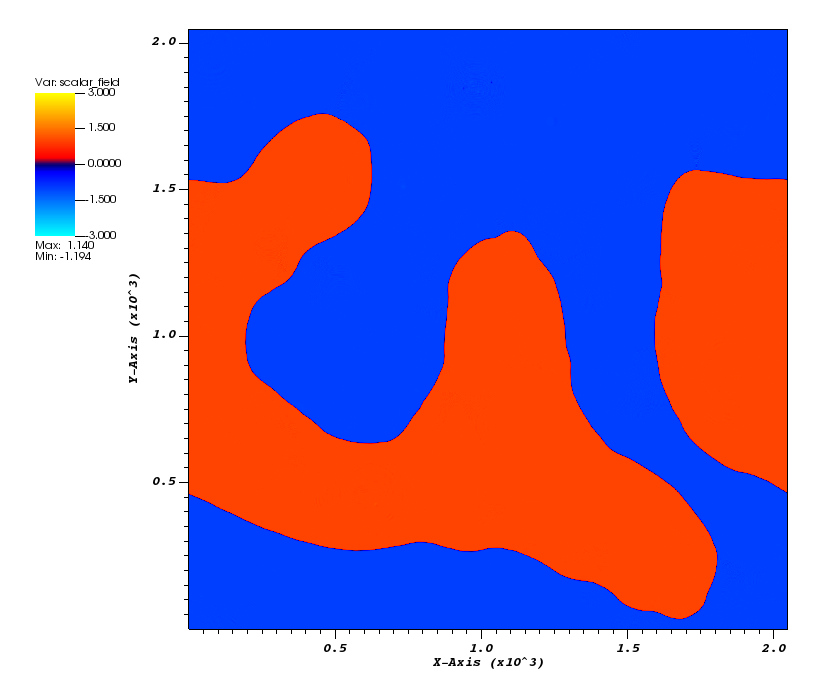}
    \includegraphics[width=1.0\columnwidth]{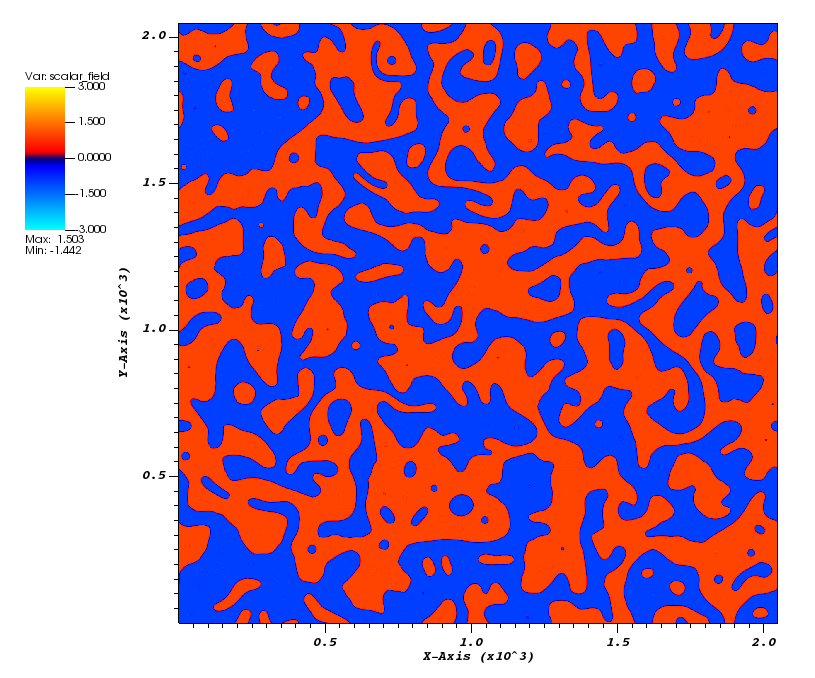}
    \includegraphics[width=1.0\columnwidth]{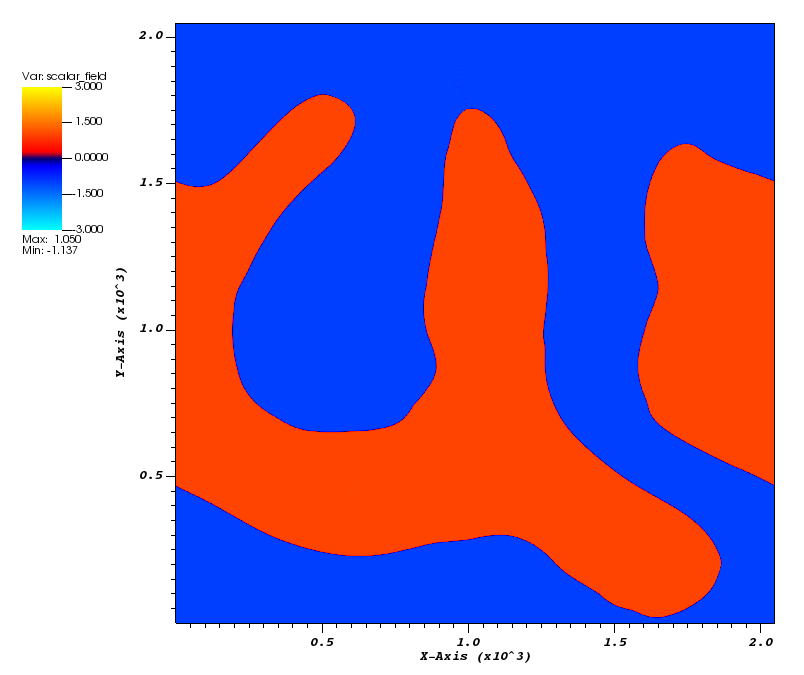}
  \caption{Snapshots of a domain wall network with a Christ-Lee potential with $\varepsilon=1$ in a $2048^2$ grid, for $\lambda=1/2$ (top) and $\lambda=2/3$ (bottom). The colormap represents the value of the field $\phi$. The snapshots were taken for conformal times $\eta=101$ (left) and $\eta=751$ (right).}\label{fig17}
\end{figure*}

Indeed, even for parameter values as low as $\varepsilon=1$ one finds results for the scaling exponents $\mu$ and $\nu$ which, within the statistical uncertainties, are in excellent agreement with those of Table~\ref{tab1}. This can be corroborated in Fig.~\ref{fig17} which shows snapshots of the evolution of the network for $\varepsilon=1$ which are visually indistinguishable from those of the quartic model---this figure can also be compared to Fig.~\ref{fig02}. Also for the asymptotic values for the density and velocity we find good agreement, with values being consistent to within one to two standard deviations.

\subsection{\label{sub6}The sextic limit}

For parameter values $\varepsilon<1/\sqrt{2}$, a third local minimum appears at $\phi=0$, which only becomes degenerate with the side minima for $\varepsilon=0$. This metastable vacuum therefore provides a scenario akin to that of biased domain walls \cite{biases1,biases2}. This implies that a relevant quantity is the difference in the potential values between the central and the outer minima $\Delta V$, which is given by
\begin{equation}
    \Delta V = \frac{\pi^2}{50} \,\frac{\varepsilon^2}{1+\varepsilon^2}\,.
\end{equation}
Naturally, this vanishes for $\varepsilon=0$. Moreover, the local maxima of the potential, which operationally can also be taken as the boundaries between the three minima, occur at
\begin{equation}
\label{phib}
\phi^2_b=\frac{1}{3}(1-2\varepsilon^2)\,.
\end{equation}
Our aim is to understand how the formation of walls depends on $\varepsilon$ (or equivalently $\Delta V$) for this symmetric triple well potential.

We begin by simulating the evolution of the system for identical initial conditions but varying values of $\varepsilon$. Our goal is to quantify the time evolution of the density of points in the central minimum, operationally defined as $|\phi|\le\phi_b$, where the value of $\phi_b$ depends on $\varepsilon$ as described by Eq.~(\ref{phib}). Specifically, as $\varepsilon \rightarrow 1 / \sqrt{2}$, $\phi_b$ approaches zero. For numerical convenience, and considering the grid size of our simulations, to compute the fraction of points in the central minimum, we set a lower bound for $\phi_b$ at 0.02, and keep it at this value for values of $\varepsilon$ near $1/\sqrt{2}$ for which the actual value would be smaller, as well as for larger values of $\varepsilon$. If domain walls form as anticipated, the density of points in the central minimum should decrease after an initial transient period.

\begin{figure*}
\centering
    \includegraphics[width=1.0\columnwidth]{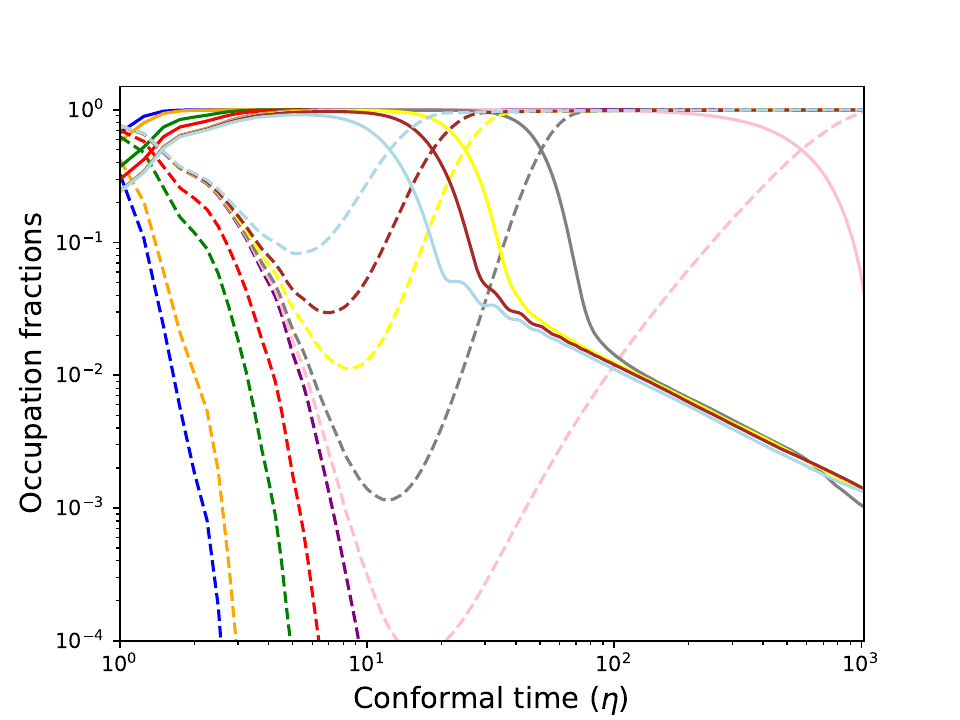}
    \includegraphics[width=1.0\columnwidth]{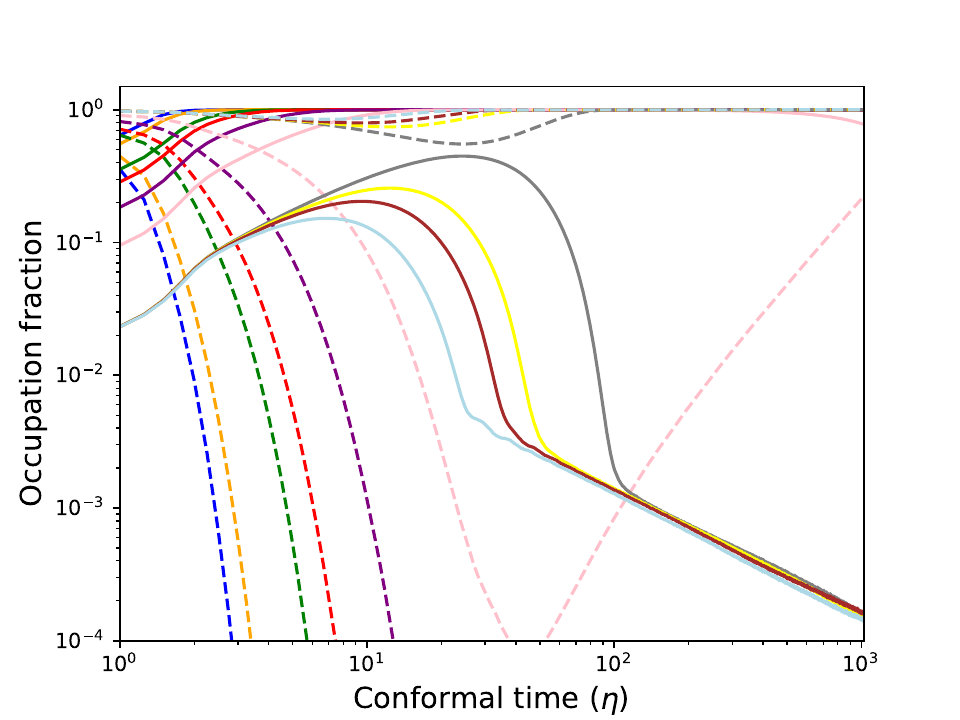}
    \caption{Evolution of the occupation fractions of the field $\phi$ at the the central minimum ($|\phi| \leq \phi_b$, shown solid lines) and the two other minima ($|\phi| > \phi_b $, shown in dashed lines) in a Christ-Lee potential as a function of conformal time. Left and right panels show the results for the radiation ($\lambda=1/2$) and matter ($\lambda=2/3$) dominated universes. Each line corresponds to a different value of $\varepsilon$: 0.20 (blue), 0.40 (orange), 0.60 (green), 0.64 (red), 0.68 (purple), 0.70 (pink), 0.72 (grey), 0.76 (yellow), 0.80 (brown) and 0.90 (light blue). This data was taken from averaging 10 simulations per $\varepsilon$ using the same initial conditions and a box size of $2048^2$.}
    \label{fig18}
\end{figure*}
\begin{figure*}
\centering
        \includegraphics[width=1.0\columnwidth]{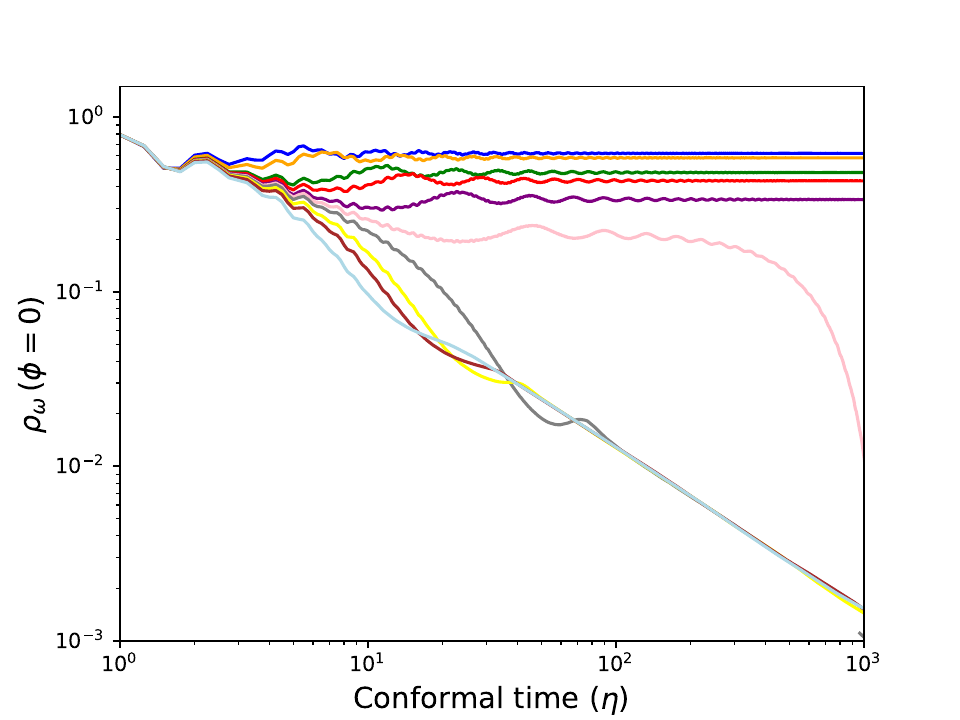}
        \includegraphics[width=1.0\columnwidth]{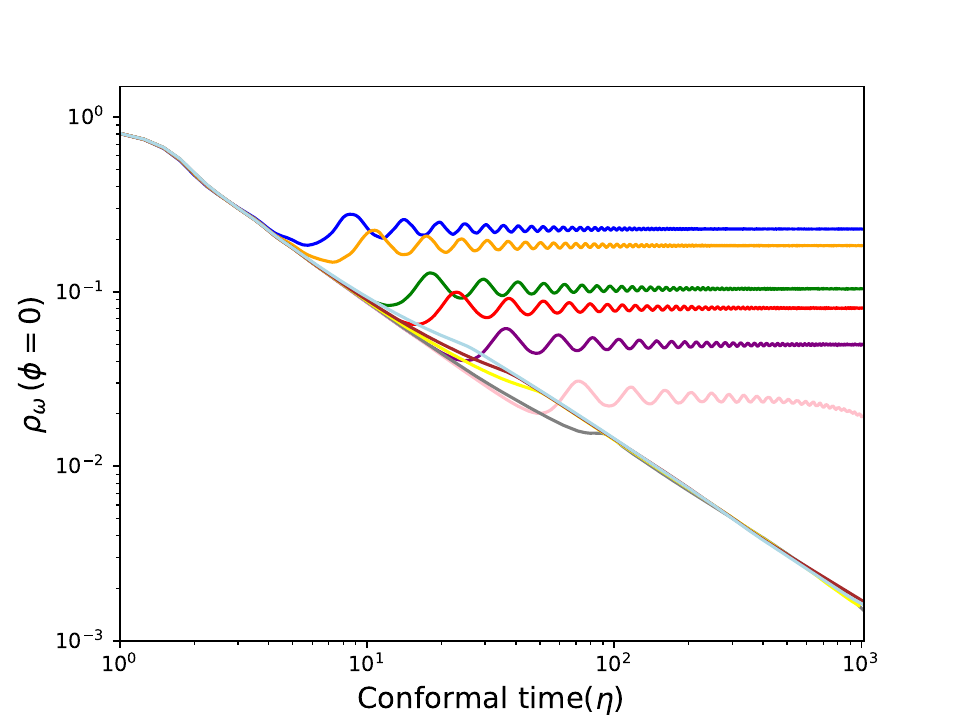}
        \caption{Evolution of the density of the field $\phi$ at $\phi = 0$ in a Christ-Lee potential as a function of conformal time. The left plot corresponds to a radiation dominated universe ($\lambda=1/2$), while the right plot represents a matter-dominated universe ($\lambda=2/3$). Each line in the plots, differentiated by color, represents a distinct value of $\varepsilon$ with each color corresponding to the same value of $\varepsilon$ as in Fig.~\ref{fig18}. Higher lines correspond to smaller values of $\varepsilon$, while lower lines correspond to larger values of $\varepsilon$. The density is computed using finite differences by evaluating changes in the field variable $\phi$ across neighboring lattice points in each spatial dimension. This approach measures the gradient magnitude at points where the field crosses zero, effectively capturing the localized density structure of the field over time.}
    \label{fig19}
\end{figure*}

The results of these simulations are shown in Fig.\ref{fig18} and \ref{fig19}, which depict the evolution for radiation and matter dominated Universes in two dimensions, for $\varepsilon$ values of 0.20, 0.40, 0.60, 0.68, 0.70, 0.72, 0.76, 0.80 and 0.90. In Fig.~\ref{fig18}, the fraction of points located in the central minimum is represented by solid lines, while the fraction of points in one of the other minima is shown by dashed lines. In the central minimum, the occupation fraction initially increases, as a result of the early numerical gradients, and for $\varepsilon < 1/\sqrt{2}$ and in the radiation era it actually reaches unity, indicating that all points remain in the central minimum and no wall formation occurs. One can envisage this behavior as the combined result of a symmetric potential with an odd number of minima and very shallow potential barriers between the minima, traded off with the expansion rate, which impacts the availability of the field to junp between minima. However, for $\varepsilon > 1/\sqrt{2}$, the density in the central minimum begins to decrease after some time, and by the end of the simulation, the densities for all $\varepsilon > 1/\sqrt{2} $ converge to one single line with a constant slope, signifying the onset of the scaling regime for the domain wall network. At this stage, the exponents $\mu$ and $\nu$, as defined in Eq.~(\ref{power-law1}), can be extracted. These exponents are consistent with the values for $\varepsilon = 1$ in Table~\ref{tab8}.

One notable difference emerges between the radiation and matter dominated cases. In the radiation era, the central minimum density always reaches unity (even if only temporarily), regardless of the value of $\varepsilon$. For the matter era, however, and for analogous initial conditions, the central minimum density fails to reach high values in the case of higher values of $\varepsilon$, which is a consequence of the enhanced damping available in this case.

Next, we examine the accumulated densities of points located in the outer two minima. As already shown in Fig.~\ref{fig14}, symmetric initial conditions lead to statistically similar dynamics in both the left and right minima. Consequently, there is no need to distinguish between them, and we focus on the combined accumulated densities of these minima, represented by dashed lines. At the start of the simulation, the densities in the outer minima are clearly non-negligible (which is expected, as the symmetric initial conditions distribute the field values uniformly across the whole lattice), but the fraction of occupied sites in the outer minima decreases rapidly after the simulation begins. For potentials with $\varepsilon<1/\sqrt{2}$, the densities in the outer minima become negligibly small already before the midpoint of the simulation time is reached. For $\varepsilon>1/\sqrt{2}$, on the other hand, the densities in the outer minima begin to increase again shortly after their initial decline, and towards the end of the simulation, nearly all points have shifted to the outer minima, again indicating the formation of a stable domain wall network.

A complementary approach to study the density evolution in the central minimum is by evaluating the changes in the field variable $\phi$ across neighboring lattice points at $\phi=0$, i.e. where the gradient magnitude crosses zero. Fig.~\ref{fig19} presents the results for radiation dominated universes (left) and matter dominated universes (right). The central density is observed to initially decrease for all types of Christ-Lee potentials, regardless of the chosen value of $\varepsilon$. After this initial decrease, the density starts oscillating and eventually reaches one of two asymptotic behaviors, depending on the value of $\varepsilon$. For $\varepsilon < 1/\sqrt{2}$, it stabilizes at a constant value for the remainder of the simulation: notably, the smaller the value of $\varepsilon$, the faster the simulation reaches this constant regime, and the higher is the final constant density. Conversely, for $\varepsilon > 1/\sqrt{2}$, the density converges to a single, decreasing branch, again indicating the scaling regime, which is already clear from Fig.~\ref{fig18}. 

A noteworthy distinction between the density of points with $\phi^2 < \phi_b^2$ (Fig.~\ref{fig18}) and the densities derived from the finite differences method at $\phi=0$ (Fig.~\ref{fig19}) lies in their asymptotic behavior for simulations using small values of $\varepsilon$. In the former method, they converge to the same final density of approximately 1. In contrast, using the finite differences method, the densities converge to a variety of different final values. To understand this difference, one has to consider that there is no wall formation for small values of $\varepsilon$. This means that the field crossing zero is not due to the presence of walls, but rather due to random field fluctuations. The lower the value of $\varepsilon$, the steeper the slope at $\phi = 0$ of the Christ-Lee potential (see Fig.~\ref{fig16}), causing a faster rolling of the field between the two maxima. Faster rolling results in a higher number of sign changes, resulting in a higher final density value as $\varepsilon$ decreases. 

All in all, Figs.~\ref{fig18} and \ref{fig19} compare two different numerical diagnostics to study the density evolution in the central minimum which result in the same physical behavior, and only show discrepancies due to numerical reasons. In other words, the two diagnostics fully agree on the physical question of whether or not domain walls form, but provide quantitatively different numbers for the simple reason that they are indeed measuring numerically different quantities. This highlights the point that the choice and interpretation of numerical diagnostics requires some care, but also shows that several independent but consistent numerical diagnostics can be extracted from the simulations.

\begin{figure}
\centering
        \includegraphics[width=1.0\columnwidth]{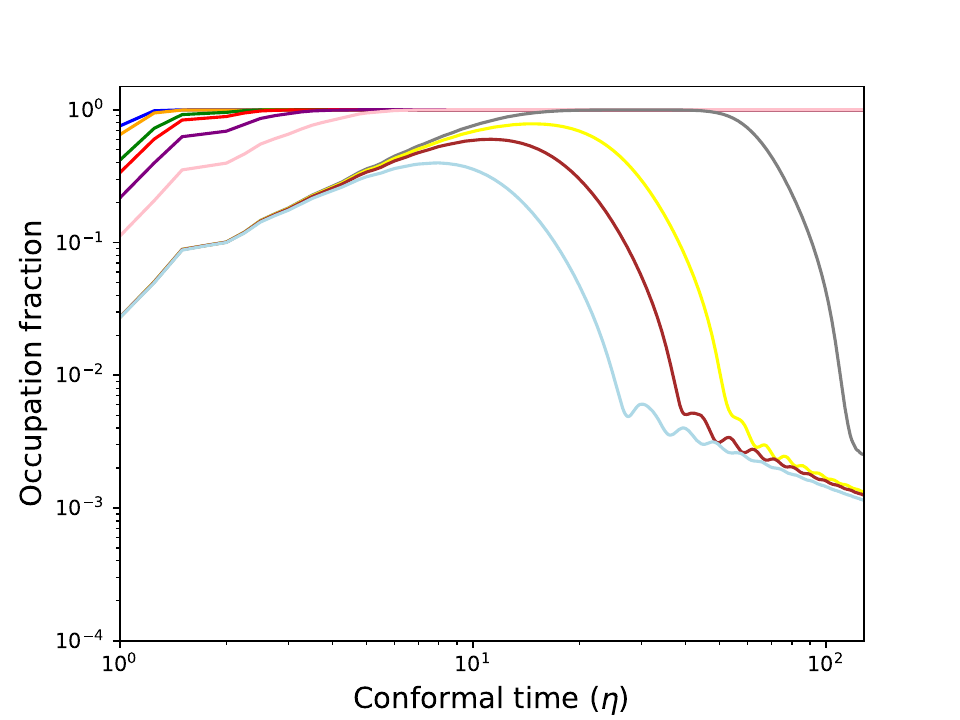}
        \caption{Evolution of the density of the field $\phi$ located at the central minimum ($|\phi| \leq \phi_b$) in a Christ-Lee potential as a function of conformal time for radiation dominated universes ($\lambda=1/2$), for a three dimensional case and a box size of $256^3$. Each line corresponds to a different value of $\varepsilon$ with assigned colors as in Fig.~\ref{fig18} and \ref{fig19}.}
        \label{fig20}
\end{figure}

For comparison, we have also done a set of simulations in three spatial dimensions for a radiation dominated Universe. The result is shown in Fig.~\ref{fig20}. Generally, one finds a similar behavior as in the two dimensional case, but due to the shorter dynamical range the final scaling regime cannot be observed in detail, as its onset occurs near the end of the simulations.

Using our simulations, we can also explore the relationship between the vacuum decay time and the parameter $\varepsilon$. We define the vacuum decay time as the conformal time at which the density of points in the central minimum drops below a specified threshold value. Here, we set the threshold at $\rho = 1 / 18 $. The main point of interest is how this relationship is affected by the cosmological damping provided by the expansion itself, and also by a further numerical damping provided by our cooling which can be introduced in a parametrically controlled way at the beginning of the simulation. The rationale for this cooling is to mitigate the potential influence of the high field gradients in the early stages of the simulation, which arise as a byproduct of our otherwise numerically convenient choice of the random initial conditions. To assess the extent of this effect, a numerical cooling mechanism is applied during the early steps of the simulations, similar to the approach discussed for cosmic strings in \cite{Cooledj}.

\begin{figure}
\centering
        \includegraphics[width=1.0\columnwidth]{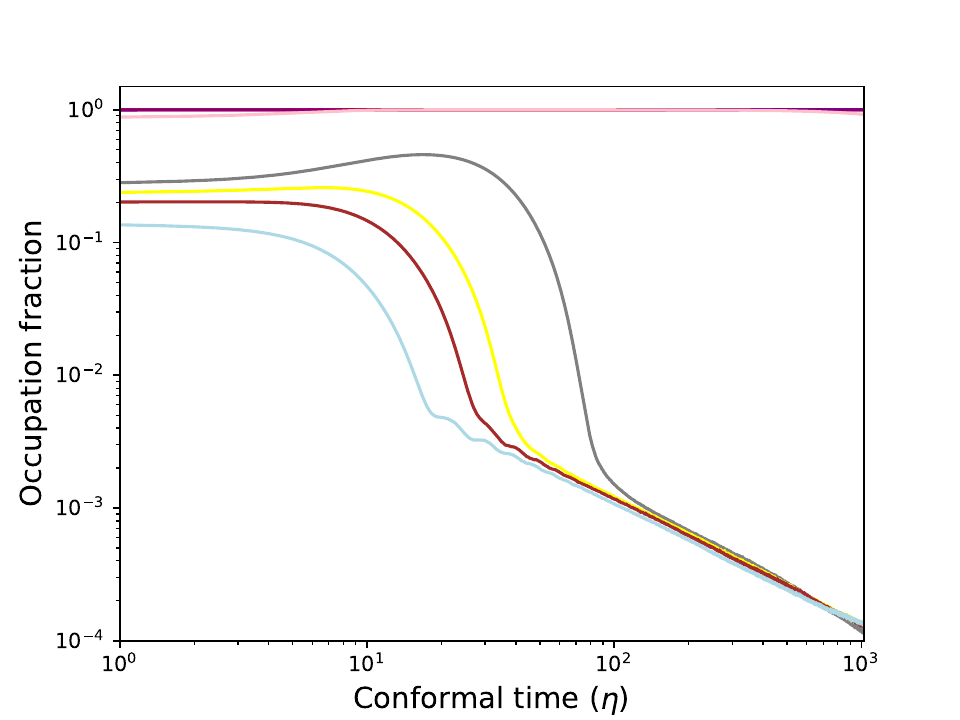}
        \caption{Evolution of the density of the field $\phi$ located at the central minimum ($|\phi| \leq \phi_b$) in a Christ-Lee potential as a function of conformal time for radiation dominated universes ($\lambda=1/2$), with a box size of $2048^2$ and after an initial cooling time of $\eta_{\mathrm{cooling}}=6$. Each line corresponds to a different value of $\varepsilon$ with assigned colors as in the previous figures.}
        \label{fig21}
\end{figure}

The effects of cooling on the evolution of these networks, specifically for an initial cooling time of $\eta_\mathrm{cooling}=6$, are illustrated in Fig.~\ref{fig21}, which can be compared with the left panels of Fig.~\ref{fig18} and \ref{fig19}. Due to the previous cooling, the initial increase in density during the beginning of the simulation does not occur in Fig.~\ref{fig21}. By $\eta=1$, when the physical evolution starts, the central densities for potentials with $\varepsilon<1/\sqrt{2}$ have already stabilized at high values near unity, where they remain throughout the rest of the simulation. For the case $\varepsilon>1/\sqrt{2}$ values, the central densities stabilize at noticeably lower values. For values of $\varepsilon$ just above $1/\sqrt{2}$, we find an intermediate situation: there is a small increase in density followed by a rapid decline towards the end of the simulations, while for larger $\varepsilon$, an immediate decrease is observed without any preceding rise. Either way, for $\varepsilon>1/\sqrt{2}$, only after a certain initial simulation time is it possible to observe a change in these densities and the domain wall network's approach to scaling, the reason being that the cooling mechanism also leads to damping of the wall velocities, and some time is therefore needed for them to become relativistic again---a relativistic speed being essential for scaling for the cosmological expansion rates under consideration.

\begin{figure*}
\centering
    \includegraphics[width=1.0\columnwidth]{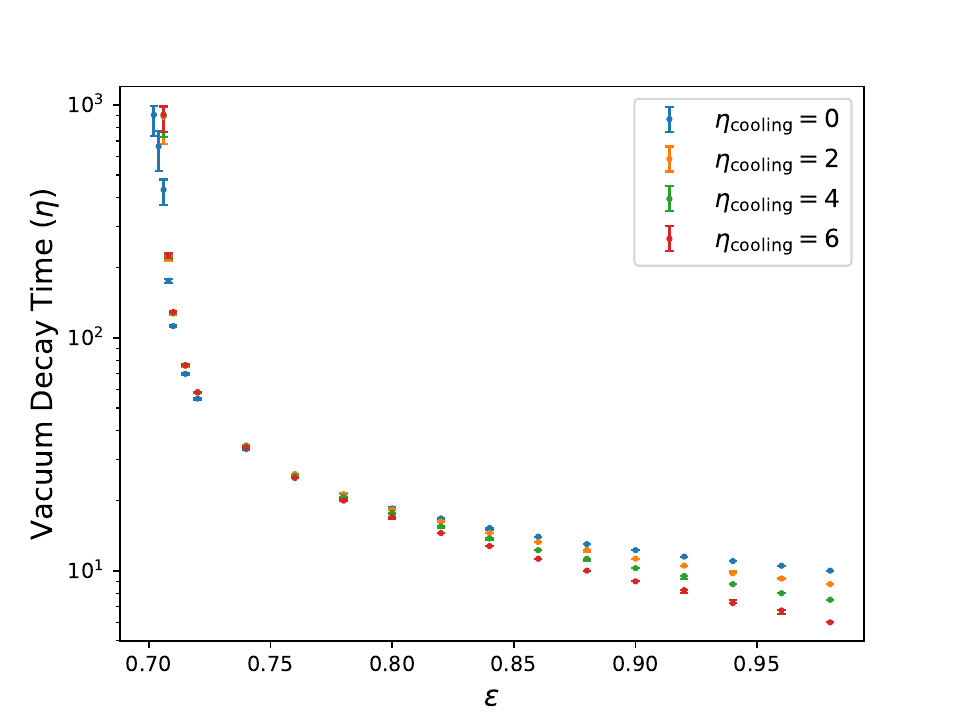}
    \includegraphics[width=1.0\columnwidth]{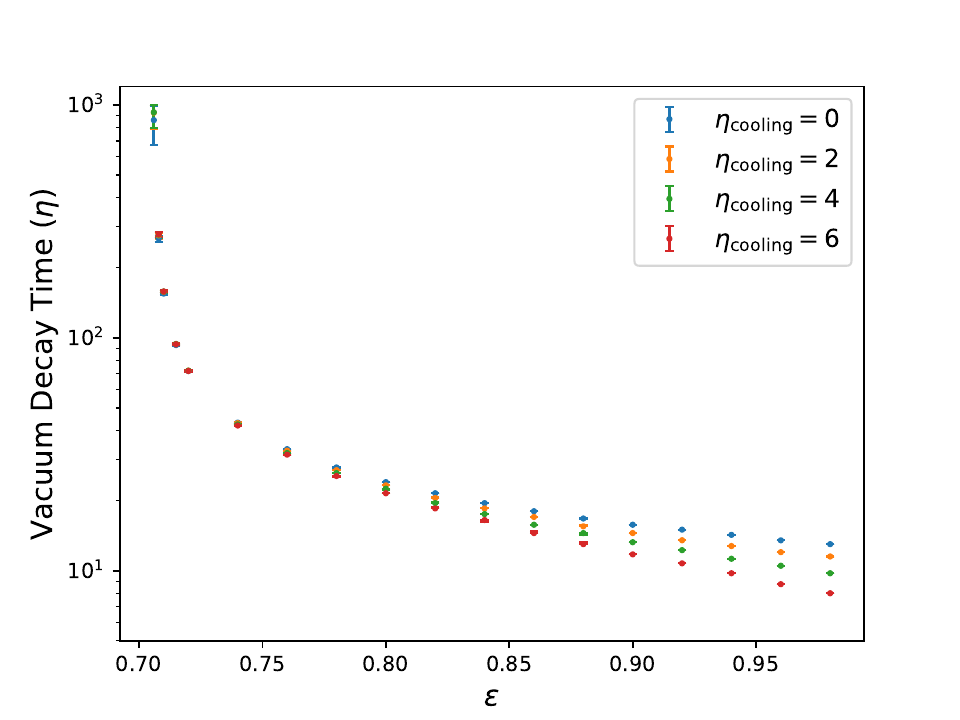}
    \caption{Dependence of the vacuum decay on $\varepsilon$ taken from the data of Fig.~\ref{fig18} using $\rho < 1/18$ as the condition for decay. The left plot shows the result for a radiation dominated universe and the right plot for a matter dominated universe. Different colors denote different initial cooling times. The plotted data points are averaged over 10 different simulations with different random initial condition seeds and error bars emerge from the standard deviation of those different simulations. The simulated potentials contain $\varepsilon$-values of: $0.700, \, 0.702, \, 0.704, \,... , \,0.708, \,0.710, \,0.720,\, 0.740,\, 0.760,\, ... , \,0.960,\, 0.98$. }
    \label{fig22}
\end{figure*}

\subsection{\label{vacuum}Vacuum decay times and wall formation}

We start with the case $\varepsilon>1/\sqrt{2}$, for which our results are summarized in Fig.~\ref{fig22}; for completeness some cases with $\varepsilon \lesssim 1/\sqrt{2}$ are also included. In this regime the potential has a double well form, enabling vacuum decay and therefore wall formation. In the absence of cooling, and for equal values of $\varepsilon$, the vacuum decay time is lower in the radiation era than in the matter era. The vacuum decay time decreases as $\varepsilon$ increases (as the shape of the potential becomes closer to the quartic limit), but the figure also shows the impact of the initial cooling time on the relationship between the two parameters: longer cooling times lead to lower vacuum decay times, and the impact is proportionally larger for larger values of $\varepsilon$.

For $\varepsilon<1/\sqrt{2}$, the potential takes on a sextic form, and in the absence of cooling mechanisms the density in the central minimum will remain above the critical value throughout the entire simulation time. Consequently, no walls form, and no vacuum decay time can be measured, except for a trivial numerical one, already seen at the far left end of each of the panels of Fig.~\ref{fig22}, which simply denotes that no vacuum decay has been observed by the end of the simulation (the latter occurring, as usual, when the horizon reaches half the box size).

The interesting question is whether this $\varepsilon<1/\sqrt{2}$ behavior can be changed in the presence of cooling---in other words, whether a period of numerical cooling could lead to the formation of a stable network of domain walls. This numerical cooling should be seen as a phenomenological proxy for possible physical damping mechanisms impacting the formation and early evolution of such networks. Our analysis of this side of the parameter space, still using $\rho < 1/18$ as the condition for decay, shows that this does not happen. Overall, our results suggest that, regardless of the amount of cooling time, and given symmetric initial conditions, a pure $\phi^6$ potential will generically not give rise to domain walls. In Appendix \ref{leeics} we briefly consider the impact of relaxing the assumptions of symmetric initial conditions and on the $\rho$ threshold.

\section{\label{chap:conc}Conclusions}

We have used our GPU-accelerated domain walls evolution code \cite{gpu-implementation} to study the evolution, for different power-law expansion rates, of these networks in several beyond-quartic potentials, and compared the results with those obtained in analogous simulations for the traditional quartic potential. Our results show that these networks do have somewhat different scaling properties, and we have verified that these results are robust to different choices of initial conditions.

For the Sine-Gordon case, these scaling differences, pertaining both to the networks' densities and velocities, stem from the fact that one can effectively have different types of walls. This does not mean that they have different tensions, but that they interpolate between different pairs of contiguous minima of the potential: even if initial conditions are such that one starts with a single type of walls, the dynamics of the field configurations can lead to the production of walls of other types, as the field explores contiguous minima of the potential. Since the conversion between different wall types requires field configurations to jump across potential maxima, this will have the side effect of introducing differences in the average velocities of the various types of walls.

One implication of our results is that the current canonical analytic model for domain wall network evolution \cite{extending-vos-walls}, which implicitly assumes that the network has a single type of wall, may not be well suited to deal with defects emerging from other potential functions which may naturally occur in the early Universe. Simple approaches to multi-tension domain wall network evolution exist, e.g \cite{Oliveira:2015xfa}, but a more robust analysis, explicitly modeling the physical effects expected in models beyond the simplest potentials, is left for future work. 

For the sextic potential and Christ-Lee potentials, our results confirm the critical role of the potential shape in determining whether wall formation takes place and the scaling regime is reached. In the pure sextic case, the formation of domain walls is suppressed as the occupation sites evolve towards the central minimum. Introducing biased initial conditions can lead, especially in the radiation era, to a temporary occupation of one of the outer minima and a significantly slower decay towards the central minimum. The end result, however, stays the same and no stable walls are formed. 

One numerically convenient feature of the Christ-Lee potential is that a single parameter thereof, denoted $\varepsilon$, interpolates between the quartic and the sextic limits and therefore enables a more detailed study. For the quartic limit ($\varepsilon>1/\sqrt{2}$), we have shown that the scaling properties of the domain wall networks, characterized by the density and velocity exponents $\mu$ and $\nu$, align with those obtained for simulations with the same box size and dynamic range in the quartic potential case. In the sextic limit ($\varepsilon<1/\sqrt{2}$), however, vacuum decay and wall formation do not take place throughout the whole simulation time. 

In the case of the sextic potential and the sextic limit of the Christ-Lee potential, the absence of domain wall formation can be understood by examining the continuous interpolation of the field between $\phi=-1$ and $\phi=1$. This interpolation results in a consistent occupation fraction at the central minimum, $\phi=0$. As a consequence, a population bias emerges early in the simulation, with the central minimum being slightly more populated than the outer minima. For the pure sextic potential, this bias rapidly leads to the full occupation of the dominant minimum at $\phi=0$. In the sextic limit of the Christ-Lee potential, however, an additional competing effect must be considered. Since the central minimum is not a true vacuum and is therefore energetically less favorable than the outer minima, a volume pressure drives the field towards the outer minima. This effect strengthens with increasing $\varepsilon$, progressively slowing the migration toward the central minimum. Once $\varepsilon>1/\sqrt{2}$, this pressure dominates the dynamics, preventing the system from settling in the central minimum and allowing wall formation to take place.

A cooling mechanism was introduced to mitigate numerical artifacts, particularly those arising from the high initial field gradients, which are a side effect of the otherwise choice computationally efficient initial conditions. This cooling mechanism did not show a significant effect on the sextic limit (i.e., it does not lead to a stable domain wall network), but does influence the behavior in the quartic limit. Specifically, in Fig.~\ref{fig22} the dependence of the vacuum decay time on various initial cooling times, the value of the $\varepsilon$ parameter and the universe's expansion rate becomes obvious.

In conclusion, this study broadens the understanding of domain wall dynamics by extending the canonical quartic potential analysis to more complex scenarios. It highlights the interplay between potential symmetry, initial conditions, and cosmological expansion rates in determining the formation and evolution of domain wall networks. These mechanisms, together with downstream effects such as the presence of conserved charges, should be incorporated into current analytic models for the evolution of such networks.

\begin{acknowledgments}
This work was financed by Portuguese funds through FCT (Funda\c c\~ao para a Ci\^encia e a Tecnologia) in the framework of the project 2022.04048.PTDC (Phi in the Sky, DOI 10.54499/2022.04048.PTDC). CJM also acknowledges FCT and POCH/FSE (EC) support through Investigador FCT Contract 2021.01214.CEECIND/CP1658/CT0001 (DOI 10.54499/2021.01214.CEECIND/CP1658/CT0001). JRCCCC acknowledges support from Research Council Finland grant 354572 and ERC grant CoCoS 101142449.

We gratefully acknowledge the support of NVIDIA Corporation with the donation of the Quadro P5000 GPU used for this research.
\end{acknowledgments}

\appendix
\section{\label{sineics}Further initial conditions for the Sine-Gordon model}

\begin{figure*}
\centering
    \includegraphics[width=1.0\columnwidth]{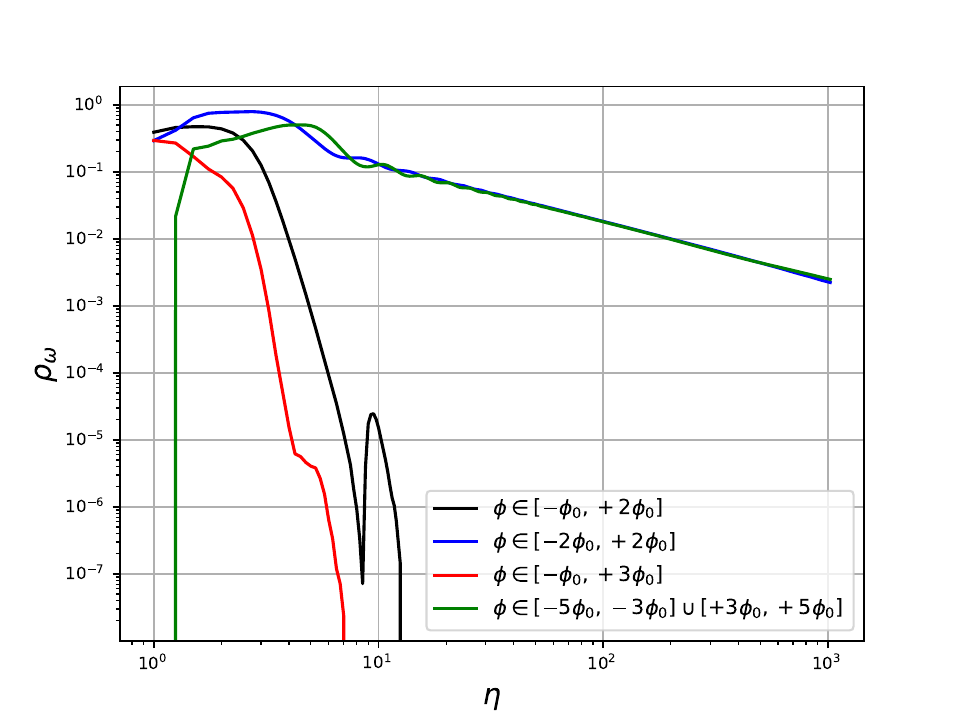}
    \includegraphics[width=1.0\columnwidth]{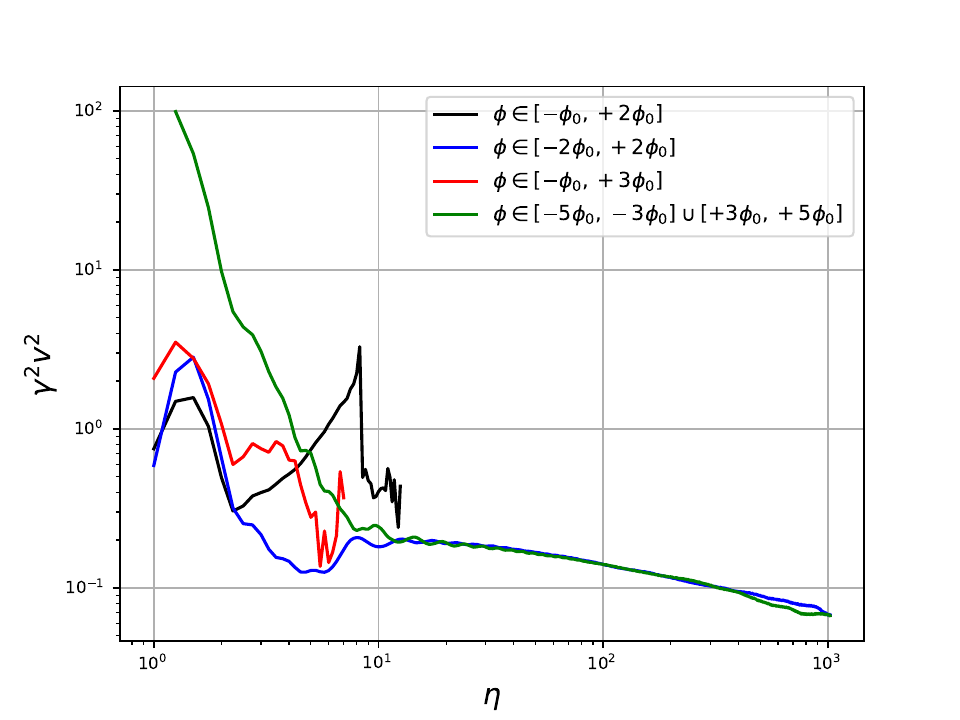}
    \includegraphics[width=1.0\columnwidth]{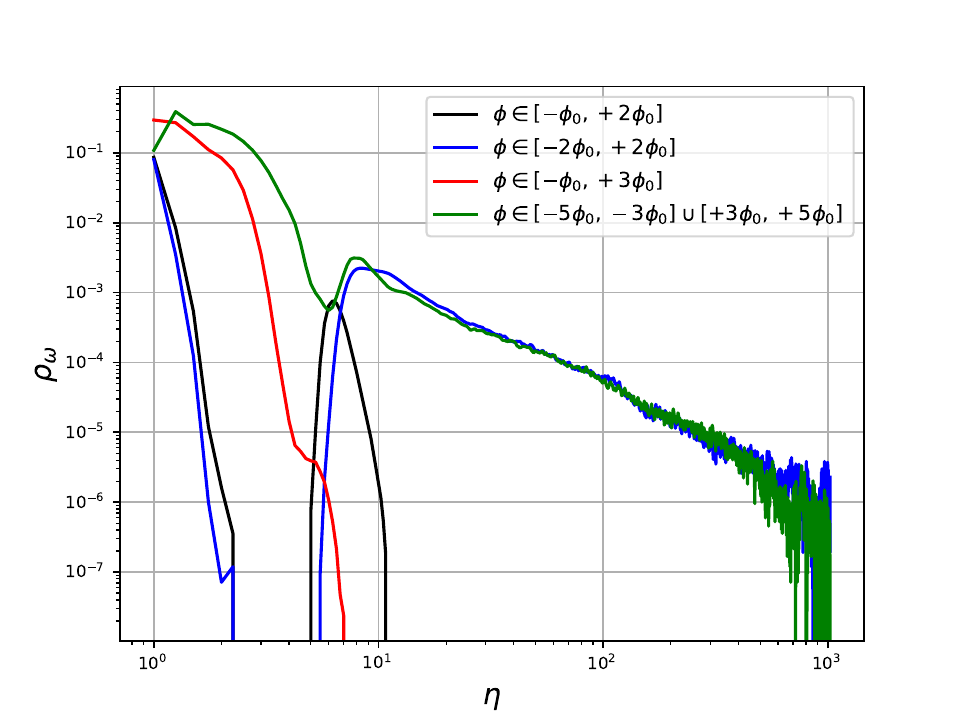}
    \includegraphics[width=1.0\columnwidth]{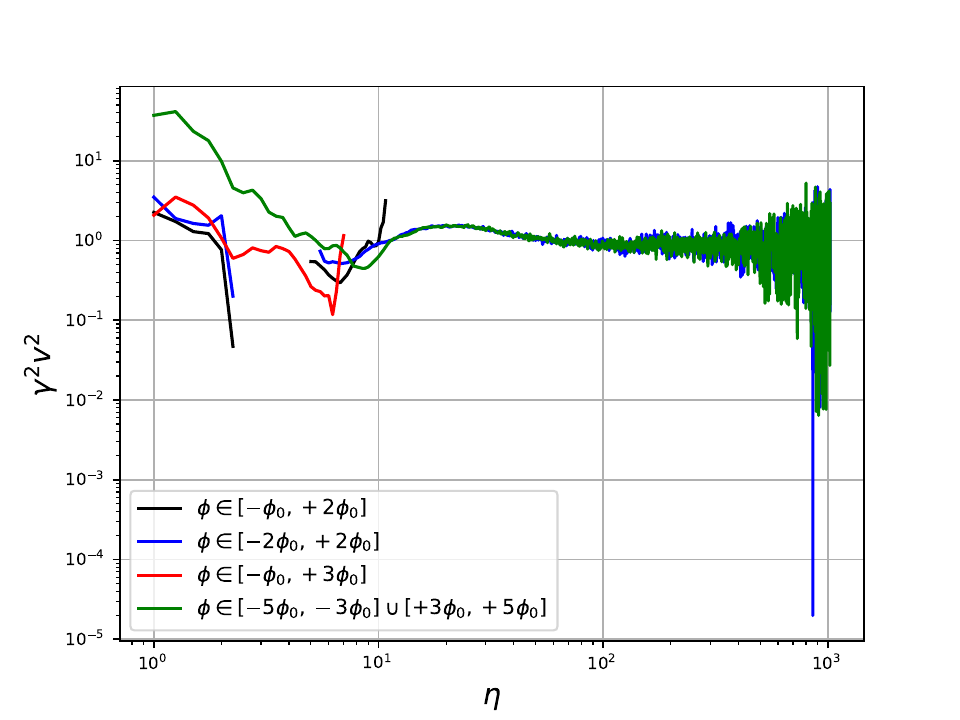}
    \includegraphics[width=1.0\columnwidth]{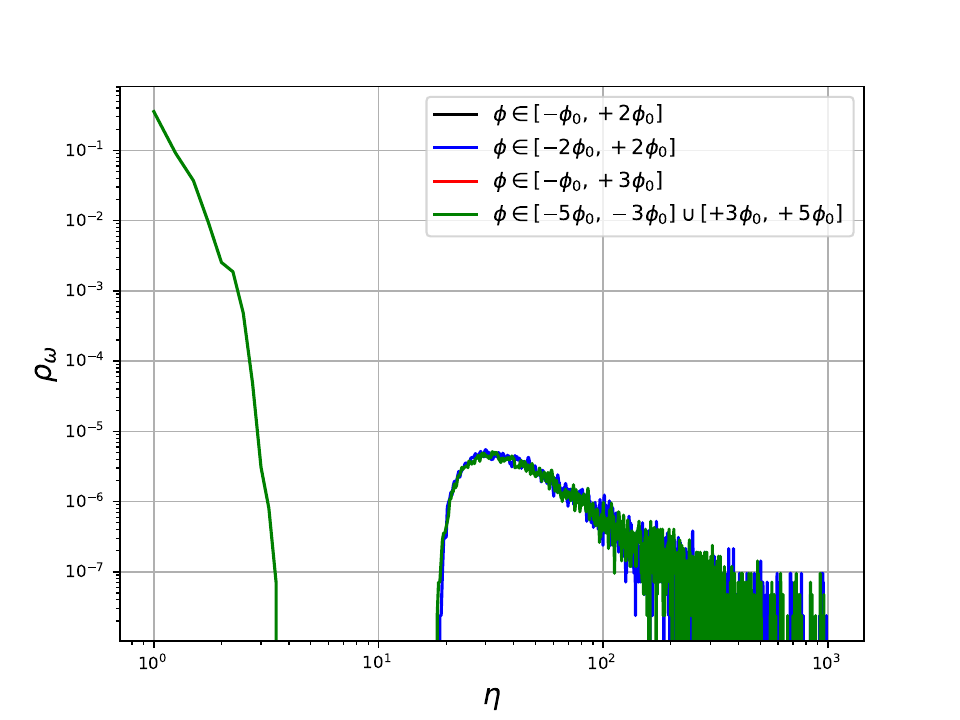}
    \includegraphics[width=1.0\columnwidth]{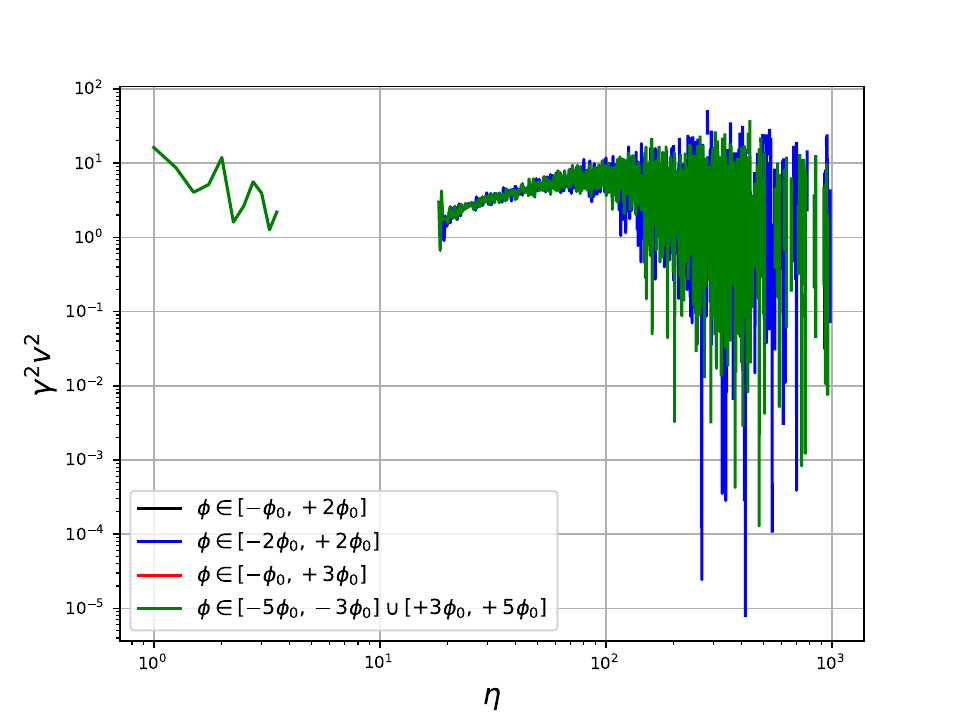}
   \caption{Evolution of the density ($\rho_\omega$, left side panels) and velocity ($\gamma^2 v^2$, right side panels) of different types of domain walls in a Sine-Gordon potential as a function of conformal time for a box size of $2048^2$ and radiation dominated epoch for different initial conditions $\phi \in [-\phi_0, +2 \phi_0]$ (black lines), $\phi \in [-2\phi_0, +2\phi_0]$ (blue lines), $\phi \in [-\phi_0, +3\phi_0]$ (red lines) and $\phi \in [-5\phi_0, -3\phi_0] \cup [+3\phi_0,+5\phi_0]$ (green lines). The top, middle and bottom plots correspond respectively to the type-I, type-II and type-III walls, as defined in Fig.~\ref{fig04}. The plotted values are averages over sets of 10 simulations, with initial conditions generated from a set of 10 different random seeds.}
\label{fig23}
\end{figure*}

As a continuation of the scaling solution robustness tests, we present results obtained from non-symmetric initial conditions and compare them to the symmetric ones. To this end, we perform simulations using four different initial conditions: $\phi$ randomly distributed within $[-\phi_0, +2\phi_0]$ (black lines), $[-2 \phi_0, +2\phi_0]$ (blue lines), $[-\phi_0, +3\phi_0]$ (red lines), and $[-5\phi_0,-3\phi_0] \cup [+3\phi_0,+5\phi_0]$, as shown in Fig.~\ref{fig23}. For each of these setups, we average over 10 runs with different random seeds. The simulations assume a radiation-dominated universe.

The left panels of Fig.~\ref{fig23} show the evolution of the domain wall density, while the right panels display the velocity evolution. We observe that the symmetric initial conditions with $\phi$ in the range $\pm2 \phi_0$ and $[-5\phi_0,-3\phi_0] \cup [+3\phi_0,+5\phi_0]$ exhibit a behavior similar to the case with broader initial conditions shown in Fig.~\ref{fig09} and split initial conditions shown in Fig.~\ref{fig10}. The networks reach the scaling regime around $\eta = 10$, and the type-II domain wall density initially decreases, then increases again before settling into their own scaling. As in the previous case, an energy transfer from type-II to type-I walls explains the velocity difference between the two wall types. However, compared to Fig.~\ref{fig09}, the type-II walls here exhibit higher velocities, indicating a weaker energy shift than before. Lastly, it is interesting to note that even for those very wide split initial conditions, we do not observe type-IV domain walls. We can therefore conclude that the positive and negative side do not evolve independently.

In contrast, for the non-symmetric initial conditions, there is no scaling. Both type-I and type-II domain wall energy densities decay rapidly during the early-time evolution, preventing domain wall formation altogether. For $\phi \in [-\phi_0,+3\phi_0]$, this behavior can be expected because the field is exposed to a potential that closely resembles the sextic potential (see Fig.~\ref{fig13}), albeit shifted to higher values of $\phi$ and with $\phi_{0,\mathrm{SG}} = 2\phi_{0,\mathrm{sextic}}$. Consequently, we expect similar dynamics to those observed in Fig.~\ref{fig14}, where the entire field evolves toward and settles in the central minimum which is now for the Sine-Gordon case located at $\phi=\phi_0$.

A similar comparison can be made for the initial condition $\phi \in [-\phi_0,+2\phi_0]$. This setup resembles simulations with the Christ-Lee potential in the far sextic limit ($\varepsilon \rightarrow 0$) with initial conditions $\phi \in [-1,\phi_b]$, as shown in the middle row of Fig.~\ref{fig25}. As with the other non-symmetric configuration, the field again relaxes into the central minimum which is again located at $\phi=\phi_0$ for the Sine-Gordon potential.

\begin{figure*}
\centering
    \includegraphics[width=1.0\columnwidth]{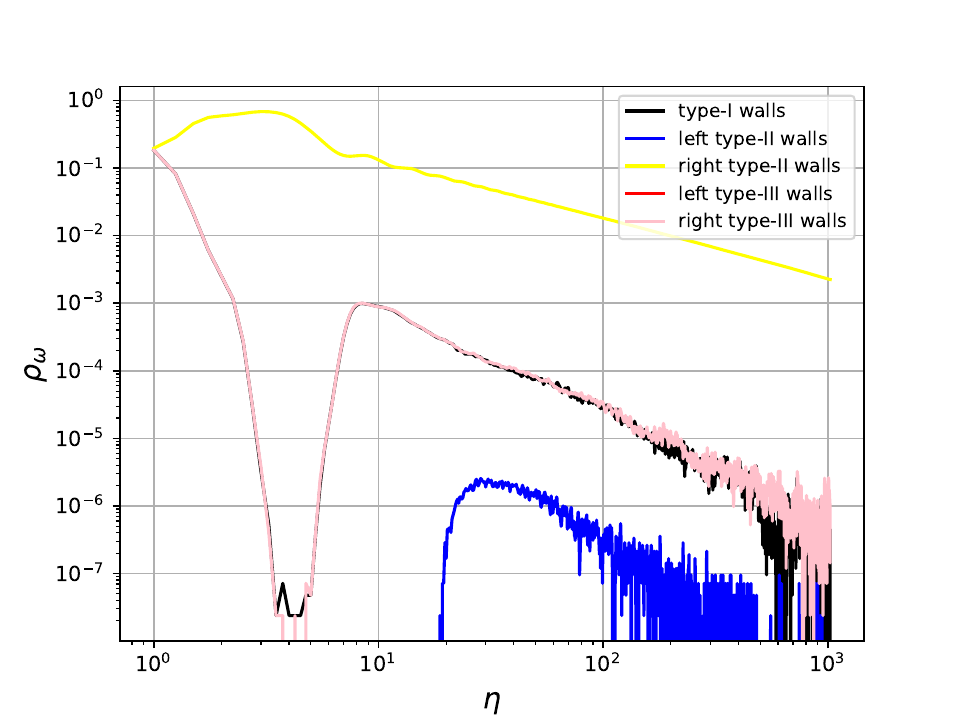}
    \includegraphics[width=1.0\columnwidth]{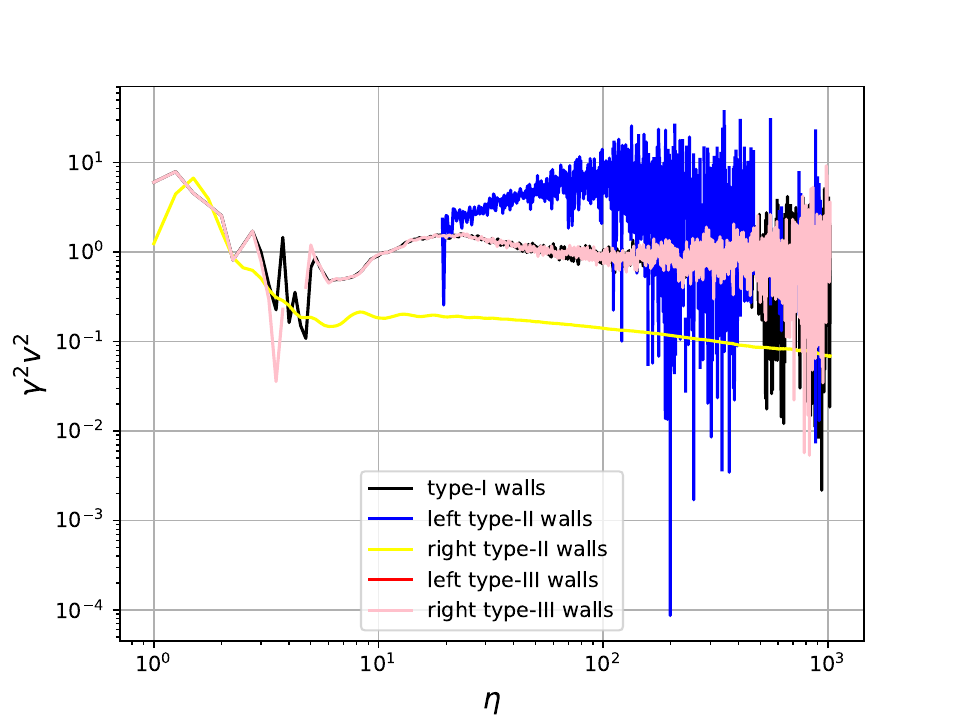}
    \includegraphics[width=1.0\columnwidth]{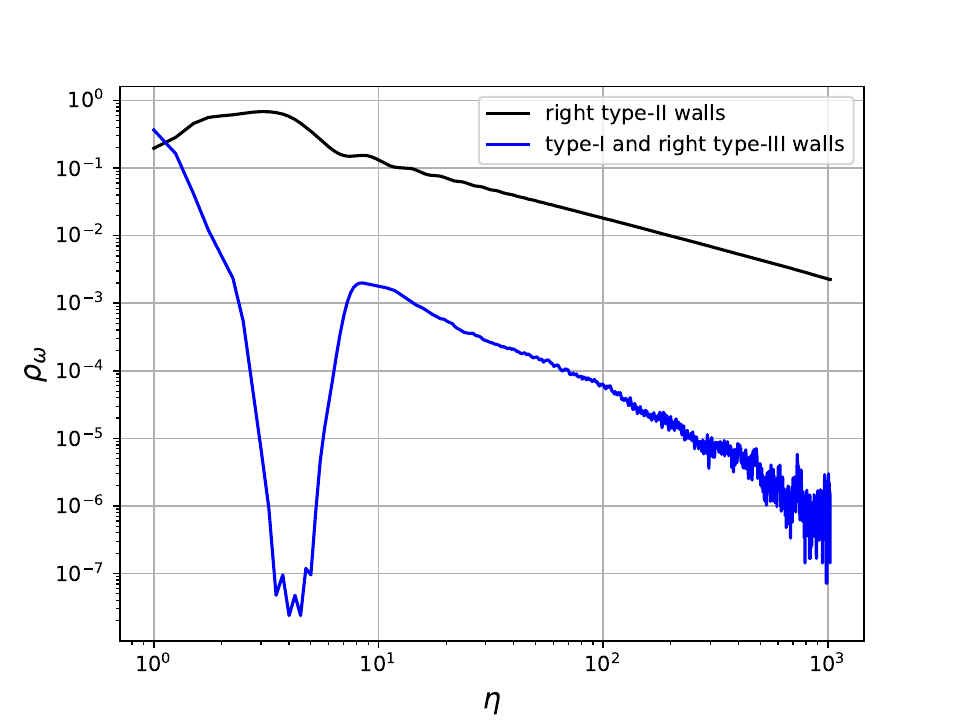}
    \includegraphics[width=1.0\columnwidth]{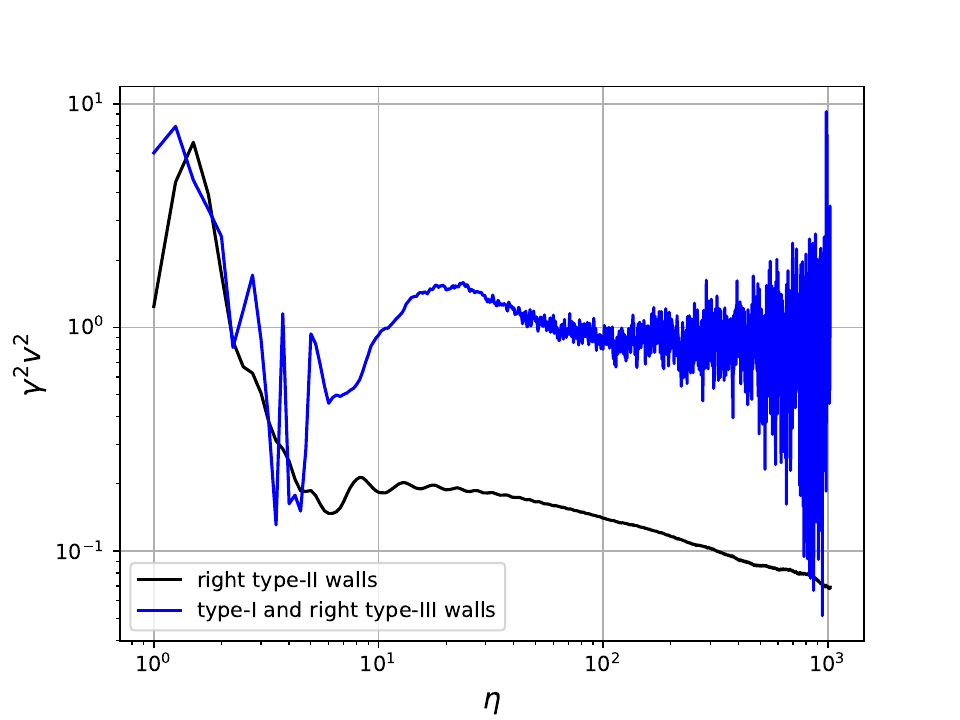}
   \caption{Evolution of the density ($\rho_\omega$, left side panels) and velocity ($\gamma^2 v^2$, right side panels) of different types of domain walls in a Sine-Gordon potential as a function of conformal time for a box size of $2048^2$ and radiation dominated epoch with $\phi$ initially distributed as $\phi \in [-\phi_0, +5 \phi_0]$ in a radiation dominated universe. The top plot represent every single wall while distinguishing whether they correspond to positive or negative values of $\phi$, in other words, corresponding to the walls on the right or on the left in Fig.~\ref{fig04} and therefore denoted as right or left walls.. The bottom panels represent the same data, but show the evolution of only the right type-II walls (black lines) and the global evolution of the right type-II walls together with the type-I walls (blue lines). The plotted values are averages over sets of 10 simulations, with initial conditions generated from a set of 10 different random seeds.}
\label{fig24}
\end{figure*}

Finally, we demonstrate that the evolution of the domain wall network in the Sine-Gordon potential is translation invariant. To illustrate this, we use an initial condition where $\phi$ is randomly and uniformly distributed within the range $[-\phi_0, +5\phi_0]$ and assuming a radiation-dominated universe. Due to translation invariance, we expect the resulting evolution to mirror the case with wide initial conditions $[-3\phi_0,+3\phi_0]$ (as shown in Fig.~\ref{fig09}), but shifted by $+2\phi_0$.

Fig.~\ref{fig24} confirms this expectation. The top panels display the evolution of both the density and velocity for each type of domain wall, distinguishing between walls located at positive and negative $\phi$ values. Given the $+2\phi_0$ shift, we anticipate the positive type-II walls to behave similarly to the type-I walls in Fig.~\ref{fig09}. Additionally, we expect the neighboring walls (according to Fig.~\ref{fig04} type-I and positive type-III walls) to follow the same dynamics as the type-II walls in Fig.~\ref{fig09}.

As a result, the black and blue lines in the bottom panels of Fig.~\ref{fig24} should match the black lines in the top and middle panels of Fig.\ref{fig09}. A direct comparison between the two figures confirms that this is indeed the case, supporting the conclusion of translation invariance.

\section{\label{leeics}Further initial conditions for the Christ-Lee model}

 \begin{figure*}
\centering
    \includegraphics[width=1.0\columnwidth]{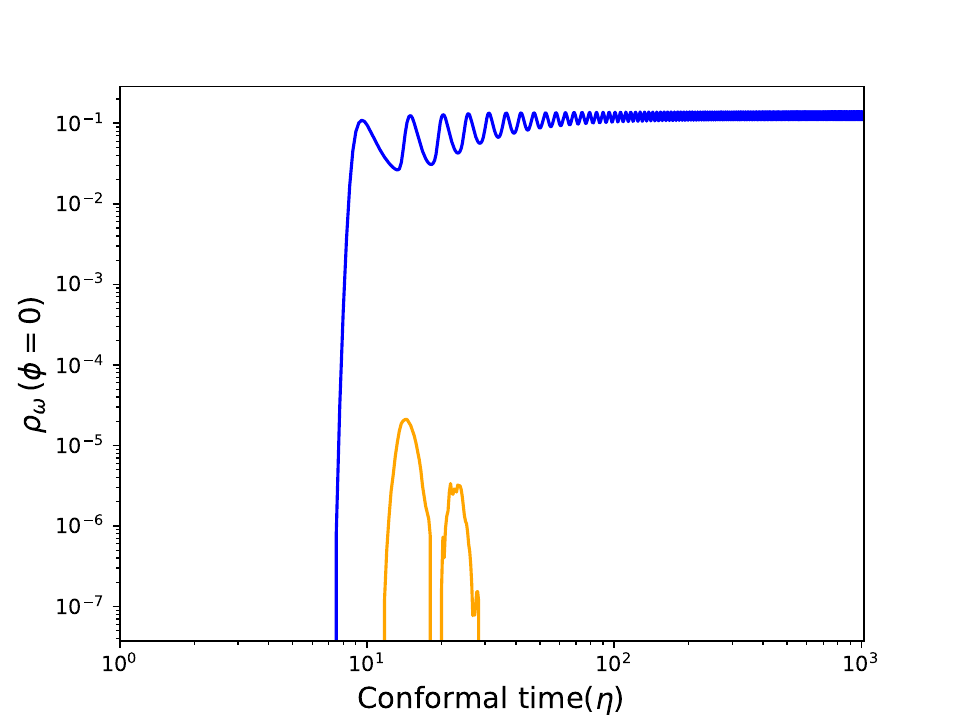}
    \includegraphics[width=1.0\columnwidth]{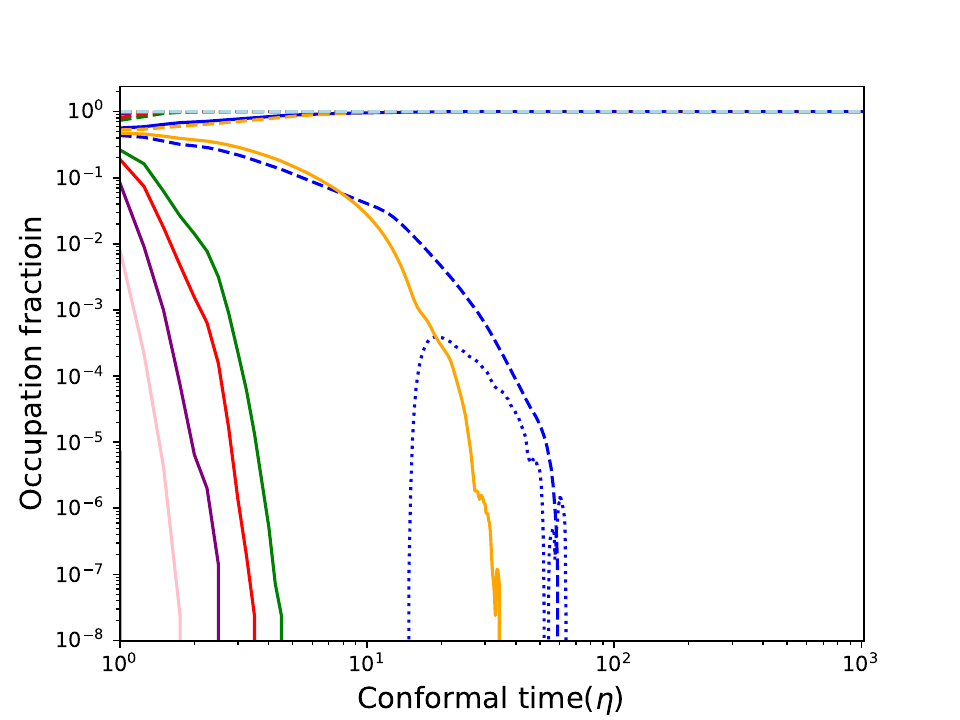}
    \includegraphics[width=1.0\columnwidth]{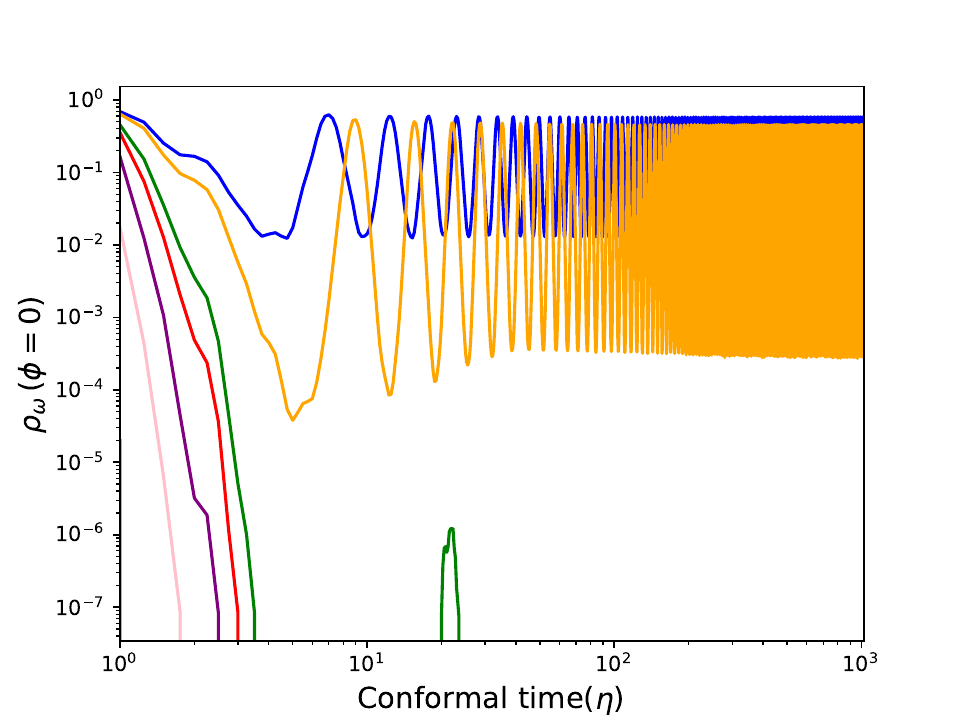}
    \includegraphics[width=1.0\columnwidth]{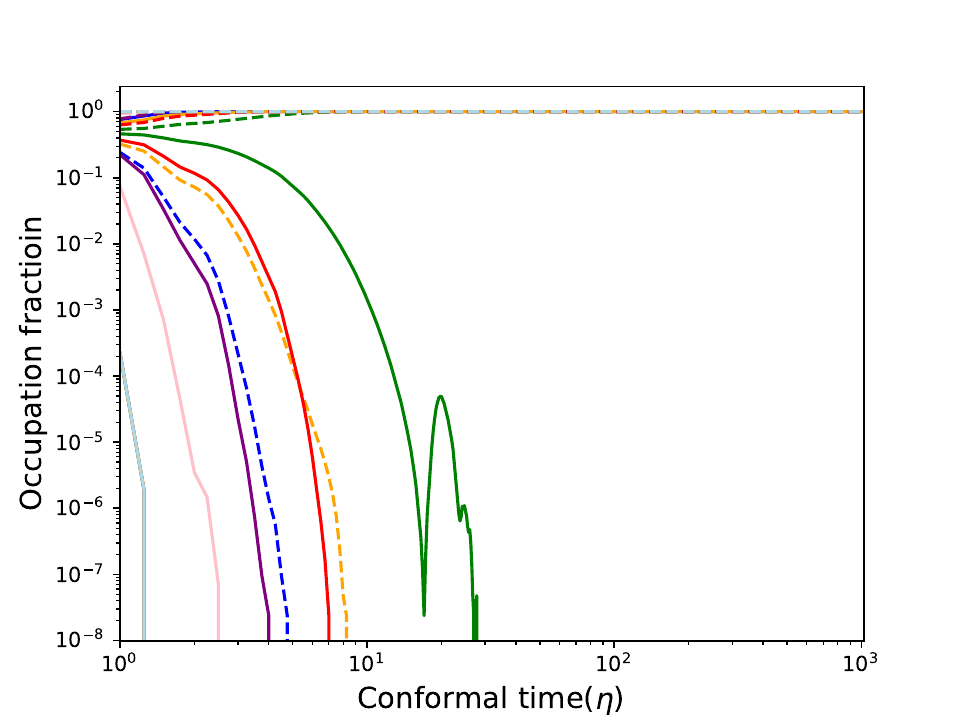}
    \includegraphics[width=1.0\columnwidth]{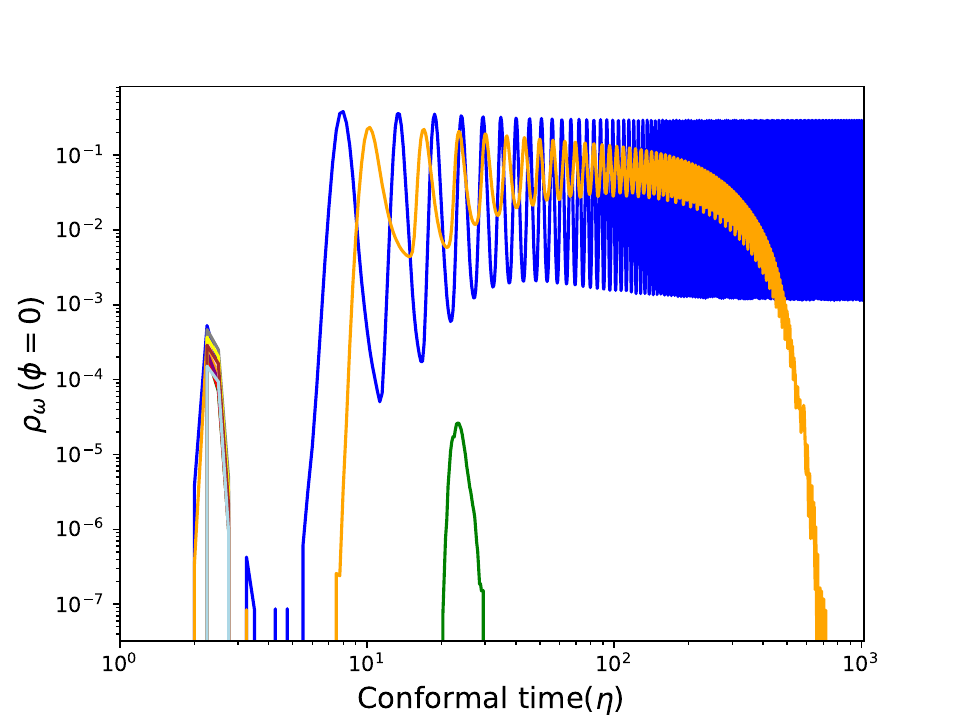}
    \includegraphics[width=1.0\columnwidth]{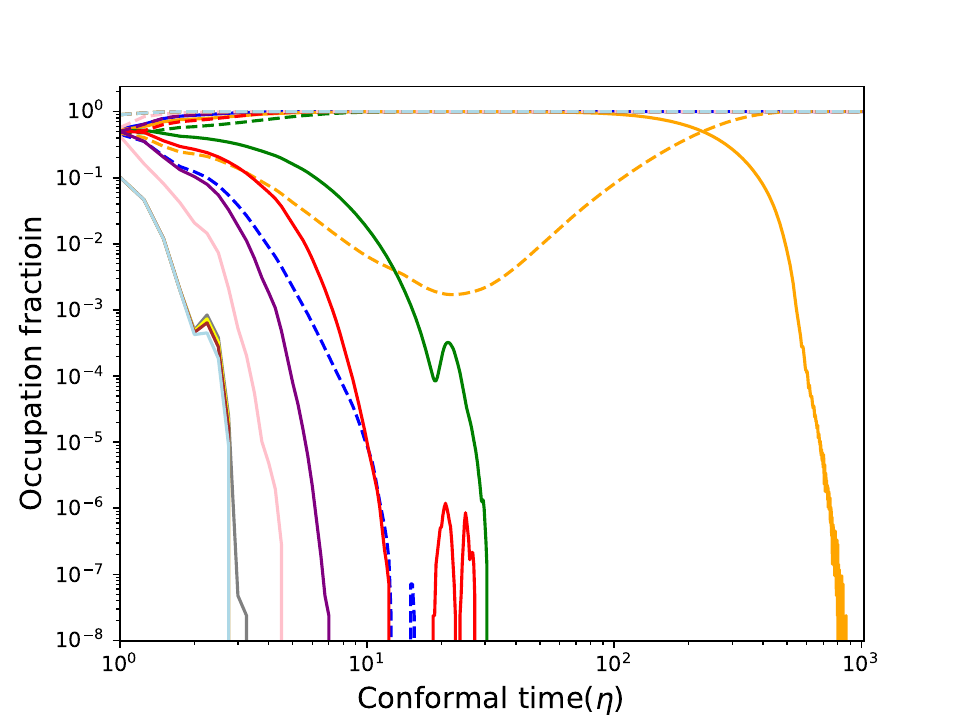}
   \caption{Evolution of the density of the field $\phi$ at $\phi = 0$ (left panels) and of the occupation fractions (right panels) at the central minimum ($|\phi| \leq \phi_b$, solid lines), the left minimum ($\phi < -\phi_b$, dashed lines) and the right minimum ($\phi > \phi_b$, dotted lines) in a Christ-Lee Potential. Each row represents different initial field distributions. The top plots represent biased initial conditions, $-1 \leq \phi \leq 0$. The middle plots represent represent wider biased initial conditions including both maxima, $-1 \leq \phi \leq \phi_b$. And the bottom blots represent split initial conditions with only the outer minima being occupied initially, $|\phi| \geq \phi_b$. Since the bottom plots represent symmetric initial conditions, we do not take a look at the occupation fractions of the outer minima separately and the dashed lines represent the sum of the occupation fractions of both, the left and the right minimum. The colors correspond to the same $\varepsilon$ values as in Fig.~\ref{fig18}. All simulations were carried out for a radiation dominated universe and without initial cooling.}
\label{fig25}
\end{figure*}

To similarly test our Christ-Lee simulations against a broader range of initial conditions, we considered three additional types of initial field distributions. Specifically, we examined two biased distributions and one symmetric split distribution. The first biased case features $\phi$ values randomly distributed between $-1$ and $0$, while the second has a broader range, with $\phi$ spanning from $-1$ to the first positive maximum, $\phi_b$. The third type of initial condition is symmetric around zero, but the field is uniformly distributed within two distinct outer regions, corresponding to the two outer minima, $\phi \geq \phi_b$.
 
The results for a radiation-dominated universe without initial cooling are shown in Fig.~\ref{fig24}. For biased initial conditions, we observe that at low $\varepsilon$, the system behaves similarly to the biased case in the sextic potential: the right minimum is occupied for a short time period, but on the long run, the whole field settles in the central minimum, as expected. At higher $\varepsilon$, the field exclusively occupies the left minimum, preventing wall formation.
 
For the broader biased distribution, the behavior is slightly different. Here, the entire field settles into the central minimum not only for $\varepsilon = 0.2$, but also for $\varepsilon = 0.4$. Notably, the right minimum remains unoccupied. For even higher $\varepsilon$, the overall behavior mirrors the biased case, with the only difference being that the field takes longer to fully settle in the left minimum.
 
In the case of split initial conditions, we initially observe very low density $\rho_\omega(\phi = 0)$ or sign changes at $\phi = 0$, as expected. However, for $\varepsilon = 0.2$ and $\varepsilon = 0.4$, the field quickly migrates to the central minimum. Interestingly, at $\varepsilon = 0.2$, it remains there, while at $\varepsilon = 0.4$, the field eventually decays into the outer minima. For even higher $\varepsilon$, the field remains split in the outer minima throughout the whole simulation.

We also performed simulations with these three initial field distributions, but with the difference of setting a cooling time of $\eta_{\mathrm{cooling}} = 6$ and also repeated them for a matter dominated universe, for both $\eta_{\mathrm{cooling}} = 0$ and $\eta_{\mathrm{cooling}}=6$. Since the results closely align with our previous findings, showing only minor variations, we do not present additional plots here.
 
Finally, we tested the robustness of our chosen threshold value $\rho < 1/18$ as the criterion for vacuum decay. This threshold must satisfy two conditions: (1) it must be high enough to ensure that no simulation has already reached scaling when its curve crosses the threshold (e.g., in the left plot of Fig.\ref{fig18}, the light blue curve at conformal time $\eta = 10.1$); and (2) it should not exceed the maximum of the lowest occupation fraction curve (e.g., the gray curve in the right plot of Fig.\ref{fig18}). The chosen value of $\rho = 1/18$ meets both criteria, though slight variations in this threshold still satisfy these conditions. Computing vacuum decay times for slightly lower or higher $\rho$ values produced consistent results with Fig.~\ref{fig22}, with lower $\rho$ values leading to generally slightly longer decay times and higher $\rho$ values resulting in slightly shorter decay times.
 
\bibliography{artigo}

\begin{thebibliography}{43}%
\makeatletter
\providecommand \@ifxundefined [1]{%
 \@ifx{#1\undefined}
}%
\providecommand \@ifnum [1]{%
 \ifnum #1\expandafter \@firstoftwo
 \else \expandafter \@secondoftwo
 \fi
}%
\providecommand \@ifx [1]{%
 \ifx #1\expandafter \@firstoftwo
 \else \expandafter \@secondoftwo
 \fi
}%
\providecommand \natexlab [1]{#1}%
\providecommand \enquote  [1]{``#1''}%
\providecommand \bibnamefont  [1]{#1}%
\providecommand \bibfnamefont [1]{#1}%
\providecommand \citenamefont [1]{#1}%
\providecommand \href@noop [0]{\@secondoftwo}%
\providecommand \href [0]{\begingroup \@sanitize@url \@href}%
\providecommand \@href[1]{\@@startlink{#1}\@@href}%
\providecommand \@@href[1]{\endgroup#1\@@endlink}%
\providecommand \@sanitize@url [0]{\catcode `\\12\catcode `\$12\catcode
  `\&12\catcode `\#12\catcode `\^12\catcode `\_12\catcode `\%12\relax}%
\providecommand \@@startlink[1]{}%
\providecommand \@@endlink[0]{}%
\providecommand \url  [0]{\begingroup\@sanitize@url \@url }%
\providecommand \@url [1]{\endgroup\@href {#1}{\urlprefix }}%
\providecommand \urlprefix  [0]{URL }%
\providecommand \Eprint [0]{\href }%
\providecommand \doibase [0]{http://dx.doi.org/}%
\providecommand \selectlanguage [0]{\@gobble}%
\providecommand \bibinfo  [0]{\@secondoftwo}%
\providecommand \bibfield  [0]{\@secondoftwo}%
\providecommand \translation [1]{[#1]}%
\providecommand \BibitemOpen [0]{}%
\providecommand \bibitemStop [0]{}%
\providecommand \bibitemNoStop [0]{.\EOS\space}%
\providecommand \EOS [0]{\spacefactor3000\relax}%
\providecommand \BibitemShut  [1]{\csname bibitem#1\endcsname}%
\let\auto@bib@innerbib\@empty
\bibitem [{\citenamefont {Kibble}(1976)}]{kibble-topology}%
  \BibitemOpen
  \bibfield  {author} {\bibinfo {author} {\bibfnamefont {T.~W.~B.}\
  \bibnamefont {Kibble}},\ }\href {\doibase 10.1088/0305-4470/9/8/029}
  {\bibfield  {journal} {\bibinfo  {journal} {J. Phys. A}\ }\textbf {\bibinfo
  {volume} {9}},\ \bibinfo {pages} {1387} (\bibinfo {year} {1976})}\BibitemShut
  {NoStop}%
\bibitem [{\citenamefont {Vilenkin}\ and\ \citenamefont
  {Shellard}(2000)}]{cosmic-strings-and-other}%
  \BibitemOpen
  \bibfield  {author} {\bibinfo {author} {\bibfnamefont {A.}~\bibnamefont
  {Vilenkin}}\ and\ \bibinfo {author} {\bibfnamefont {E.~P.~S.}\ \bibnamefont
  {Shellard}},\ }\href@noop {} {\emph {\bibinfo {title} {{Cosmic Strings and
  Other Topological Defects}}}}\ (\bibinfo  {publisher} {Cambridge University
  Press},\ \bibinfo {year} {2000})\BibitemShut {NoStop}%
\bibitem [{\citenamefont {Kibble}(1985)}]{kibble-onescale}%
  \BibitemOpen
  \bibfield  {author} {\bibinfo {author} {\bibfnamefont {T.~W.~B.}\
  \bibnamefont {Kibble}},\ }\href {\doibase
  https://doi.org/10.1016/0550-3213(85)90439-0} {\bibfield  {journal} {\bibinfo
   {journal} {Nuclear Physics B}\ }\textbf {\bibinfo {volume} {252}},\ \bibinfo
  {pages} {227 } (\bibinfo {year} {1985})}\BibitemShut {NoStop}%
\bibitem [{\citenamefont {Martins}(2016)}]{VOS-book}%
  \BibitemOpen
  \bibfield  {author} {\bibinfo {author} {\bibfnamefont {C.~J. A.~P.}\
  \bibnamefont {Martins}},\ }\href
  {https://books.google.pt/books?id=Qx34DAAAQBAJ} {\emph {\bibinfo {title}
  {Defect Evolution in Cosmology and Condensed Matter: Quantitative Analysis
  with the Velocity-Dependent One-Scale Model}}},\ SpringerBriefs in Physics\
  (\bibinfo  {publisher} {Springer International Publishing},\ \bibinfo {year}
  {2016})\BibitemShut {NoStop}%
\bibitem [{\citenamefont {Martins}\ and\ \citenamefont
  {Shellard}(1996)}]{vos-strings1}%
  \BibitemOpen
  \bibfield  {author} {\bibinfo {author} {\bibfnamefont {C.~J. A.~P.}\
  \bibnamefont {Martins}}\ and\ \bibinfo {author} {\bibfnamefont {E.~P.~S.}\
  \bibnamefont {Shellard}},\ }\href {\doibase 10.1103/PhysRevD.54.2535}
  {\bibfield  {journal} {\bibinfo  {journal} {Phys. Rev. D}\ }\textbf {\bibinfo
  {volume} {54}},\ \bibinfo {pages} {2535} (\bibinfo {year} {1996})},\ \Eprint
  {http://arxiv.org/abs/hep-ph/9602271} {arXiv:hep-ph/9602271} \BibitemShut
  {NoStop}%
\bibitem [{\citenamefont {Avelino}\ \emph {et~al.}(2005)\citenamefont
  {Avelino}, \citenamefont {Martins},\ and\ \citenamefont
  {Oliveira}}]{VOS-walls}%
  \BibitemOpen
  \bibfield  {author} {\bibinfo {author} {\bibfnamefont {P.~P.}\ \bibnamefont
  {Avelino}}, \bibinfo {author} {\bibfnamefont {C.~J. A.~P.}\ \bibnamefont
  {Martins}}, \ and\ \bibinfo {author} {\bibfnamefont {J.~C. R.~E.}\
  \bibnamefont {Oliveira}},\ }\href {\doibase 10.1103/PhysRevD.72.083506}
  {\bibfield  {journal} {\bibinfo  {journal} {Phys. Rev. D}\ }\textbf {\bibinfo
  {volume} {72}},\ \bibinfo {pages} {083506} (\bibinfo {year}
  {2005})}\BibitemShut {NoStop}%
\bibitem [{\citenamefont {Martins}\ and\ \citenamefont
  {Ach\'ucarro}(2008)}]{vos-monopoles}%
  \BibitemOpen
  \bibfield  {author} {\bibinfo {author} {\bibfnamefont {C.~J. A.~P.}\
  \bibnamefont {Martins}}\ and\ \bibinfo {author} {\bibfnamefont
  {A.}~\bibnamefont {Ach\'ucarro}},\ }\href {\doibase
  10.1103/PhysRevD.78.083541} {\bibfield  {journal} {\bibinfo  {journal} {Phys.
  Rev. D}\ }\textbf {\bibinfo {volume} {78}},\ \bibinfo {pages} {083541}
  (\bibinfo {year} {2008})}\BibitemShut {NoStop}%
\bibitem [{\citenamefont {Martins}\ \emph {et~al.}(2016)\citenamefont
  {Martins}, \citenamefont {Rybak}, \citenamefont {Avgoustidis},\ and\
  \citenamefont {Shellard}}]{extending-vos-walls}%
  \BibitemOpen
  \bibfield  {author} {\bibinfo {author} {\bibfnamefont {C.~J. A.~P.}\
  \bibnamefont {Martins}}, \bibinfo {author} {\bibfnamefont {I.~Y.}\
  \bibnamefont {Rybak}}, \bibinfo {author} {\bibfnamefont {A.}~\bibnamefont
  {Avgoustidis}}, \ and\ \bibinfo {author} {\bibfnamefont {E.~P.~S.}\
  \bibnamefont {Shellard}},\ }\href {\doibase 10.1103/PhysRevD.93.043534}
  {\bibfield  {journal} {\bibinfo  {journal} {Phys. Rev. D}\ }\textbf {\bibinfo
  {volume} {93}},\ \bibinfo {pages} {043534} (\bibinfo {year} {2016})},\
  \Eprint {http://arxiv.org/abs/1602.01322} {arXiv:1602.01322 [hep-ph]}
  \BibitemShut {NoStop}%
\bibitem [{\citenamefont {Correia}\ and\ \citenamefont
  {Martins}(2017)}]{gpu-implementation}%
  \BibitemOpen
  \bibfield  {author} {\bibinfo {author} {\bibfnamefont {J.~R. C. C.~C.}\
  \bibnamefont {Correia}}\ and\ \bibinfo {author} {\bibfnamefont {C.~J. A.~P.}\
  \bibnamefont {Martins}},\ }\href {\doibase 10.1103/PhysRevE.96.043310}
  {\bibfield  {journal} {\bibinfo  {journal} {Phys. Rev. E}\ }\textbf {\bibinfo
  {volume} {96}},\ \bibinfo {pages} {043310} (\bibinfo {year} {2017})},\
  \Eprint {http://arxiv.org/abs/1710.10420} {arXiv:1710.10420
  [physics.comp-ph]} \BibitemShut {NoStop}%
\bibitem [{\citenamefont {Correia}\ and\ \citenamefont
  {Martins}(2021{\natexlab{a}})}]{GPUstrings}%
  \BibitemOpen
  \bibfield  {author} {\bibinfo {author} {\bibfnamefont {J.~R. C. C.~C.}\
  \bibnamefont {Correia}}\ and\ \bibinfo {author} {\bibfnamefont {C.~J. A.~P.}\
  \bibnamefont {Martins}},\ }\href {\doibase 10.1016/j.ascom.2020.100438}
  {\bibfield  {journal} {\bibinfo  {journal} {Astron. Comput.}\ }\textbf
  {\bibinfo {volume} {34}},\ \bibinfo {pages} {100438} (\bibinfo {year}
  {2021}{\natexlab{a}})},\ \Eprint {http://arxiv.org/abs/2005.14454}
  {arXiv:2005.14454 [physics.comp-ph]} \BibitemShut {NoStop}%
\bibitem [{\citenamefont {Correia}\ and\ \citenamefont
  {Martins}(2021{\natexlab{b}})}]{Correia2}%
  \BibitemOpen
  \bibfield  {author} {\bibinfo {author} {\bibfnamefont {J.~R. C. C.~C.}\
  \bibnamefont {Correia}}\ and\ \bibinfo {author} {\bibfnamefont {C.~J. A.~P.}\
  \bibnamefont {Martins}},\ }\href {\doibase 10.1103/PhysRevD.104.063511}
  {\bibfield  {journal} {\bibinfo  {journal} {Phys. Rev. D}\ }\textbf {\bibinfo
  {volume} {104}},\ \bibinfo {pages} {063511} (\bibinfo {year}
  {2021}{\natexlab{b}})},\ \Eprint {http://arxiv.org/abs/2108.07513}
  {arXiv:2108.07513 [astro-ph.CO]} \BibitemShut {NoStop}%
\bibitem [{\citenamefont {Widrow}(1989)}]{Widrow}%
  \BibitemOpen
  \bibfield  {author} {\bibinfo {author} {\bibfnamefont {L.~M.}\ \bibnamefont
  {Widrow}},\ }\href {\doibase 10.1103/PhysRevD.40.1002} {\bibfield  {journal}
  {\bibinfo  {journal} {Phys. Rev. D}\ }\textbf {\bibinfo {volume} {40}},\
  \bibinfo {pages} {1002} (\bibinfo {year} {1989})}\BibitemShut {NoStop}%
\bibitem [{\citenamefont {Hill}\ and\ \citenamefont
  {Ross}(1988)}]{schizon-model}%
  \BibitemOpen
  \bibfield  {author} {\bibinfo {author} {\bibfnamefont {C.~T.}\ \bibnamefont
  {Hill}}\ and\ \bibinfo {author} {\bibfnamefont {G.~G.}\ \bibnamefont
  {Ross}},\ }\href {\doibase https://doi.org/10.1016/0370-2693(88)91583-3}
  {\bibfield  {journal} {\bibinfo  {journal} {Physics Letters B}\ }\textbf
  {\bibinfo {volume} {203}},\ \bibinfo {pages} {125 } (\bibinfo {year}
  {1988})}\BibitemShut {NoStop}%
\bibitem [{\citenamefont {Peyrard}\ and\ \citenamefont {Campbell}(1983)}]{SG1}%
  \BibitemOpen
  \bibfield  {author} {\bibinfo {author} {\bibfnamefont {M.}~\bibnamefont
  {Peyrard}}\ and\ \bibinfo {author} {\bibfnamefont {D.~K.}\ \bibnamefont
  {Campbell}},\ }\href@noop {} {\bibfield  {journal} {\bibinfo  {journal}
  {Physica D}\ }\textbf {\bibinfo {volume} {9}},\ \bibinfo {pages} {33}
  (\bibinfo {year} {1983})}\BibitemShut {NoStop}%
\bibitem [{\citenamefont {Babelon}\ and\ \citenamefont {Bernard}(1993)}]{SG2}%
  \BibitemOpen
  \bibfield  {author} {\bibinfo {author} {\bibfnamefont {O.}~\bibnamefont
  {Babelon}}\ and\ \bibinfo {author} {\bibfnamefont {D.}~\bibnamefont
  {Bernard}},\ }\href {\doibase 10.1016/0370-2693(93)91009-C} {\bibfield
  {journal} {\bibinfo  {journal} {Phys. Lett. B}\ }\textbf {\bibinfo {volume}
  {317}},\ \bibinfo {pages} {363} (\bibinfo {year} {1993})},\ \Eprint
  {http://arxiv.org/abs/hep-th/9309154} {arXiv:hep-th/9309154} \BibitemShut
  {NoStop}%
\bibitem [{\citenamefont {Kobayashi}\ and\ \citenamefont {Nitta}(2013)}]{SG3}%
  \BibitemOpen
  \bibfield  {author} {\bibinfo {author} {\bibfnamefont {M.}~\bibnamefont
  {Kobayashi}}\ and\ \bibinfo {author} {\bibfnamefont {M.}~\bibnamefont
  {Nitta}},\ }\href {\doibase 10.1103/PhysRevD.87.085003} {\bibfield  {journal}
  {\bibinfo  {journal} {Phys. Rev. D}\ }\textbf {\bibinfo {volume} {87}},\
  \bibinfo {pages} {085003} (\bibinfo {year} {2013})},\ \Eprint
  {http://arxiv.org/abs/1302.0989} {arXiv:1302.0989 [hep-th]} \BibitemShut
  {NoStop}%
\bibitem [{\citenamefont {Alzaleq}\ and\ \citenamefont
  {Manoranjan}(2021)}]{SG4}%
  \BibitemOpen
  \bibfield  {author} {\bibinfo {author} {\bibfnamefont {L.}~\bibnamefont
  {Alzaleq}}\ and\ \bibinfo {author} {\bibfnamefont {V.}~\bibnamefont
  {Manoranjan}},\ }\href {\doibase 10.1088/1402-4896/abe678} {\bibfield
  {journal} {\bibinfo  {journal} {Phys. Scripta}\ }\textbf {\bibinfo {volume}
  {96}},\ \bibinfo {pages} {055218} (\bibinfo {year} {2021})}\BibitemShut
  {NoStop}%
\bibitem [{\citenamefont {Campos}\ and\ \citenamefont {Mohammadi}(2021)}]{SG5}%
  \BibitemOpen
  \bibfield  {author} {\bibinfo {author} {\bibfnamefont {J.~a. G.~F.}\
  \bibnamefont {Campos}}\ and\ \bibinfo {author} {\bibfnamefont
  {A.}~\bibnamefont {Mohammadi}},\ }\href {\doibase 10.1007/JHEP09(2021)067}
  {\bibfield  {journal} {\bibinfo  {journal} {JHEP}\ }\textbf {\bibinfo
  {volume} {09}},\ \bibinfo {pages} {067} (\bibinfo {year} {2021})},\ \Eprint
  {http://arxiv.org/abs/2103.04908} {arXiv:2103.04908 [hep-th]} \BibitemShut
  {NoStop}%
\bibitem [{\citenamefont {Gomes}\ and\ \citenamefont {Simas}(2024)}]{SG6}%
  \BibitemOpen
  \bibfield  {author} {\bibinfo {author} {\bibfnamefont {A.~R.}\ \bibnamefont
  {Gomes}}\ and\ \bibinfo {author} {\bibfnamefont {F.~C.}\ \bibnamefont
  {Simas}},\ }\href@noop {} {\  (\bibinfo {year} {2024})},\ \Eprint
  {http://arxiv.org/abs/2410.10445} {arXiv:2410.10445 [hep-th]} \BibitemShut
  {NoStop}%
\bibitem [{\citenamefont {Chen}\ and\ \citenamefont {Luhrmann}(2024)}]{SG7}%
  \BibitemOpen
  \bibfield  {author} {\bibinfo {author} {\bibfnamefont {G.}~\bibnamefont
  {Chen}}\ and\ \bibinfo {author} {\bibfnamefont {J.}~\bibnamefont
  {Luhrmann}},\ }\href@noop {} {\  (\bibinfo {year} {2024})},\ \Eprint
  {http://arxiv.org/abs/2411.07004} {arXiv:2411.07004 [math.AP]} \BibitemShut
  {NoStop}%
\bibitem [{\citenamefont {Bountis}\ \emph {et~al.}(2024)\citenamefont
  {Bountis}, \citenamefont {Cantis\'an}, \citenamefont {Cuevas-Maraver},
  \citenamefont {Mac\'\i{}as-D\'\i{}az},\ and\ \citenamefont
  {Kevrekidis}}]{SG8}%
  \BibitemOpen
  \bibfield  {author} {\bibinfo {author} {\bibfnamefont {T.}~\bibnamefont
  {Bountis}}, \bibinfo {author} {\bibfnamefont {J.}~\bibnamefont {Cantis\'an}},
  \bibinfo {author} {\bibfnamefont {J.}~\bibnamefont {Cuevas-Maraver}},
  \bibinfo {author} {\bibfnamefont {J.~E.}\ \bibnamefont
  {Mac\'\i{}as-D\'\i{}az}}, \ and\ \bibinfo {author} {\bibfnamefont {P.~G.}\
  \bibnamefont {Kevrekidis}},\ }\href@noop {} {\  (\bibinfo {year} {2024})},\
  \Eprint {http://arxiv.org/abs/2411.18600} {arXiv:2411.18600 [nlin.PS]}
  \BibitemShut {NoStop}%
\bibitem [{\citenamefont {Zamolodchikov}(1986)}]{phi6-6}%
  \BibitemOpen
  \bibfield  {author} {\bibinfo {author} {\bibfnamefont {A.~B.}\ \bibnamefont
  {Zamolodchikov}},\ }\href@noop {} {\bibfield  {journal} {\bibinfo  {journal}
  {Sov. J. Nucl. Phys.}\ }\textbf {\bibinfo {volume} {44}},\ \bibinfo {pages}
  {529} (\bibinfo {year} {1986})}\BibitemShut {NoStop}%
\bibitem [{\citenamefont {Kim}\ \emph {et~al.}(1996)\citenamefont {Kim},
  \citenamefont {Maeda},\ and\ \citenamefont {Sakai}}]{phi6-8}%
  \BibitemOpen
  \bibfield  {author} {\bibinfo {author} {\bibfnamefont {Y.}~\bibnamefont
  {Kim}}, \bibinfo {author} {\bibfnamefont {K.-i.}\ \bibnamefont {Maeda}}, \
  and\ \bibinfo {author} {\bibfnamefont {N.}~\bibnamefont {Sakai}},\ }\href
  {\doibase 10.1016/S0550-3213(96)90151-0} {\bibfield  {journal} {\bibinfo
  {journal} {Nucl. Phys. B}\ }\textbf {\bibinfo {volume} {481}},\ \bibinfo
  {pages} {453} (\bibinfo {year} {1996})},\ \Eprint
  {http://arxiv.org/abs/gr-qc/9604030} {arXiv:gr-qc/9604030} \BibitemShut
  {NoStop}%
\bibitem [{\citenamefont {Arnold}\ and\ \citenamefont {Wright}(1997)}]{phi6-5}%
  \BibitemOpen
  \bibfield  {author} {\bibinfo {author} {\bibfnamefont {P.~B.}\ \bibnamefont
  {Arnold}}\ and\ \bibinfo {author} {\bibfnamefont {D.}~\bibnamefont
  {Wright}},\ }\href {\doibase 10.1103/PhysRevD.55.6274} {\bibfield  {journal}
  {\bibinfo  {journal} {Phys. Rev. D}\ }\textbf {\bibinfo {volume} {55}},\
  \bibinfo {pages} {6274} (\bibinfo {year} {1997})},\ \Eprint
  {http://arxiv.org/abs/hep-ph/9610226} {arXiv:hep-ph/9610226} \BibitemShut
  {NoStop}%
\bibitem [{\citenamefont {do~Amaral}(1998)}]{phi6-2}%
  \BibitemOpen
  \bibfield  {author} {\bibinfo {author} {\bibfnamefont {M.~G.}\ \bibnamefont
  {do~Amaral}},\ }\href {\doibase 10.1088/0954-3899/24/6/002} {\bibfield
  {journal} {\bibinfo  {journal} {J. Phys. G}\ }\textbf {\bibinfo {volume}
  {24}},\ \bibinfo {pages} {1061} (\bibinfo {year} {1998})}\BibitemShut
  {NoStop}%
\bibitem [{\citenamefont {Lu}\ \emph {et~al.}(1998)\citenamefont {Lu},
  \citenamefont {Ni},\ and\ \citenamefont {Wang}}]{phi6-7}%
  \BibitemOpen
  \bibfield  {author} {\bibinfo {author} {\bibfnamefont {W.-F.}\ \bibnamefont
  {Lu}}, \bibinfo {author} {\bibfnamefont {G.-J.}\ \bibnamefont {Ni}}, \ and\
  \bibinfo {author} {\bibfnamefont {Z.-G.}\ \bibnamefont {Wang}},\ }\href
  {\doibase 10.1088/0954-3899/24/3/016} {\bibfield  {journal} {\bibinfo
  {journal} {J. Phys. G}\ }\textbf {\bibinfo {volume} {24}},\ \bibinfo {pages}
  {673} (\bibinfo {year} {1998})}\BibitemShut {NoStop}%
\bibitem [{\citenamefont {Flores}\ \emph {et~al.}(1999)\citenamefont {Flores},
  \citenamefont {Ramos},\ and\ \citenamefont {Svaiter}}]{phi6-3}%
  \BibitemOpen
  \bibfield  {author} {\bibinfo {author} {\bibfnamefont {G.~H.}\ \bibnamefont
  {Flores}}, \bibinfo {author} {\bibfnamefont {R.~O.}\ \bibnamefont {Ramos}}, \
  and\ \bibinfo {author} {\bibfnamefont {N.~F.}\ \bibnamefont {Svaiter}},\
  }\href {\doibase 10.1142/S0217751X99001718} {\bibfield  {journal} {\bibinfo
  {journal} {Int. J. Mod. Phys. A}\ }\textbf {\bibinfo {volume} {14}},\
  \bibinfo {pages} {3715} (\bibinfo {year} {1999})},\ \Eprint
  {http://arxiv.org/abs/hep-th/9903009} {arXiv:hep-th/9903009} \BibitemShut
  {NoStop}%
\bibitem [{\citenamefont {Joy}\ and\ \citenamefont {Kuriakose}(2003)}]{phi6-4}%
  \BibitemOpen
  \bibfield  {author} {\bibinfo {author} {\bibfnamefont {M.}~\bibnamefont
  {Joy}}\ and\ \bibinfo {author} {\bibfnamefont {V.~C.}\ \bibnamefont
  {Kuriakose}},\ }\href {\doibase 10.1142/S0217732303009502} {\bibfield
  {journal} {\bibinfo  {journal} {Mod. Phys. Lett. A}\ }\textbf {\bibinfo
  {volume} {18}},\ \bibinfo {pages} {937} (\bibinfo {year} {2003})},\ \Eprint
  {http://arxiv.org/abs/hep-th/0102177} {arXiv:hep-th/0102177} \BibitemShut
  {NoStop}%
\bibitem [{\citenamefont {Bergner}\ and\ \citenamefont
  {Bettencourt}(2003)}]{phi6-1}%
  \BibitemOpen
  \bibfield  {author} {\bibinfo {author} {\bibfnamefont {Y.}~\bibnamefont
  {Bergner}}\ and\ \bibinfo {author} {\bibfnamefont {L.~M.~A.}\ \bibnamefont
  {Bettencourt}},\ }\href {\doibase 10.1103/PhysRevD.68.025014} {\bibfield
  {journal} {\bibinfo  {journal} {Phys. Rev. D}\ }\textbf {\bibinfo {volume}
  {68}},\ \bibinfo {pages} {025014} (\bibinfo {year} {2003})},\ \Eprint
  {http://arxiv.org/abs/hep-ph/0206053} {arXiv:hep-ph/0206053} \BibitemShut
  {NoStop}%
\bibitem [{\citenamefont {Bodeker}\ \emph {et~al.}(2005)\citenamefont
  {Bodeker}, \citenamefont {Fromme}, \citenamefont {Huber},\ and\ \citenamefont
  {Seniuch}}]{phi6-0}%
  \BibitemOpen
  \bibfield  {author} {\bibinfo {author} {\bibfnamefont {D.}~\bibnamefont
  {Bodeker}}, \bibinfo {author} {\bibfnamefont {L.}~\bibnamefont {Fromme}},
  \bibinfo {author} {\bibfnamefont {S.~J.}\ \bibnamefont {Huber}}, \ and\
  \bibinfo {author} {\bibfnamefont {M.}~\bibnamefont {Seniuch}},\ }\href
  {\doibase 10.1088/1126-6708/2005/02/026} {\bibfield  {journal} {\bibinfo
  {journal} {JHEP}\ }\textbf {\bibinfo {volume} {02}},\ \bibinfo {pages} {026}
  (\bibinfo {year} {2005})},\ \Eprint {http://arxiv.org/abs/hep-ph/0412366}
  {arXiv:hep-ph/0412366} \BibitemShut {NoStop}%
\bibitem [{\citenamefont {Dorey}\ \emph {et~al.}(2011)\citenamefont {Dorey},
  \citenamefont {Mersh}, \citenamefont {Romanczukiewicz},\ and\ \citenamefont
  {Shnir}}]{Dorey0}%
  \BibitemOpen
  \bibfield  {author} {\bibinfo {author} {\bibfnamefont {P.}~\bibnamefont
  {Dorey}}, \bibinfo {author} {\bibfnamefont {K.}~\bibnamefont {Mersh}},
  \bibinfo {author} {\bibfnamefont {T.}~\bibnamefont {Romanczukiewicz}}, \ and\
  \bibinfo {author} {\bibfnamefont {Y.}~\bibnamefont {Shnir}},\ }\href
  {\doibase 10.1103/PhysRevLett.107.091602} {\bibfield  {journal} {\bibinfo
  {journal} {Phys. Rev. Lett.}\ }\textbf {\bibinfo {volume} {107}},\ \bibinfo
  {pages} {091602} (\bibinfo {year} {2011})},\ \Eprint
  {http://arxiv.org/abs/1101.5951} {arXiv:1101.5951 [hep-th]} \BibitemShut
  {NoStop}%
\bibitem [{\citenamefont {Weigel}(2014)}]{Weigel}%
  \BibitemOpen
  \bibfield  {author} {\bibinfo {author} {\bibfnamefont {H.}~\bibnamefont
  {Weigel}},\ }\href {\doibase 10.1088/1742-6596/482/1/012045} {\bibfield
  {journal} {\bibinfo  {journal} {J. Phys. Conf. Ser.}\ }\textbf {\bibinfo
  {volume} {482}},\ \bibinfo {pages} {012045} (\bibinfo {year} {2014})},\
  \Eprint {http://arxiv.org/abs/1309.6607} {arXiv:1309.6607 [nlin.PS]}
  \BibitemShut {NoStop}%
\bibitem [{\citenamefont {Christ}\ and\ \citenamefont {Lee}(1975)}]{chris-lee}%
  \BibitemOpen
  \bibfield  {author} {\bibinfo {author} {\bibfnamefont {N.~H.}\ \bibnamefont
  {Christ}}\ and\ \bibinfo {author} {\bibfnamefont {T.~D.}\ \bibnamefont
  {Lee}},\ }\href {\doibase 10.1103/PhysRevD.12.1606} {\bibfield  {journal}
  {\bibinfo  {journal} {Phys. Rev. D}\ }\textbf {\bibinfo {volume} {12}},\
  \bibinfo {pages} {1606} (\bibinfo {year} {1975})}\BibitemShut {NoStop}%
\bibitem [{\citenamefont {Demirkaya}\ \emph {et~al.}(2017)\citenamefont
  {Demirkaya}, \citenamefont {Decker}, \citenamefont {Kevrekidis},
  \citenamefont {Christov},\ and\ \citenamefont {Saxena}}]{Demirkaya}%
  \BibitemOpen
  \bibfield  {author} {\bibinfo {author} {\bibfnamefont {A.}~\bibnamefont
  {Demirkaya}}, \bibinfo {author} {\bibfnamefont {R.}~\bibnamefont {Decker}},
  \bibinfo {author} {\bibfnamefont {P.~G.}\ \bibnamefont {Kevrekidis}},
  \bibinfo {author} {\bibfnamefont {I.~C.}\ \bibnamefont {Christov}}, \ and\
  \bibinfo {author} {\bibfnamefont {A.}~\bibnamefont {Saxena}},\ }\href
  {\doibase 10.1007/JHEP12(2017)071} {\bibfield  {journal} {\bibinfo  {journal}
  {JHEP}\ }\textbf {\bibinfo {volume} {12}},\ \bibinfo {pages} {071} (\bibinfo
  {year} {2017})},\ \Eprint {http://arxiv.org/abs/1706.01193} {arXiv:1706.01193
  [nlin.PS]} \BibitemShut {NoStop}%
\bibitem [{\citenamefont {Dorey}\ \emph {et~al.}(2023)\citenamefont {Dorey},
  \citenamefont {Gorina}, \citenamefont {Roma\'nczukiewicz},\ and\
  \citenamefont {Shnir}}]{Dorey}%
  \BibitemOpen
  \bibfield  {author} {\bibinfo {author} {\bibfnamefont {P.}~\bibnamefont
  {Dorey}}, \bibinfo {author} {\bibfnamefont {A.}~\bibnamefont {Gorina}},
  \bibinfo {author} {\bibfnamefont {T.}~\bibnamefont {Roma\'nczukiewicz}}, \
  and\ \bibinfo {author} {\bibfnamefont {Y.}~\bibnamefont {Shnir}},\ }\href
  {\doibase 10.1007/JHEP09(2023)045} {\bibfield  {journal} {\bibinfo  {journal}
  {JHEP}\ }\textbf {\bibinfo {volume} {09}},\ \bibinfo {pages} {045} (\bibinfo
  {year} {2023})},\ \Eprint {http://arxiv.org/abs/2304.11710} {arXiv:2304.11710
  [hep-th]} \BibitemShut {NoStop}%
\bibitem [{\citenamefont {Correia}\ and\ \citenamefont
  {Martins}(2020{\natexlab{a}})}]{MCMC-calib}%
  \BibitemOpen
  \bibfield  {author} {\bibinfo {author} {\bibfnamefont {J.~R. C. C.~C.}\
  \bibnamefont {Correia}}\ and\ \bibinfo {author} {\bibfnamefont {C.~J. A.~P.}\
  \bibnamefont {Martins}},\ }\href {\doibase 10.1103/PhysRevD.102.043503}
  {\bibfield  {journal} {\bibinfo  {journal} {Phys. Rev. D}\ }\textbf {\bibinfo
  {volume} {102}},\ \bibinfo {pages} {043503} (\bibinfo {year}
  {2020}{\natexlab{a}})},\ \Eprint {http://arxiv.org/abs/2007.12008}
  {arXiv:2007.12008 [astro-ph.CO]} \BibitemShut {NoStop}%
\bibitem [{\citenamefont {Press}\ \emph {et~al.}(1989)\citenamefont {Press},
  \citenamefont {Ryden},\ and\ \citenamefont
  {Spergel}}]{dynamical-evolution-of-domain-walls}%
  \BibitemOpen
  \bibfield  {author} {\bibinfo {author} {\bibfnamefont {W.~H.}\ \bibnamefont
  {Press}}, \bibinfo {author} {\bibfnamefont {B.~S.}\ \bibnamefont {Ryden}}, \
  and\ \bibinfo {author} {\bibfnamefont {D.~N.}\ \bibnamefont {Spergel}},\
  }\href {\doibase 10.1086/168151} {\bibfield  {journal} {\bibinfo  {journal}
  {Astrophys. J.}\ }\textbf {\bibinfo {volume} {347}},\ \bibinfo {pages} {590}
  (\bibinfo {year} {1989})}\BibitemShut {NoStop}%
\bibitem [{\citenamefont {Munshi}(2012)}]{openCL12}%
  \BibitemOpen
  \bibfield  {author} {\bibinfo {author} {\bibfnamefont {A.}~\bibnamefont
  {Munshi}},\ }\href@noop {} {\emph {\bibinfo {title} {OpenCL 1.2
  Specification}}},\ \bibinfo {organization} {Khronos Group Working Group},\
  \bibinfo {edition} {19th}\ ed. (\bibinfo {year} {2012})\BibitemShut {NoStop}%
\bibitem [{\citenamefont {Leite}\ and\ \citenamefont
  {Martins}(2011)}]{scaling-properties-of-domain-walls}%
  \BibitemOpen
  \bibfield  {author} {\bibinfo {author} {\bibfnamefont {A.~M.~M.}\
  \bibnamefont {Leite}}\ and\ \bibinfo {author} {\bibfnamefont {C.~J. A.~P.}\
  \bibnamefont {Martins}},\ }\href {\doibase 10.1103/PhysRevD.84.103523}
  {\bibfield  {journal} {\bibinfo  {journal} {Phys. Rev. D}\ }\textbf {\bibinfo
  {volume} {84}},\ \bibinfo {pages} {103523} (\bibinfo {year} {2011})},\
  \Eprint {http://arxiv.org/abs/1110.3486} {arXiv:1110.3486 [hep-ph]}
  \BibitemShut {NoStop}%
\bibitem [{\citenamefont {Correia}\ \emph {et~al.}(2014)\citenamefont
  {Correia}, \citenamefont {Leite},\ and\ \citenamefont {Martins}}]{biases1}%
  \BibitemOpen
  \bibfield  {author} {\bibinfo {author} {\bibfnamefont {J.~R. C. C.~C.}\
  \bibnamefont {Correia}}, \bibinfo {author} {\bibfnamefont {I.~S. C.~R.}\
  \bibnamefont {Leite}}, \ and\ \bibinfo {author} {\bibfnamefont {C.~J. A.~P.}\
  \bibnamefont {Martins}},\ }\href {\doibase 10.1103/PhysRevD.90.023521}
  {\bibfield  {journal} {\bibinfo  {journal} {Phys. Rev. D}\ }\textbf {\bibinfo
  {volume} {90}},\ \bibinfo {pages} {023521} (\bibinfo {year} {2014})},\
  \Eprint {http://arxiv.org/abs/1407.3905} {arXiv:1407.3905 [hep-ph]}
  \BibitemShut {NoStop}%
\bibitem [{\citenamefont {Correia}\ \emph {et~al.}(2018)\citenamefont
  {Correia}, \citenamefont {Leite},\ and\ \citenamefont {Martins}}]{biases2}%
  \BibitemOpen
  \bibfield  {author} {\bibinfo {author} {\bibfnamefont {J.~R. C. C.~C.}\
  \bibnamefont {Correia}}, \bibinfo {author} {\bibfnamefont {I.~S. C.~R.}\
  \bibnamefont {Leite}}, \ and\ \bibinfo {author} {\bibfnamefont {C.~J. A.~P.}\
  \bibnamefont {Martins}},\ }\href {\doibase 10.1103/PhysRevD.97.083521}
  {\bibfield  {journal} {\bibinfo  {journal} {Phys. Rev. D}\ }\textbf {\bibinfo
  {volume} {97}},\ \bibinfo {pages} {083521} (\bibinfo {year}
  {2018})}\BibitemShut {NoStop}%
\bibitem [{\citenamefont {Correia}\ and\ \citenamefont
  {Martins}(2020{\natexlab{b}})}]{Cooledj}%
  \BibitemOpen
  \bibfield  {author} {\bibinfo {author} {\bibfnamefont {J.~R. C. C.~C.}\
  \bibnamefont {Correia}}\ and\ \bibinfo {author} {\bibfnamefont {C.~J. A.~P.}\
  \bibnamefont {Martins}},\ }\href {\doibase 10.1103/PhysRevD.102.043503}
  {\bibfield  {journal} {\bibinfo  {journal} {Phys. Rev. D}\ }\textbf {\bibinfo
  {volume} {102}},\ \bibinfo {pages} {043503} (\bibinfo {year}
  {2020}{\natexlab{b}})},\ \Eprint {http://arxiv.org/abs/2007.12008}
  {arXiv:2007.12008 [astro-ph.CO]} \BibitemShut {NoStop}%
\bibitem [{\citenamefont {Oliveira}\ and\ \citenamefont
  {Martins}(2015)}]{Oliveira:2015xfa}%
  \BibitemOpen
  \bibfield  {author} {\bibinfo {author} {\bibfnamefont {M.~F.}\ \bibnamefont
  {Oliveira}}\ and\ \bibinfo {author} {\bibfnamefont {C.~J. A.~P.}\
  \bibnamefont {Martins}},\ }\href {\doibase 10.1103/PhysRevD.91.043527}
  {\bibfield  {journal} {\bibinfo  {journal} {Phys. Rev. D}\ }\textbf {\bibinfo
  {volume} {91}},\ \bibinfo {pages} {043527} (\bibinfo {year} {2015})},\
  \Eprint {http://arxiv.org/abs/1503.00234} {arXiv:1503.00234 [hep-ph]}
  \BibitemShut {NoStop}%
\end{thebibliography}%
\end{document}